\documentclass[sigconf]{acmart}

\renewcommand\footnotetextcopyrightpermission[1]{}
\newcommand{\sysname}{\textsc{Papers-to-Posts}}
\usepackage{caption}
\usepackage{subcaption}
\usepackage{listings}

\AtBeginDocument{%
  }

\begin{document}

\title{Papers-to-Posts: Supporting Detailed Long-Document Summarization with an Interactive LLM-Powered Source Outline}

\author{Marissa Radensky}
\authornote{Work done during internship at the Allen Institute for AI.}
\email{radensky@cs.washington.edu}
\orcid{0000-0002-5045-8269}
\affiliation{%
  \institution{University of Washington}
  \city{Seattle}
  \state{WA}
  \country{USA}
}
\author{Daniel S. Weld}
\email{danw@allenai.org}
\orcid{0000-0002-3255-0109}
\affiliation{%
  \institution{Allen Institute for AI \& University of Washington}
  \city{Seattle}
  \state{WA}
  \country{USA}
}
\author{Joseph Chee Chang}
\email{josephc@allenai.org}
\orcid{0000-0002-0798-4351}
\affiliation{%
  \institution{Allen Institute for AI}
  \city{Seattle}
  \state{WA}
  \country{USA}
}
\author{Pao Siangliulue}
\email{paos@allenai.org}
\orcid{}
\affiliation{%
  \institution{Allen Institute for AI}
  \city{Seattle}
  \state{WA}
  \country{USA}
}
\author{Jonathan Bragg}
\email{jbragg@allenai.org}
\orcid{0000-0001-5460-9047}
\affiliation{%
  \institution{Allen Institute for AI}
  \city{Seattle}
  \state{WA}
  \country{USA}
}

\renewcommand{\shortauthors}{Radensky et al.}
\begin{abstract}
Compressing long and technical documents (e.g., >10 pages) into shorter-form articles (e.g., <2 pages) is critical for communicating information to different audiences, for example, blog posts of scientific research paper or legal briefs of dense court proceedings. 
While large language models (LLMs) are powerful tools for condensing large amounts of text, current interfaces to these models lack support for understanding and controlling what content is included in a detailed summarizing article.
Such capability is especially important for detail- and technical-oriented domains, in which tactical selection and coherent synthesis of key details is critical for effective communication to the target audience. 
For this, we present \textit{interactive reverse source outlines}, a novel mechanism for controllable long-form summarization featuring outline bullet points with automatic point selections that the user can iteratively adjust to obtain an article with the desired content coverage. 
We implement this mechanism in \sysname, a new LLM-powered system for authoring research-paper blog posts. 
Through a within-subjects lab study (n=20) and a between-subjects deployment study (n=37 blog posts, 26 participants), we compare \sysname\ to a strong baseline tool that provides an LLM-generated draft and access to free-form prompting. 
Under time constraints, \sysname\ significantly increases writer satisfaction with blog post quality, particularly with respect to content coverage. 
Furthermore, quantitative results showed an increase in editing power (change in text for an amount of time or writing actions) while using \sysname, and qualitative results showed that participants found incorporating key research-paper insights in their blog posts easier while using \sysname.
\end{abstract}
\maketitle

\begin{figure}[H]
  \includegraphics[width=.8\columnwidth]{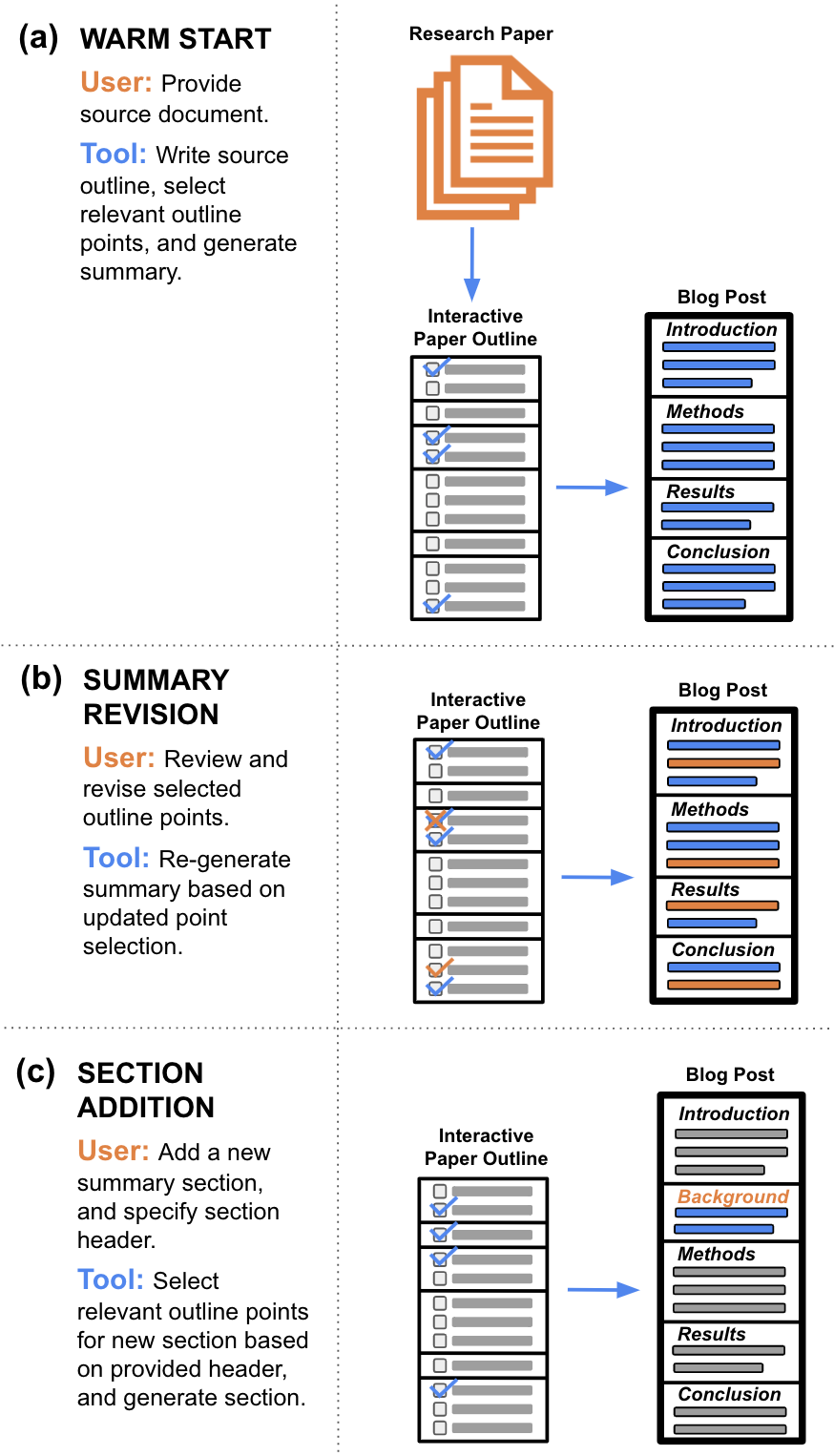}
  \centering
  \caption{\textit{Interactive reverse source outlines} in the \sysname\ system.
  Users input a long-form source document (research paper) and the system provides (a) a WARM START for the summarization task by generating a reverse document outline. The system then produces a draft summary article (blog post) with sections based on outline bullet points (and associated source paragraphs) selected by the LLM. Given the draft, users can perform two main actions: (b) SUMMARY REVISION, where users adjust the system's bullet point selection for the summary, triggering system re-generation of the summary; and (c) SECTION ADDITION, where users provide a header for a new desired section, based on which the system selects bullet points and generates a draft of the section content. Users may also edit the draft manually.}
  \label{fig:teaser}
  \Description{The figure has three rows and two columns. Each row describes an action supported by the interactive reverse source outlines. The first column describes the action, and the second column illustrates the action. The first row, labeled "a," is titled "Warm Start." It says the following. "User [in orange]: Provide source document. Tool [in blue]: Write source outline, select relevant outline points, and generate summary." Meanwhile, the illustration shows a set of orange papers labeled "research paper." A blue arrow points from the research paper to an image (labeled "interactive paper outline") of an outline with sets of 2 or 3 bullet points, some checked off with a blue check-mark and some not. A blue arrow points from the outline to an image labeled "blog post." It contains four sections labeled "introduction," "methods," "results," and "conclusion." There are blue lines in all four section of the blog post. The second row, labeled "b," is titled "Summary Revision." It says the following. "User [in orange]: Review and revise selected outline points. Tool [in blue]: Re-generate summary based on updated point selection." The illustration shows the interactive paper outline updated so that an orange "X" covers one of the blue checkmarks, and an orange checkmark is added for one of the bullet points previously unchecked. A blue arrow points from the outline to the blog post, which is updated to have some of its blue lines turned orange. The length of the conclusion section also changed. The third row, labeled "c," is titled "Section Addition." It says the following. "User [in orange]: Add a new summary section header. Tool [in blue]: Select relevant outline points for new section based on provided header and generate section. In the illustration, the outline is updated to have no orange marks and blue checkmarks on different bullet points than previously. A blue arrow points from the outline to the blog post, which contains a new section added as the second section. It is titled "Background" in orange and its lines are in blue. The other blog post sections' lines representing writing are in gray.}
\end{figure}

\section{Introduction}
Condensing long and technical documents into shorter forms for different audiences is crucial for many applications, from blog posts about scientific papers to legal briefs about court proceedings.
In science, communicating scientific ideas from research papers to broader audiences increases awareness and impact \cite{jucan2014power,burns2003science}. Scientists often seek different ways to publicize their work, such as through tweets or blog posts~\cite{Gero2021WhatMT,Long2023TweetorialHG,Williams2022AnHR, holmberg2014disciplinary}; indeed, tweeting about research papers has been shown to raise their citation count \cite{luc2021does}.
While some tools have been created to help author short-form derivatives like tweets~\cite{Long2023TweetorialHG}, significant challenges remain for scientists seeking to create more detailed derivatives like blog posts. Summarizing long documents (10k+ words) in detailed articles (500-1000 words) takes considerable time and effort, even with LLM support. Detailed articles require the writer to identify large amounts of relevant content from the source document. Given a long document, identifying all the important information to include can be daunting. While LLMs have impressive abilities to synthesize information from a long piece of text, what the LLM chooses to leave out is unclear to the user, and quickly iterating on the selected content is difficult. In domains like science, law, and business, attention to detail is critical; a summary missing key content could lead to misrepresented work and misinformed readers.

While some prior work has investigated fully automatic summarization of long documents \cite{koh2022empirical},
a mixed-initiative approach allows users to have more control over their summaries, which is important in detail-oriented domains like scientific research. Prior work in human-AI text summarization has often focused on helping create \textit{short-form} summaries around a paragraph in length, which are easier to control through selection of a few source-document sentences \cite{zhang2011idvs,wang2023podreels}. However, when writing longer summaries, selecting source content at the sentence level becomes unwieldy. Other work in human-AI text summarization has looked into other interactions that may support longer-form summaries, but these works involve interactions detached from the source narrative, like post-editing of an automatically generated summary \cite{lim2024co,lai2022exploration,moramarco2021preliminary} or selecting concepts to include in the summary \cite{zhang2023concepteva,bayatmakou2022interactive,avinesh2017joint, avinesh2018sherlock, ding2023harnessing}. Without sufficient context from the source document, making sure that important details are not missing in a longer summary is difficult.

Inspired by prior work on LLM-supported reverse outlining for general writing~\cite{dang2022beyond}, we present
\textit{interactive reverse source outlines}, a new interaction mechanism for controlling summary content selection with an interactive LLM-generated source outline, which maintains the source’s narrative structure and simplifies the task of selecting relevant content (Figure~\ref{fig:teaser}).
To the best of our knowledge, this is the first
mixed-initiative approach for \textit{detailed} and \textit{grounded} summarization of long documents (Figure \ref{fig:rwgraph}). 

We implement this mechanism in \sysname, a mixed-initiative system that helps researchers write research-paper blog posts. The system follows a plan-draft-revise writing workflow \cite{Flower1981ACP}. In the planning phase, the user may review the system's initial blog post draft and adjust the system's selection of key bullet points within an LLM-generated paper outline. In the drafting phase, the user may customize preset instructions (e.g., include a hook, write one paragraph) to the LLM for each blog post section. The LLM then uses the selected bullet points and adjusted instructions to generate new text for the given section. In the revising phase, the user may further refine text with LLM-powered macros that encapsulate frequent editing transformations for scientific blog post writing (e.g., to reduce jargon, to better capture attention).

We evaluated \sysname\ through two studies in which participants wrote blog posts for their own papers. In these studies, we compared our tool to a strong baseline tool--an LLM-generated draft and access to free-form LLM prompting. In a within-subjects lab study (N=20 participants) in which researchers wrote blog posts for two papers (one per tool), we found that \sysname\ led to significantly higher writer satisfaction with the final blog post, and participants were particularly more satisfied with the content covered in the final blog post. Furthermore, participants had higher confidence that all essential information was in the final blog post when using \sysname. 
To study \sysname\ in a more realistic setting, we also conducted an unmonitored, between-subjects deployment study (N=37 blog posts, 26 participants). 
Across both studies, participants found \sysname\ more helpful. They found it easier to incorporate research-paper content in the blog post and to iterate on that content. Moreover, participants demonstrated increased editing power (change to the provided blog post draft for a given amount of time or for a given number of writing actions) without an increase in cognitive load.

In summary, we make the following contributions:
\begin{itemize}
    \item \textit{Interactive reverse source outlines}, a novel mechanism for more user control in human-LLM detailed summarization of long documents, consisting of an LLM-generated reverse outline of the source document with pre-selected bullet points for an initial draft of the summary, with which the writer iteratively interacts to control content selection
    \item \sysname, a tool that implements this mechanism for the purpose of writing research-paper blog posts, a common long-document summarization task that requires attention to detail
    \item Findings from a within-subjects lab study (N=20 participants) and a between-subjects deployment study (N=26 participants) showing that \sysname...
    \begin{itemize}
        \item Under time constraints, significantly increases writers’ satisfaction with their summaries’ quality (primarily with respect to content coverage)
        \item Is considered more helpful by writers, making incorporating source content in a detailed summary easier and iterating on the summary easier, with increased editing power (change in writing within an amount of time or writing actions) and no increase in cognitive load
    \end{itemize}
\end{itemize}

\section{Related Work}
\subsection{Human-LLM Text Summarization}
\label{sec:haiTextSummarization}
\begin{figure}[tb]
  \includegraphics[width=\columnwidth]{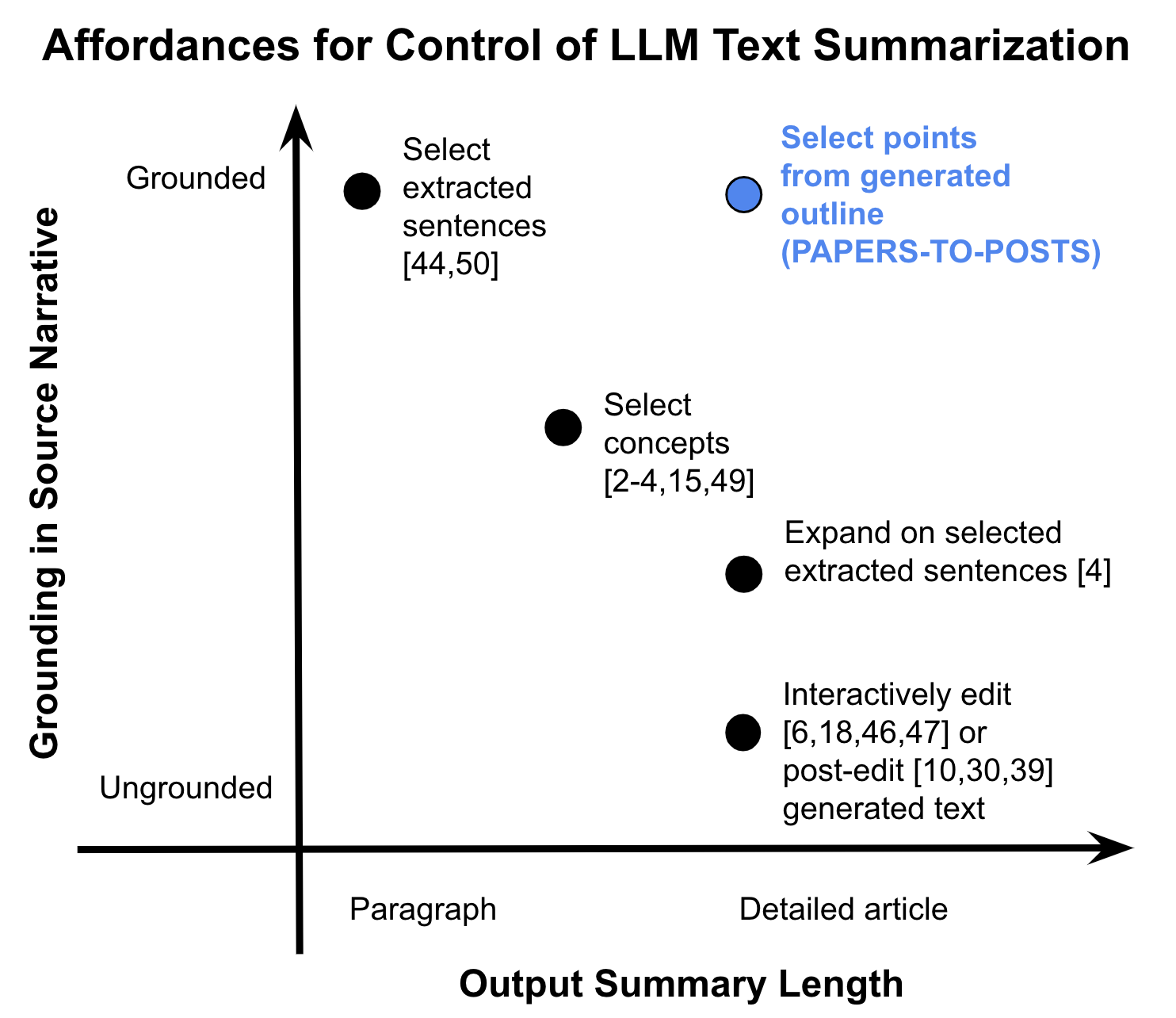}
  \centering
  \caption{Comparing interactive reverse source outlines in  \sysname\ (blue, upper right) with other affordances for control of LLM summarization, in terms of 1) how grounded the affordance is in the source narrative and 2) the length of the output summary that is supported. \sysname\ is the first LLM-powered tool to provide writers with highly grounded control over a detailed summary.}
  \label{fig:rwgraph}
  \Description{A scatter plot titled "Affordances for Control of LLM Text Summarization." The x-axis is an arrow pointing right, titled "output summary length." The left side of the arrow is labeled "paragraph" and the right side is labeled "detailed article." The y-axis is an arrowing pointing upward, titled "grounding in source narrative." The bottom of the arrow is labeled "ungrounded," and the top is labeled "grounded." The scatterplot has five dots: 1) a dot in the upper left labeled "select extracted sentences [44,50]"; 2) a dot near the center labeled "select concepts [2-4,15,49]"; 3) a dot in the bottom right labeled "interactively edit [6,18,46,47] or post-edit [10,30,39] generated text"; 4) a dot around the middle right labeled "expand on selected extracted sentences [4]"; and 5) a dot in the upper right labeled "select points from generated outline (PAPERS-TO-POSTS)."}
\end{figure}

Although there is much work on automatic text summarization \cite{koh2022empirical,el2021automatic}, issues remain with the output summaries, such as inaccuracies and style misalignment \cite{lu2023hybrid,koh2022empirical}. To address these issues, other works have looked into supporting humans in working together with LLMs to generate a summary \cite{cheng2022mapping}. These tools may be divided into groups in terms of 1) the summary length that they support and 2) how grounded the interaction is in the context of the source document (Figure \ref{fig:rwgraph}).

Some prior work in human-LLM text summarization supports an interaction in which the human selects sentences extracted directly from the source \cite{wang2023podreels,zhang2011idvs} (Figure \ref{fig:rwgraph}, top left). For a short summary, identifying the few key sentences from the source to include may be manageable, but for a detailed summary, this would require considerable time and effort. One work enables more detailed summaries by generating expanded text based on the user selecting a few source sentences~\cite{bayatmakou2022interactive} (Figure \ref{fig:rwgraph}, middle right), but what source content has been included versus neglected then becomes unclear.
Other prior work provides users with automatically generated summaries that they can adjust through interactive editing (e.g., via ratings, gaze) \cite{bohn2021hone,gao2020preference,xie2023interactive,yan2011summarize} or post-editing \cite{moramarco2021preliminary,cai2022generation,lai2022exploration} (Figure \ref{fig:rwgraph}, bottom right). While these interactions are flexible in terms of the length of summary generated, they do not support users in understanding what source content has or has not been represented.
Yet another group of prior works strikes a balance between generating longer summaries and grounding them in the source document, by having the user select concepts or keywords to include in the summary \cite{zhang2023concepteva,bayatmakou2022interactive,avinesh2017joint, avinesh2018sherlock, ding2023harnessing} (Figure \ref{fig:rwgraph}, center). Concepts are easier to sort through than full sentences, which makes generating a longer summary more feasible, and selecting concepts provides users some sense of what source content is utilized overall. Nevertheless, concepts are high-level and abstract, limiting their utility for generating detailed summaries and easily identifying what source details are and are not included. 
Our novel interactive reverse source outlines, implemented in \sysname\ (Figure \ref{fig:rwgraph}, top right), support both grounded and long-form summarization by allowing users to select outline bullet points directly tied to the source document in order to generate a detailed summary.

\subsection{Outlines in Human-LLM Writing}
There are some works in human-LLM writing that have utilized LLM-generated outlines for writing. Some focus on outlining output content, rather than outlining input content for summarization \cite{zhang2023visar,lu2024corporate}. For instance, Zhang et al. introduce the tool VISAR, a human-AI argumentative writing assistant that allows users to iterate on their draft with adjustments to a visual outline of the draft \cite{zhang2023visar}. Other works focus on outlining input content (as we do), but the outline is not utilized to synthesize a detailed summary of the input \cite{petridis2023anglekindling,yu2024bnotehelper}. As an example, AngleKindling generates a few main points from an input press release as a quick overview for journalists to then ideate about potential angles for reporting on the press release \cite{petridis2023anglekindling}. However, there is no support for interacting with the main points to generate a summary of the press release.

Most related to our work, Dang et al. present a text editor that generates summaries of each paragraph written in the editor to comprise a reverse outline of the writing \cite{dang2022beyond}. The summaries then become affordances to help writers think about how to revise their writing. Our work also generates a reverse outline, but it is an outline of the \textit{input} document for users to interactively guide summary generation. For summarizing a long source document, an outline is particularly useful, as reviewing the entire source itself requires substantial time and effort.

\subsection{Human-LLM Scientific Writing}
Several works have investigated how to support human-LLM writing. Most of these works focus on tools for creative writing \cite{Singh2022ASS,Chung2022TaleBrushSS,mirowski2023co}, but a few cater to scientific writing. A few works have investigated human-LLM scientific writing for broader science communication tasks that do not involve a source document. Gero et al. present a system for generating scientific ``sparks,'' or inspiring sentences \cite{gero2022sparks}. Long et al. examine how LLM scaffolding can help people generate relatable hooks for complex scientific topics \cite{Long2023TweetorialHG}, and Kim et al. investigate how science writers can generate extended metaphors for scientific ideas with the help of an LLM \cite{Kim2023MetaphorianLL}. The few prior works that have explored human-AI scientific summarization \cite{lim2024co,zhang2023concepteva,bayatmakou2022interactive} do not support fully grounded and detailed summaries like \sysname, as described in Section \ref{sec:haiTextSummarization}. Also of note, Google's tool NotebookLM for general human-AI document summarization does not permit any content selection \cite{notebooklm}. 


\subsection{Scientific Blog Posts}
Scientists can engage with broader communities through various channels, including press releases, magazines, journals, and tweets \cite{august2020writing,gero2022sparks}. This work focuses on scientific blog posts, which offer increased recognition and transparency of one's work, editorial freedom, and public engagement \cite{jarreau2015all,blanchard20113,zou2019reworking}. There are several forms of scientific blog posts, such as academic commentary, free commentary, and mediation of research to laypersons \cite{mahrt2014science}. Here, we focus on blog posts with the primary goal of communicating information from a research paper to people unfamiliar with the paper's specific research topic.

\begin{figure*}[htb]
    \centering
    \includegraphics[width=\textwidth]{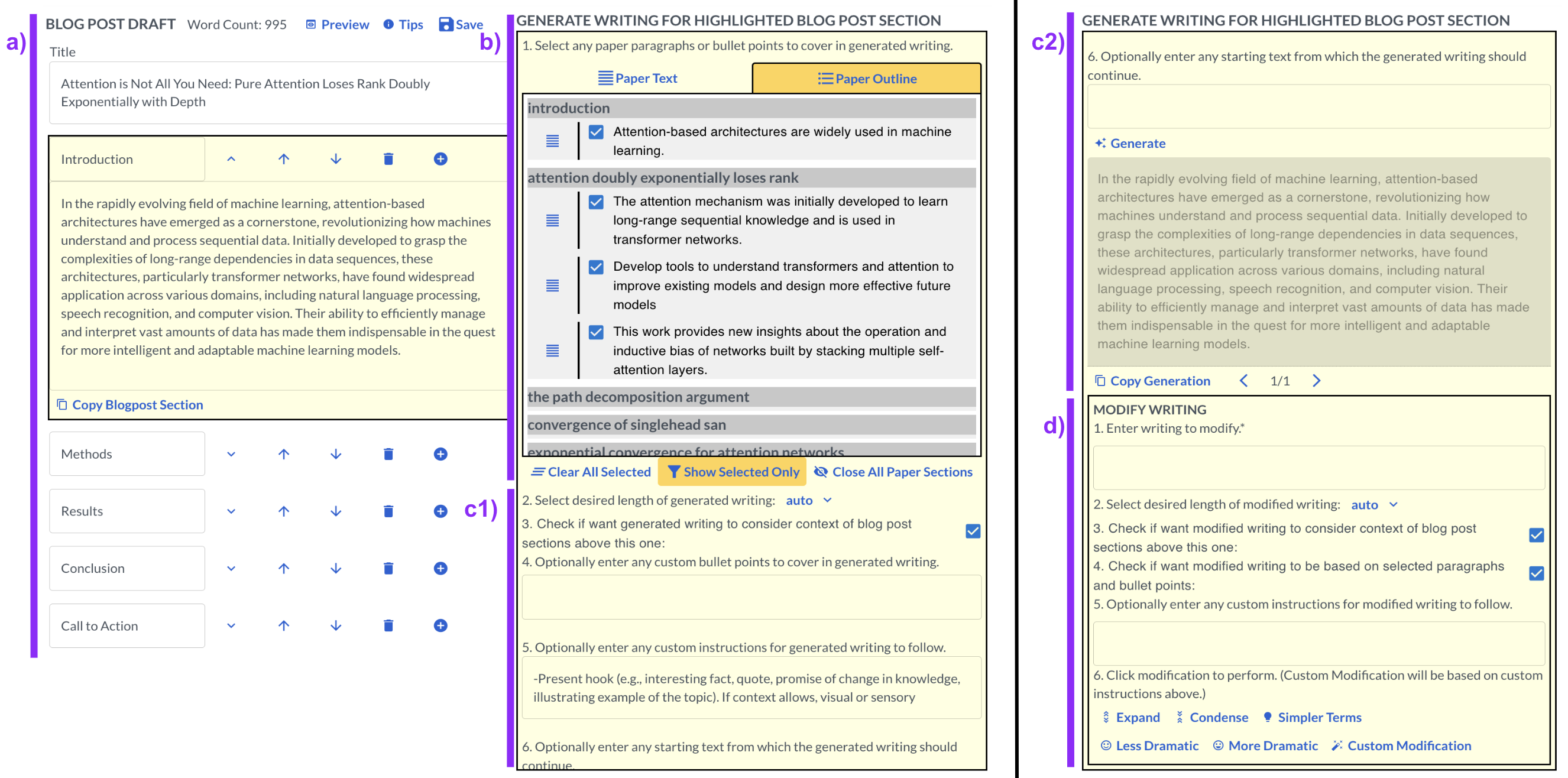}
    \caption{\sysname' user interface. a) The Blog-Post Area, where the user writes the blog post in sections. b) The Planning-Support Area, which contains both the interactive paper outline and original paper text. c1) The Drafting-Support Area, which contains inputs and outputs for generating text. c2) The Drafting-Support Area continued, seen if one scrolls below the c1 area. d) The Revising-Support Area, which contains inputs and outputs for modifying text and is located below the Drafting-Support Area. (Not Pictured: The modification output appears below the modification buttons, with a button to copy the output text and buttons to view previous modifications and their inputs.)}
    \Description{There are two images next to each other. The left image has two panels. The left panel is labeled "a" and titled "Blog Post Draft." It shows the blog post draft with the introduction section open and highlighted in light yellow, while the rest of the blog post sections are collapsed and have a white background. The draft shows a box for the title at the top followed by each initial draft section (introduction, methods, results, conclusion, call to action). Each section has a few buttons next to its header: a short arrow for collapsing/expanding the section, up and down arrows for moving the sections to a new location in the blog post, a trash icon, and a plus icon for adding a new section. Meanwhile, the right panel of the left image is light yellow and titled "Generate writing for highlighted blog post section." The top half of the panel is labeled "b" and the bottom half is labeled "c1." The top half contains the instruction "1. Select any paper paragraphs or bullet points to cover in generated writing." Below, it shows two tab buttons labeled "Paper Text" and "Paper Outline." The Paper Outline tab is clicked, and beneath there is a light gray box with paper section headers and a few checked bullet points under some of the headers. The last section header visible is cut off, indicating there are unseen sections. Underneath, there are three buttons: "Clear All Selected," "Show Selected Only," and "Close All Paper Sections." In the bottom half of the panel, from top to bottom, there is the following: 1) text saying "2. Selected desired length of generated writing:" followed by a drop-down menu set to "auto" to its right; 2) text saying "3. Check if want generated writing to consider context of blog post sections above this one: followed by a checked box to its right"; 3) text saying "4. Optionally enter any custom bullet points to cover in generated writing." followed by an empty text box; 4) text saying "5. Optionally enter any custom instructions for generated writing to follow." followed by a text box with text saying "-Present hook (e.g., interesting fact, quote, promise of change in knowledge, illustrating example of the topic). If context allows, visual or sensory"; 5) text saying "6. Optionally enter any starting text from which the generated writing should continue." The second image of the figure shows the continuation of the right panel (if one was to scroll down). The top half of the panel shown is labeled "c2" and the bottom half is labeled "d." In the top half, from top to bottom, there is the following: 1) text saying "6. Optionally enter any starting text from which the generated writing should continue." followed by an empty text box; 2) a button labeled "Generate"; 3) A paragraph of text with a light gray background that starts with "In the rapidly evolving field of machine-learning, attention-based architectures..." and continues; 4) a button that says "Copy Generation" followed by a left arrow, "1/1" and a right arrow to its right. The bottom half of the panel is enclosed in a black border. It is titled "Modify Writing." From top to bottom, it contains the following: 1) text saying "1. Enter writing to modify.*" followed by an empty text box; 2) text saying "2. Select desired length of modified writing:" followed by a drop-down menu set to "auto"; 3) text saying "3. Check if want modified writing to consider context of blog post sections above this one:" followed by a checked box to its right; 4) text saying "4. check if want modified writing to be based on selected paragraphs and bullet point:" followed by a checked box to the right; 5) text saying "5. optionally enter any custom instructions for modified writing to follow." followed by an empty text box; 6) text saying "6. click modification to perform. (Custom Modification will be based on custom instructions above.); 7) Buttons labeled "Expand," "Condense," and "Simpler Terms"; 8) Buttons labeled "Less Dramatic," "More Dramatic," and "Custom Modification."}
    \label{fig:treatTool}
\end{figure*}

\section{The \sysname\ System}
\label{sec:system}
This section describes the design goals for \sysname\ and provides a walkthrough of the resulting system's four main panel areas, which show the blog post draft and enable executing the Plan-Draft-Revise workflow on individual blog post sections.\footnote{A walkthrough video of the tool may be found in the supplementary materials.} We developed \sysname\ based on key design goals and pilot study feedback. The pilot study feedback helped us to identify system features to add for increased transparency, flexibility, and ease of use. 

\subsection{Design Goals}
\label{sec:dgs}
When communicating scientific research in a detailed summary, making sure all important details are covered is key. Otherwise, the research may be misinterpreted, leading to misinformed readers. Indeed, in a closely related work on human-LLM scientific text summarization, a featured design goal based on their preliminary survey of professors was to provide provenance for summary content by showing ``direct or indirect contributors to a summary to help the user verify whether the summary reflects the structure and key components of the original document'' \cite{zhang2023concepteva}.
This also applies to similar domains, such as law and business, in which delivering information with accuracy and precision is critical. 
However, existing text summarization tools (including the one just referenced) either 1) do not fully ground the summary content selection in the source, which makes understanding what content is versus is not included difficult, or 2) only support selection of summary content at the level of source sentences, which is unwieldy when writing a detailed summary (see Section \ref{sec:haiTextSummarization}). 

Furthermore, we identified and synthesized guidelines for academic blog posts from popular resources \cite{kudos,uwaterloo,lse,cmu}, prior work \cite{mehlenbacher2019science}, and feedback from three communications experts, recruited from our institutions. There are multiple guidelines on the content, format, and language for academic blog posts. \sysname\ seeks to address these guidelines by default but does not enforce them.\footnote{The synthesized guidelines may be found in the appendix.}

Thus, \sysname' design goals are as follows (science domain-specific details in parentheses):
\begin{itemize}
  \item \textbf{DG1}: Help users (researchers) to understand what long-document (research-paper) content is versus is not included in its detailed summary (blog post).
  \item \textbf{DG2}: Help users (researchers) to control and iterate on what long-document (research-paper) content is included in a detailed summary (blog post).
  \item \textbf{DG3}: Help users (researchers) to follow domain-specific guidelines (guidelines for academic blog posts), as a starting point.
\end{itemize}

\subsection{Implementation Details}
The frontend of \sysname\ was developed using React and TypeScript, while the backend was developed with Python. Unless otherwise noted, the LLM was GPT-4 in the lab study and GPT-4-0125-preview in the deployment study.

\subsection{Walkthrough of Warm Start Step}
The \textbf{Blog Post Area} (Figure \ref{fig:treatTool}a) is the user's main working area for reviewing the target blog post artifact.
At the top, the user views the blog post's word count, a preview of the full blog post, and tips for writing blog posts (described in Section \ref{sec:dgs}).
The user then reviews the blog post draft, divided into sections.
To mitigate cognitive overload, the user selects one section at a time to focus on. The selected section is the target of subsequent Plan-Draft-Revise workflow actions, described in the other step descriptions below.

In order to obtain the initial blog post draft for the user's input research paper, we developed an LLM-powered pipeline that first generates a reverse source outline, selects relevant points, and then generates a draft (Figure \ref{fig:teaser}a).

\subsubsection{Preprocessing}
We used a public tool for converting scientific PDFs to HTML\footnote{\url{https://papertohtml.org}} to extract the paper text from the paper PDF. For the lab study, we then manually adjusted the parsed HTML so that the section headers were correct, no section or subsection was completely missing, and only the main paper text was included (i.e., no appendix, references, or footnotes). We also manually retrieved the tables and figures from the parsing. For the deployment study, the text parsing was automatic to better align with a real-world setting, and there was no table or figure retrieval. 

\subsubsection{Outline generation}
We used GPT-3.5 Turbo to generate between one and three bullet points to summarize each paragraph in the parsed paper, depending on the length of the paragraph. While we used GPT-4 (GPT-4-0125-preview in the deployment study) for all of \sysname' other tasks, we decided to use GPT-3.5-Turbo for this task for two reasons: 1) it produced shorter bullet points that were quicker to read, and 2) it was many times cheaper for this token-intensive generation task. All together, the generated bullet points formed a paper outline.

\subsubsection{Outline point selection}
We provided the LLM with all of the paper's bullet points along with the request to select 10 bullet points relevant to the initial four blog post sections: introduction, methods, results, and conclusion. We requested 10 bullet points per blog post section in order to provide the model sufficient content without overwhelming it.

\subsubsection{Summary generation}
We prompted the LLM to generate each section in succession using the selected bullet points for that section, the paragraphs associated with the selected bullet points (to better ground the model), the portion of the blog post that had already been generated, and section-specific guidelines (see Section \ref{sec:dgs}). We observed that this process of generating one draft section at a time with the context of prior sections helped to make the draft more coherent while avoiding overwhelming the model with instructions for multiple sections. The model was prompted to make each section around 125 to 250 words and around one to three paragraphs.\footnote{The prompts for generating the initial draft are in Appendix \ref{sec:startupPrompts}. An example initial draft and associated selected bullet points are in Appendices \ref{sec:treatmentDraftExample} and \ref{sec:treatmentBulletpointsExample}.}

\subsection{Walkthrough of Planning Step}
In the \textbf{Planning-Support Area} (Figure \ref{fig:treatTool}b), the user works with \sysname's interative reverse source outline to plan the blog post section on which they are currently working, addressing \textbf{DG1} and \textbf{DG2}. 
All of the user's planning actions (as well as drafting and revising actions) are connected to the user's selected blog post section. In other words, the state and history of all inputs and outputs for text generation are stored separately for each section. This is somewhat similar to the fragmented document history proposed by Buschek \cite{buschek2024collage}.

The user sees the interactive reverse source outline, divided into sets of bullet points, each corresponding to a paragraph from the original text, which can be viewed by clicking the button to the left of the bullet point set.
\footnote{In the lab study, the tool presented figures/tables in a final collapsible portion of the paper outline. Their captions could be copied for pasting in the editor to help the user plan where to put figures in the final blog post.}
They review the tool's pre-selected bullet points for the current blog post section by scrolling through the full outline or clicking the ``Show Selected Only'' button.\footnote{When this filter button is activated and the user is under the Paper Text tab, if a paragraph is not selected but a corresponding bullet point is selected, the tool notes that ``A corresponding bulletpoint(s) is still selected''; the opposite is true for the Paper Outline tab.} 
The user adjusts the selection of bullet points (or paragraphs) for the LLM to use in generating text for the current blog post section (Figure~\ref{fig:teaser}b). Selecting a paragraph does not affect the selection of associated bullet points, and vice versa, from the user's perspective. The system, however, does receive associated paragraphs with selected bullet points for grounding.

\subsection{Walkthrough of Drafting Step}
\label{sec:drafting-walkthrough}
In the \textbf{Drafting-Support Area} (Figure \ref{fig:treatTool}c), the user works with the tool to synthesize text for the blog post section on which they are currently working, addressing \textbf{DG2} and \textbf{DG3}. 

\subsubsection{Section Revision}
At the top of the Drafting Support Area, the user changes the desired length of the generated text to their liking. The length defaults to ``auto'' and may be updated to one sentence, one paragraph, or a few paragraphs. 
Below, the user sees the toggle is on to provide the LLM with the context of blog post sections above the one on which they are working.
In the ``Custom Bulletpoints'' text field, the user enters any additional bullet points that they want their generated text for the blog post section to cover. 
In the ``Custom Instructions'' text field, the user adjusts the instructions for the section's generated text. For each of the initial draft sections, this box is pre-filled with section-specific instructions, which are based on blog post guidelines we collected on content and formatting for different academic blog post sections \cite{kudos,uwaterloo,lse,cmu,mehlenbacher2019science}. 
In the text field labeled ``Starting Text to Continue from'', the user optionally provides text that should be at the start of the generated text. 

The LLM then generates revised text for the section at hand with the user's adjusted selection of bullet points and paragraphs as well as their adjusted section-specific instructions (Figure \ref{fig:teaser}b).\footnote{The prompts for drafting text may be found in Appendix \ref{sec:revisePrompts}, and example outputs are in Appendices \ref{sec:treatmentDraftExample} and \ref{sec:treatmentBulletpointsExample}.} With respect to the selected bullet points, the model receives the context of the associated paragraph for each selected bullet point, even if the associated paragraph was not selected by the user. This is to ensure that the model's generation is grounded in the actual paper text. When the user selects the ``auto'' length for generated text, the model is instructed to generate text that is 125 to 250 words and between one and three paragraphs. 

The user copies and pastes all or part of the generated text into the highlighted blog post section and edits it as they see fit. If the user wants to remember what they generated previously for this section, they may peruse a history of their inputs (paper content selection, LLM instructions) and outputs (text generations). 

\subsubsection{Section Addition}
The user can also add new sections to the blog post (Figure \ref{fig:teaser}c). In the Blog Post Area, the user clicks the plus-sign button next to an existing blog post section, and the tool places the new section below the existing one. The tool gives the user the option to either create a blank section or generate a section based on the section header that they provide. 
To generate a new section, the LLM selects 10 bullet points for the section based on the provided header and generates text based on those bullet points and associated paragraphs.


\subsection{Walkthrough of Revising Step}
\label{sec:revising-walkthrough-frontend}
In the \textbf{Revising-Support Area} (Figure \ref{fig:treatTool}d), the user works with the tool to revise the blog post section on which they are currently working, addressing \textbf{DG3}. At the top of this area, the user pastes the text that they want to modify in the text field. This text may be anything, including text from the blog post draft, the paper, or a model generation. The user then adjusts the desired length of the modified text if need be, as in the Drafting-Support Area (Section \ref{sec:drafting-walkthrough}).\footnote{The desired length ``auto'' is unique to each modifying action.}
The user leaves the second toggle on, indicating that the LLM will utilize the currently selected paragraphs and bullet points to generate the modified text. 
The user considers the five preset modification options: 1) expand the text, 2) condense the text, 3) rewrite the text in simpler terms to be more understandable to a layperson, 4) rewrite the text in a less dramatic tone to align better with the unadorned language of scientific writing, and 5) rewrite the text in a more dramatic tone to better capture readers' attention. The preset modification buttons called prompts based on our synthesized guidelines for academic blog posts (see \ref{sec:dgs}), which suggested avoiding jargon \cite{kudos,uwaterloo,lse,cmu}, sensationalism (noted by a communications expert), and wordiness \cite{kudos,uwaterloo,lse}. There is also a ``custom modification'' button that will follow specified custom instructions.\footnote{If the user utilizes a preset modification button, they may still provide custom instructions for the LLM to take into consideration.} The user copies and pastes the modified text into the highlighted blog post section and edits it as needed. If the user wants to remember their previous text modifications for this section, they may peruse a history of their instructions and modifications. 
With the user's instructions, the LLM modifies the user's provided text snippet.\footnote{The prompts for revising text may be found in Appendix \ref{sec:revisePrompts}, and example outputs are in Appendix \ref{sec:modificationExamples}.}

\section{Evaluation Studies}
\subsection{Lab Study}
We conducted a within-subjects lab study to determine if and how \sysname\ may support researchers in writing blog posts about their research papers.

\subsubsection{Hypotheses}
\label{sec:hypos}
We hypothesized that, compared to a baseline consisting of an LLM draft and the ability to prompt the LLM, \sysname\ would provide participants a better experience and outcome. Our individual hypotheses were as follows:
\begin{itemize}
    \item \textbf{H1}: \sysname\ leads to higher satisfaction with the output blog post than the baseline.    
    \item \textbf{H2}: \sysname\ leads to higher satisfaction with the tool than the baseline.
    \item \textbf{H3}: \sysname\ leads to lower cognitive load than the baseline.
    \item \textbf{H4}: \sysname\ leads to less time spent to generate a blog post that the writer would be comfortable posting publicly, in comparison to the baseline.
\end{itemize}

\subsubsection{Participants}
We recruited 20 participants (M: 14, W: 6) via academic social networks and institutional mailing lists and compensated them with \$83.33 USD for 2.5 hours of their time. Participants were predominantly early career researchers (PhD student: 17, master's student: 1, postdoc: 1, industry researcher: 1), who are often lead authors tasked with writing blog posts. All participants conducted research in computer science. 
Participants had a broad range of experience in terms of LLM use (>10 sittings: 14, 6-10 sittings: 3, 2-5 sittings: 2, <2 sittings: 1). Most had not authored a research-paper blog post before (0 posts: 14, 1-2 posts: 5, 3-10 posts: 1). In order to simulate a naturalistic setting, participants were required to have authored at least 2 research papers for use in the study (2 papers: 1, 3-10 papers: 14, >10 papers: 5) and to be more interested than not 
in writing blog posts for their research papers.

\subsubsection{Study Conditions}
\begin{figure}
    \centering
    \includegraphics[width=\linewidth]{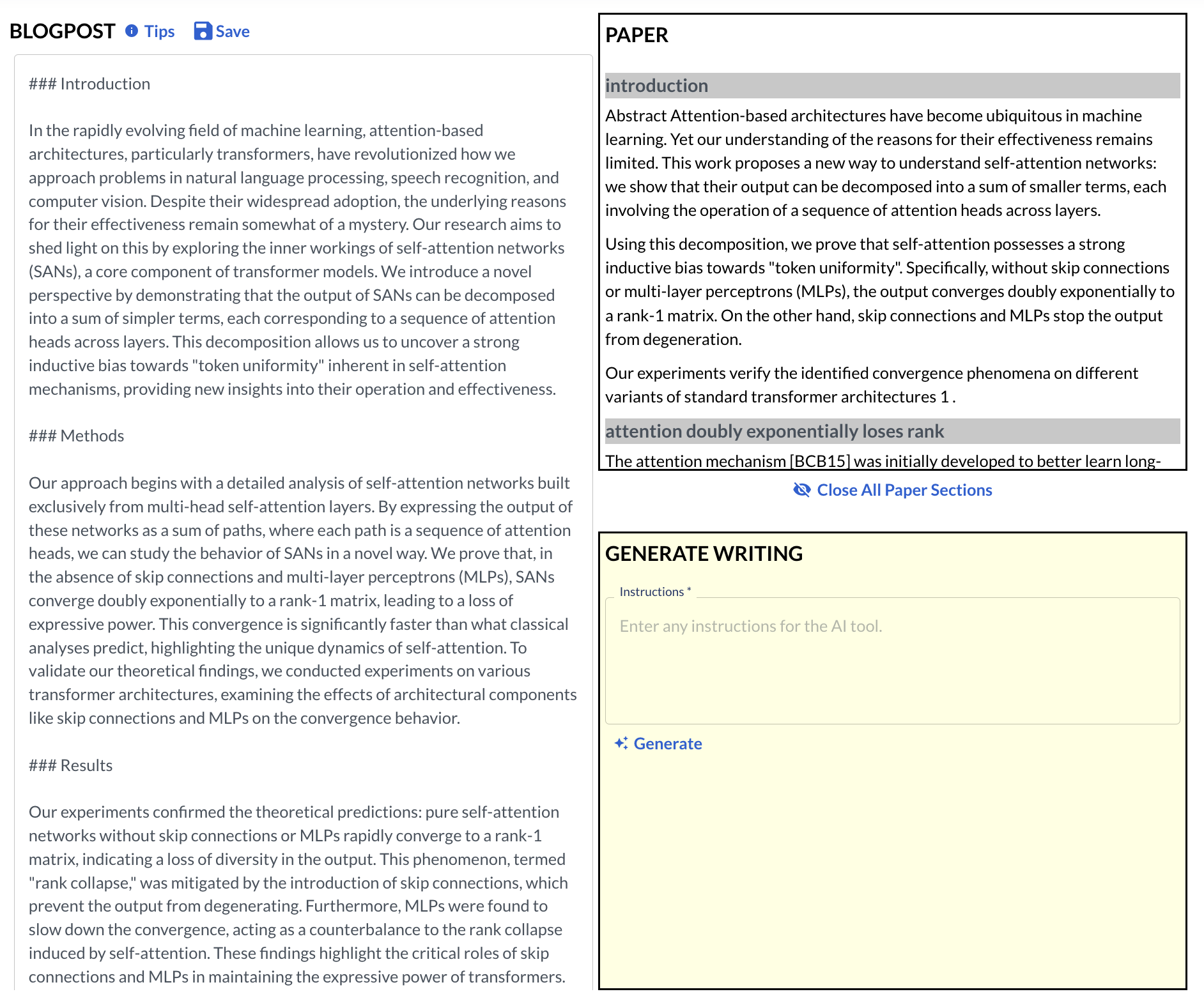}
  \caption{The baseline tool's user interface, consisting of areas for writing the blog post (left), viewing the paper (top right), and providing instructions to the LLM for generating writing (bottom right).}
  \label{fig:baseTool}
  \Description{The image may be divided into three areas: one on the left, one on the top right, and one on the bottom right. The left area is titled "Blogpost." There is a white text box filling the entire area. It contains text with three sections (introduction, methods, and results) visible. The top right area is titled "Paper" and shows two section header with a gray background. The text of the sections has a white background. The bottom right area is yellow and titled "Generate Writing." It contains a text box labeled "Instructions*" with placeholder text that says "Enter any instructions for the AI tool." Below the text box, there is a button labeled "Generate."}
\end{figure}

In the treatment condition, participants interacted with \sysname\, described in Section \ref{sec:system}.\footnote{The system figures above show how the tool looked for the deployment study, after minor usability updates (see Section \ref{sec:depStudyConds}. Appendix \ref{fig:firstTool} shows how the tool looked for the lab study.} The tool in the baseline condition was designed to be a simplified version of \sysname\ without affordances developed to address the design goals (Figure~\ref{fig:baseTool}). The baseline provided the input paper's text, sourced from the PDF and divided into paragraphs and collapsible sections as in \sysname. The baseline also provided the ability to prompt GPT-4 with any instructions and copy its output. A history of generations and their associated instructions was retained for the user's perusal. Lastly, the baseline had an editor for writing the blog post, which was pre-populated with an LLM draft generated by GPT-4-32k using the entirety of the input paper.\footnote{An example of a baseline initial draft is in Appendix \ref{sec:baselineDraftExample}.} GPT-4-32k was used rather than GPT-4 because the context window for GPT-4 was insufficient for longer research papers. The prompt was designed to mirror the prompt for \sysname' initial draft sections (Appendix \ref{sec:baselinePrompts}), so the resulting draft also had an introduction, methods, results, and conclusion section. However, the prompt did not contain section-specific instructions.

In each condition, the tool was set up for one of the participant's two research papers submitted in the recruitment survey. Each paper was required to have been published online between 2018 and the study period in 2023, with the participant as an author. Only one paper was published in 2018, and the rest were published during or after 2020. There was no significant difference in publication dates between conditions (Paired-Samples \emph{t}-Test, t(19)=0.90, p=n.s.). The average input paper length across conditions differed by less than 10\% (baseline: 8452 words, treatment: 7742 words).

\subsubsection{Study Procedure}
\label{sec:labProcedure}
The study was within-subjects, so each participant completed two study sessions, one in the treatment condition and one in the baseline, in randomized order and following a counter-balanced design. The participants' selected papers were also assigned to conditions in randomized order. Each study session lasted up to about 75 minutes. For sufficient flexibility, participants selected two study sessions within a week of each other, but the majority of sessions were a day apart (1 day: 13, 2 days: 3, 3 days: 3, 4 days: 1).\footnote{The condition order was counter-balanced within the group of participants with sessions one day apart and within the group of participants with sessions over one day apart.}

Participants completed a participation agreement before the study. We recorded and transcribed each session in Google Meet.\footnote{The lab study session script may be found in the supplementary materials.} To start, the session coordinator introduced the task and presented a video tutorial on the assigned tool.\footnote{The tutorial scripts are in the supplementary materials.} Participants then had two minutes to interact with and ask questions about the tool in the context of a sample paper that was not their own. Next, participants opened a tool link initialized with one of their research papers. The coordinator asked them to assume that their audience was people in their broad area (i.e., computer science), which was the most common audience for whom potential study recruits wanted to write blog posts (52/60). The coordinator instructed participants to work as if they were not there, but to let them know if any questions about the tool arose. They also instructed participants not to use any external tools or information.\footnote{We made an exception twice when participants wanted to view their paper's PDF due to issues with its parsing in the tool.} Lastly, the coordinator asked participants to let them know if they reached a point at which they would feel comfortable publicly publishing the blog post in its current state. 

Each participant had up to one hour to work on the blog post. At the 20-minute and 40-minute marks, they were reminded to let the coordinator know if they reached the point of comfort with publication. If the participant reached that point, the coordinator asked them to save their draft and continue revising the blog post. After the first four study sessions, we noticed that participants were sometimes done early and did not desire to revise further. Thus, we added an additional rule that the participant would let the coordinator know if they reached a point at which they were proud of the blog post and saw no reason to revise further. If the participant reached that point, or if an hour had passed, they submitted their final blog post and moved on to a survey regarding their experience with the tool. The survey consisted of Likert-type questions regarding perceived efficiency, cognitive load, ease of using the tool in relation to the design goals, satisfaction with the tool, satisfaction with the final blog post, perceived control, and perceived ownership.\footnote{See Appendix \ref{sec:control} for a discussion of the exploratory results regarding participants' perceived control and ownership.} There were also short answer questions regarding the difficulties and benefits of using the tool in relation to the design goals.\footnote{The survey may be found in the supplementary materials.}
At the session's end, the coordinator told the participant to close the tool and not to access it before the next session if they had one. Each time the participant prompted the LLM or submitted the blog post, a log of the action along with the state of the blog post was saved for analysis.

\subsection{Deployment Study}
We conducted a deployment study to understand if and how the results of the lab study translate to a less controlled and more realistic setting. To make the deployment study more realistic than the lab study, participants wrote blog posts on their own time with the explicit goal of sharing the blog post with others (e.g., on social media, in an email to colleagues).

\subsubsection{Research Questions}
\label{sec:rqs}
Having observed that the lab study participants had significantly more editing power when using \sysname\ rather than the baseline tool, we investigated if \sysname\ leads participants to have greater editing power in this real-world context (\textbf{RQ1}). As in the lab study, we also evaluated satisfaction with the final blog post, satisfaction with the tool, and cognitive load (\textbf{RQ2}). We did not evaluate task completion time in the deployment study, as the lab study did not demonstrate any interesting difference between the conditions.
\begin{itemize}
    \item \textbf{RQ1}: How is participants' editing power impacted by \sysname\ versus the baseline tool in a real-world setting?
    \item \textbf{RQ2}: How do participants perceive the outcome (blog post satisfaction) and experience (tool satisfaction, cognitive load) of interacting with \sysname\ versus the baseline tool in a real-world setting?
\end{itemize}

For the deployment study, we posed research questions rather than evaluated hypotheses for statistical significance because power analyses indicated that we would need a larger sample size than we could obtain to see the same significant results from the lab study.\footnote{The significant editing power result would have required 29 participants per condition (power=0.95, alpha=0.05, effect size=0.45), and the significant blog post satisfaction result would have required 83 participants per condition (power=0.95, alpha=0.05, effect size=0.98).}

\subsubsection{Participants}
We recruited 26 participants (M: 17, W: 8, undisclosed: 1) through academic social networks and institutional mailing lists. They were compensated with \$50 over PayPal for each blog post that they wrote, with a maximum of two blog posts. Participants were predominantly early career researchers (undergraduate student: 1, master's student: 4, PhD student: 13, postdoc: 2, professor: 2, industry researcher: 4). All participants conducted research in an area of computer science, except for one who studied cognitive science. The majority of the participants had interacted with an LLM several times (>10 sittings: 21, 6-10 sittings: 4, 2-5 sittings: 1). The participants had varied experience in writing blog posts (0 posts: 16, 1-2 posts: 9, 3-10 posts: 1). They were required to have authored at least one research paper (1-2 papers: 6, 3-10 papers: 11, >10 research papers: 9) and to be more interested than not 
in writing research-paper blog posts. We also required participants to be willing to share their final blog posts with others if they were able to reach the point at which they would feel comfortable doing so. Lastly, the papers that the participants wanted to write about had to be compatible with our tools' PDF parser.

\subsubsection{Study Conditions}
\label{sec:depStudyConds}
While the treatment and baseline conditions were largely the same as in the lab study, several minor usability issues identified in the lab study were addressed before proceeding with the deployment study. For example, there were better instructions and less intrusive highlight colors, as depicted in the system figures above.
We also replaced GPT-4 and GPT-4-32k with the newer GPT-4-0125-preview model, which had a larger context window and became available after the lab study. This enabled both conditions to support longer inputs. In the deployment study, participants could freely close and reopen their assigned tools in their browsers to work on their drafts when convenient, and they could save their progress across sessions with a newly added save button. Furthermore, the treatment tool now automatically parsed the input paper. We note that, due to the manner in which text was logged in the deployment study, the tool in both conditions had a lag if participants pasted a chunk of text more then a couple paragraphs long. 

The majority of the 37 papers about which participants wrote were published in 2023 or 2024 (26 papers). The two oldest papers were from 2019 and 2008. The average parsed paper length across conditions differed by less than 5\% (baseline: 37064 characters, treatment: 35845 characters).

\subsubsection{Study Procedure}
The task for the deployment study was to share a blog post with others, which can require substantially higher effort than \textit{voicing} comfort in publicly posting the blog post, as in the lab study. Considering this difference for participant recruitment, we opted for a between-subjects design, so each participant only had to write one blog post. Still, we allowed participants who wished to participate twice (N=11/26 participants) to use both tools in randomized, counter-balanced order. Their selected papers were assigned randomly to each condition. When analyzing these participants' results, we treated them as unpaired data. Eighteen participants completed the baseline, and 19 completed the treatment. Participants had up to six days to complete a paper blog post using the assigned tool on their own time. We provided a link to the tool, a video tutorial on the tool\footnote{The scripts for the video tutorials are in the supplementary materials.}, and a link to the paper about which to write. We noted that participants should aim to share the final blog post with others, but if they felt that they could not reach that point, they would still be compensated. Participants were required to spend at least 20 minutes using the tool. We recorded participants' interactions, including prompting and text edits, in logs. In addition, we also took snapshots of the blog post whenever they (re)opened the systems in their browsers and every 60 seconds during active usage.

Once participants were satisfied with their draft, they shared it. We provided instructions on how to create blog posts on the Medium platform,\footnote{\url{https://medium.com/}} but they were allowed to use other publishing platforms if they preferred. They then emailed us the blog post link and a screenshot or link showcasing evidence that they shared the blog post with others (e.g., tweet, email to co-authors, Slack channel message, LinkedIn post). Only 2 of the 26 participants (one under both conditions and one under the treatment) could not reach the point at which they felt comfortable sharing the blog post with others in the allotted time.\footnote{We note that some participants indicated that the blog post was part of an experiment when sharing with others, which may have increased their willingness to share it; regardless, they were willing to share a post at the end of the study.} Participants then completed the same survey from the lab study regarding their experience (see Section \ref{sec:labProcedure}).\footnote{We note that four participants who we realized had not completed 20 minutes of interaction were asked to do so after completing the survey.} The only difference was the removal of a few questions irrelevant to this study.
Lastly, the first author engaged willing participants (15) in a 15-minute semi-structured interview to discuss their experience and survey responses. 

\begin{figure*}[tb]
    \centering
    \includegraphics[width=\textwidth]{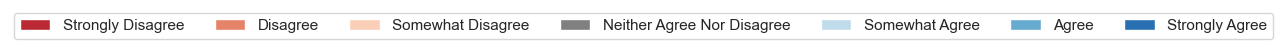} 
    
    \begin{subfigure}[b]{0.49\textwidth}
        \centering
        \vspace{-\abovecaptionskip}
        \captionsetup{labelformat=empty}
        \caption{Lab Study}
        \includegraphics[clip,width=\linewidth]{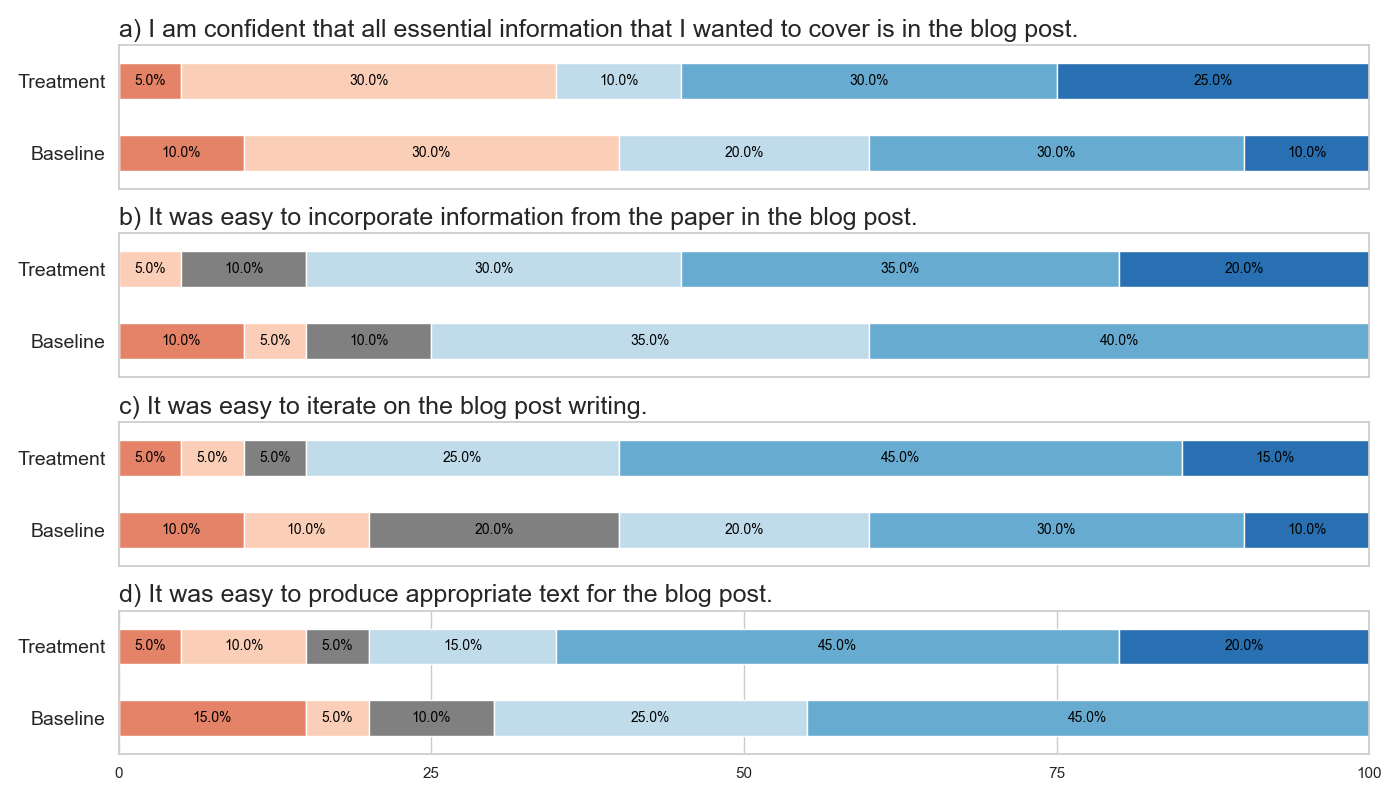}
    \end{subfigure}
    \hfill
    \begin{subfigure}[b]{0.49\textwidth}
        \centering
        \vspace{-\abovecaptionskip}
        \captionsetup{labelformat=empty}
        \caption{Deployment Study}
        \includegraphics[clip,width=\linewidth]{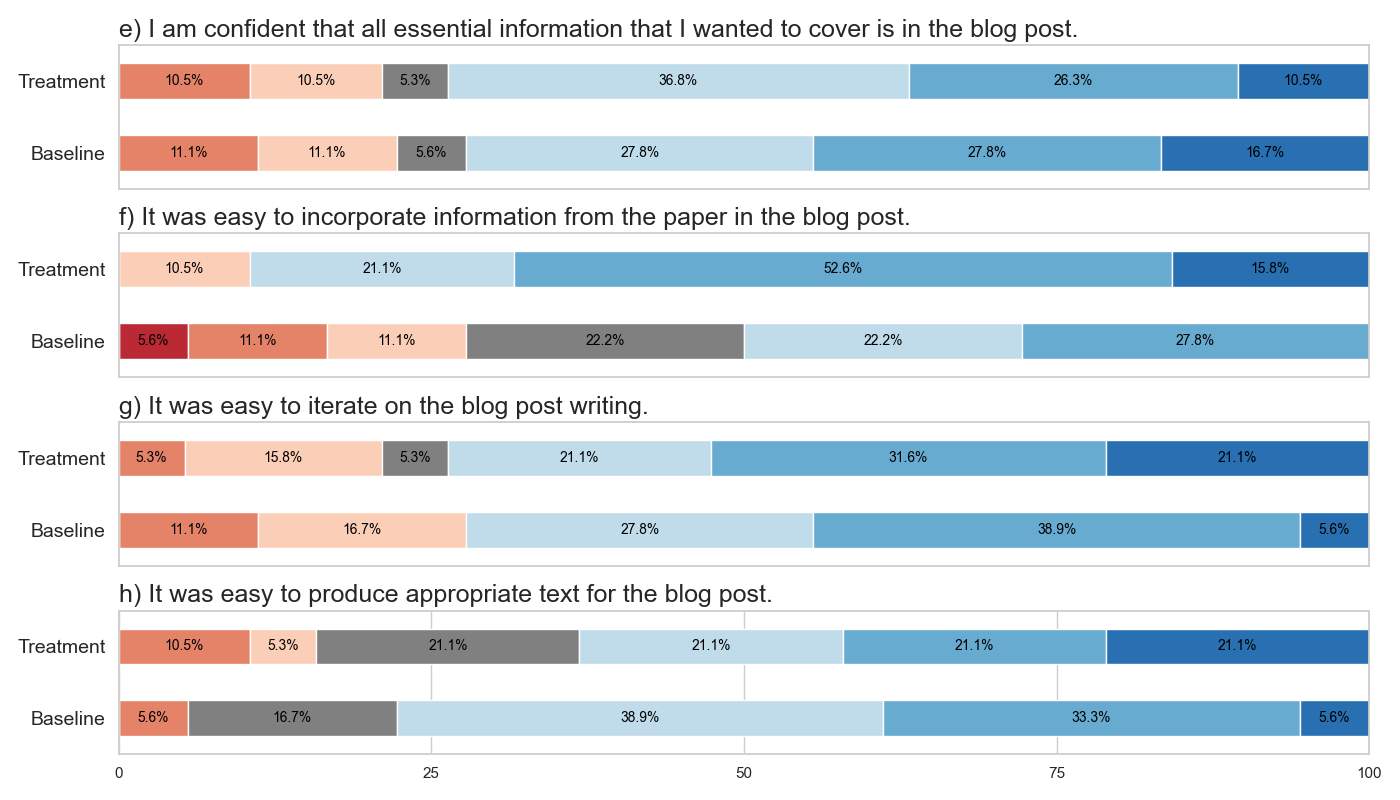}
    \end{subfigure}
    
    \caption{Survey responses to 7-point Likert-type questions regarding design goals in the a-d) lab study and e-h) deployment study. Responses are shown for both the treatment and baseline conditions.}
    \Description{The top of the figure shows a legend with the following mapping: dark red = strongly disagree, medium red = disagree, light red = somewhat disagree, gray = neither agree nor disagree, light blue = somewhat agree, medium blue = agree, and dark blue = strongly agree. Below, there are two areas. The left area is labeled "Lab Study," and the right area is labeled "Deployment Study." In each area, there are four questions labeled "a" through "d" for the lab study and "e" through "h" for the deployment study. The questions are: 1) "I am confident that all essential information that wanted to cover is in the blog post."; 2) "It was easy to incorporate information from the paper in the blog post."; 3) "It was easy to iterate on the blog post writing."; 4) "It was easy to produce appropriate text for the blog post." Below each question, there are two horizontal stacked bars, one labeled "treatment" and the other "baseline." In the lab study area, treatment is shown to perform somewhat better than baseline for each question. For all of the questions, the medians for treatment followed by baseline are agree vs. somewhat agree. In the deployment study area, treatment is shown to perform somewhat better than baseline for the second and third question. It appears to do slightly worse overall for the first question, and which condition did better is unclear for the last question. For each question, the medians for treatment followed by baseline are: somewhat agree vs. somewhat agree, agree vs between neither agree nor disagree and somewhat agree, agree vs somewhat agree, and somewhat agree vs. somewhat agree.}
    \label{fig:dg}
\end{figure*}

\section{Evaluation Studies' Results}
In this section, we present the quantitative and qualitative results of the lab and deployment studies in parallel. To obtain the qualitative results, the first author conducted inductive thematic analysis \cite{braun2006using} of the participants' interview and short-answer responses across the two studies. Please note that the hypotheses from the within-subjects lab study were not re-evaluated for significance in the between-subjects deployment study because there were not enough participants to do so (see Section \ref{sec:rqs}).

\subsection{Design Goals 1 and 2}
\label{sec:dg1}
We analyzed participants' Likert-type responses to questions related to fulfilling our first two design goals, which focused on helping researchers understand and control what content is included in their research-paper blog posts (Figure \ref{fig:dg}a-c,e-g).

\subsubsection{Lab Study - Exploratory}
Based on the median results, using \sysname, participants were only slightly more confident that all essential information was in the blog post (treatment: M=6.00, Q1=3.00, Q3=6.25; baseline: M=5.00, Q1=3.00, Q3=6.00). They found it easier to incorporate information from the paper (treatment: M=6.00, Q1=5.00, Q3=6.00; baseline: M=5.00, Q1=4.75, Q3=6.00), and they found it easier to iterate on the blog post text (treatment: M=6.00, Q1=5.00, Q3=6.00; baseline: M=5.00, Q1=4.00, Q3=6.00) (Figure \ref{fig:dg}a-c). The interaction logs showed that participants utilized \sysname' affordance designed to support these outcomes, selecting relevant paper bullet points and paragraphs from the LLM-generated source outline to incorporate in the blog post. All participants adjusted the LLM's pre-selected content to some degree. On average, participants added or removed 28 bullet points and 13 paragraphs.
\footnote{Some added bullet points come from automatic selection of bullet points used to generate a new section rather than manual selection. These averages do not include 2 participants due to a logging error.} Participants also added their own custom bullet points an average of 1.70 times. 

\subsubsection{Deployment Study - Exploratory}
Looking at the median results, participants did not indicate an improvement in feeling confident that all essential information was covered with \sysname\ (M=5.00, Q1=4.50, Q3=6.00) as opposed to the baseline (M=5.00, Q1=4.25, Q3=6.00). Nevertheless, participants still indicated an increased ease in incorporating information from the paper in the blog post with \sysname\ (M=6.00, Q1=5.00, Q3=6.00) compared to the baseline (M=4.50, Q1=3.25, Q3=5.75). In addition, the median participant still indicated that \sysname\ (M=6.00, Q1=4.50, Q3=6.00) facilitated iterating on the blog post more so than the baseline (M=5.00, Q1=3.50, Q3=6.00). 

\subsubsection{Summary}
In both studies, the median participant using \sysname\ indicated increased ease in incorporating paper content in the blog post and iterating on the blog post text. However, the two studies presented little evidence of a difference between the conditions with respect to participants' confidence in all essential information being covered in the blog post. That said, participants were still more satisfied with their blog posts' content coverage under time constraints (see Section \ref{sec:bpSatLab}). Perhaps participants were happier with the content they were able to include under time constraints but still needed more assurance that all essential information was covered.

\subsection{Editing Power}
As DG2 focuses on helping researchers to iterate on the content of their blog posts, we conducted a related investigation of how participants' editing power (change in writing for a given amount of time or writing actions) changed between conditions.

\subsubsection{Lab Study - Exploratory}
\label{sec:editingPowerLab}
Lab study participants made greater change to the initial draft blog post when using \sysname\ as compared to the baseline tool. For each condition, we calculated the Levenshtein distance between the initial draft blog posts and participants' blog posts at the point at which they would feel comfortable publicly publishing the blog post. Levenshtein distance measures how many character insertions, deletions, and replacements would be needed to transform one text into another. For statistical testing, we employed a paired-samples \emph{t}-test and observed that this Levenshtein distance is significantly higher with \sysname\ (M=3977.05, SD=1146.46) compared to the baseline (M=2764.15, SD=1042.16) (Paired-Samples \emph{t}-Test, t(19)=4.31, p<.0005).\footnote{We note that this significance test was not part of our initial hypotheses and was therefore exploratory in nature.}
The difference in Levenshtein distance does not appear attributable to the time spent on the task or the change in blog post length. Participants did not spend substantially more time to reach the point at which they were comfortable publicly publishing the blog post using \sysname\ as compared to the baseline (see Section \ref{sec:timeSpent}). Figure \ref{fig:levs}a shows Levenshtein distance plotted against the time taken to get to a publishable blog post, with the removal of one outlier. Furthermore, participants changed the length of their blog posts, measured in characters, less with \sysname\ (M=-162.00, Q1=-1126.25, Q3=297.00) than with the baseline (M=1971.00, Q1=965.25, Q3=2514.25).\footnote{As you can see in Appendix \ref{fig:lablengths}, the initial length of treatment drafts was significantly higher than the initial length of baseline drafts.}

\subsubsection{Deployment Study - RQ1}
\label{sec:editingPowerDeployment}
In the deployment study, we investigated editing power in terms of not only amount of time taken but also amount of writing actions taken. We define writing actions as insertions and deletions of a character or text span in the draft or LLM instructions.\footnote{We did not count when participants moved, inserted, or deleted a blog post section in \sysname\ as a writing action because these actions involved clicking a button rather than manipulating text.} To analyze editing power, we excluded the three study runs in which the participant made no change to the initial draft.
We also excluded six study runs in which the participant claimed to have made edits to the blog post outside the tool, beyond small edits like those related to grammar, formatting, and adding images. The resulting analysis thus included 28 study runs (treatment: 14; baseline: 14).

\begin{figure}
    \centering
    \begin{subfigure}[b]{.33\textwidth}
        \centering
        \includegraphics[width=\textwidth]{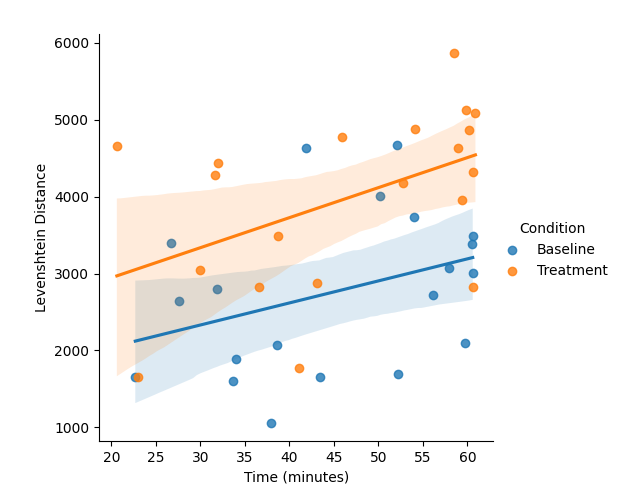}
        \caption{}
        \label{fig:levsVsTimesEnd}
    \end{subfigure}
    \begin{subfigure}[b]{.33\textwidth}
        \centering
        \includegraphics[width=\textwidth]{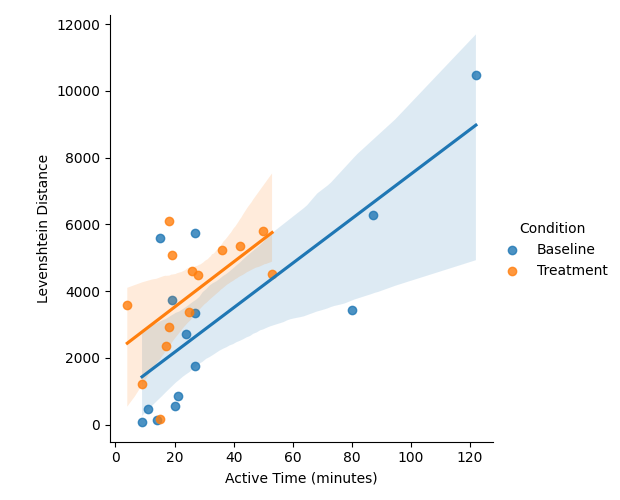}
        \caption{}
        \label{fig:deployLevsVsTimes}
    \end{subfigure}
    \begin{subfigure}[b]{.33\textwidth}
        \centering
        \includegraphics[width=\textwidth]{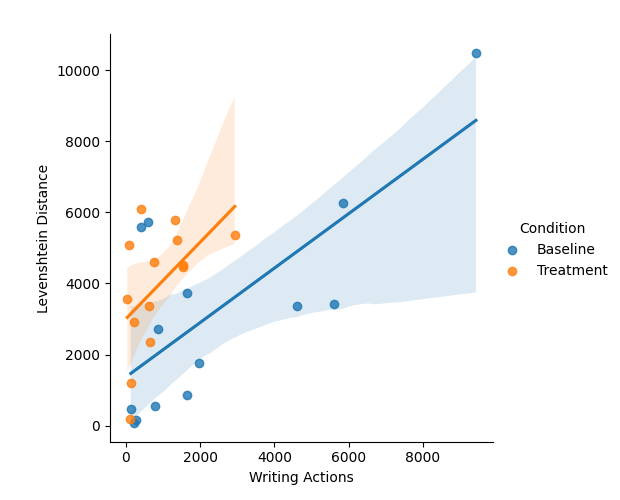}
        \caption{}
        \label{fig:deployLevsActions}
    \end{subfigure}
    \caption{Per condition, Levenshtein distance between a) initial blog post and blog post comfortable with publishing publicly as a function of time spent using the tool (lab study), b) initial and final blog posts as a function of active time spent using the tool (deployment study), and c) initial and final blog posts as a function of number of writing actions taken (deployment study). Across the two studies, participants using \sysname\ consistently showed more editing power, or change in the blog post for a given amount of time or writing actions. Note that the lab study duration was controlled (up to 60 minutes), whereas deployment study participants wrote on their own time.}
    \label{fig:levs}
    \Description{There are three scatterplots labeled "a," "b," and "c." For each scatterplot, there is a legend titled "Condition" with a blue dot labeled "baseline" and an orange dot labeled "treatment." The "a" scatterplot has an x-axis labeled "Time (minutes)" with tick marks for 20 through 60 and a y-axis labeled "Levenshtein Distance" with tick marks for 1000 through 8000. The plot shows an orange line going from around (20,3000) on the left to around (60,4250) on the right and a blue line going from around (22,2000) on the left to around (60,3000) on the right. Shading around each line overlaps more towards the start of the lines. The shading of one line overlaps with another line in only one area: the orange shading overlaps with the start of the blue line. The "b" scatterplot has an x-axis labeled "Active Time (minutes)" with tick marks for 0 through 120 and a y-axis labeled "Levenshtein Distance" with tick marks for 0 through 12000. The plot shows an orange line going from around (5,2500) on the left to around (60,5500) on the right and a blue line going from around (10,1500) on the left to around (120,9000) on the right. Shading around each line overlaps more towards the start of the lines. The shading of one line slightly overlaps with another line in two areas: the orange shading overlaps with the start of the blue line and the blue shading overlaps with near the start of the orange line. The "c" scatterplot has an x-axis labeled "Writing Actions" with tick marks for 0 through 8000 and a y-axis labeled "Levenshtein Distance" with tick marks for 0 through 10000. The plot shows an orange line going from around (0,3000) on the left to around (3000,6000) on the right and a blue line going from around (0,1500) on the left to around (10000,8500) on the right. Shading around each line overlaps towards the start of the lines. The shading of one line slightly overlaps with another line in only one area: the blue shading overlaps with near the start of the orange line.}
\end{figure}

Participants using \sysname\ took fewer writing actions to create one unit of change in Levenshtein distance to the blog post (Figure \ref{fig:deployLevsActions}). \sysname\ also led to a greater final Levenshtein distance for a given amount of active time, which we define as time during which participants were taking any actions in the tool (Figure \ref{fig:deployLevsVsTimes}). Once more, this is despite the fact that participants did not spend substantially more active time using \sysname\ (M=25.71 minutes, SD=14.72) than the baseline (M=35.93 minutes, SD=34.38) (Appendix \ref{fig:deploytimespent}).\footnote{Considering that there were three outliers with a high amount of active time in the baseline condition, we can also look at the median behavior. We again see little difference between the conditions (treatment: M=22.00, Q1=17.25, Q3=34.00; baseline: M=22.50, Q1=16.00, Q3=27.00).} Furthermore, participants 
changed the length of their blog posts, measured in characters, less using \sysname\ (M=-5.50, Q1=-525.25, Q3=961.00) than using the baseline (M=710.00, Q1=47.25, Q3=4261.25).
\footnote{As you can see in Appendix \ref{fig:deploylengths}, the initial length of treatment drafts was significantly higher than the initial length of baseline drafts.}

\subsubsection{Summary}
In the lab study, our exploratory analysis revealed that \sysname\ led to increased editing power, in the sense that participants made more change to their blog posts in a given amount of time. In the deployment study, participants experienced the same outcome in a more realistic setting, and we further found that participants made more change to their blog posts for a given number of writing actions. This aligns with the fact that participants found iterating on their blog posts easier with \sysname.

\subsection{Design Goal 3}
We analyzed participants' Likert-type responses to questions related to fulfilling our third design goal, which focuses on supporting researchers in following guidelines for academic blog posts (Figure \ref{fig:dg}d,h).

\subsubsection{Lab Study - Exploratory}
Participants' found it easier to produce appropriate blog post text (treatment: M=6.00, Q1=5.00, Q3=6.00; baseline: M=5.00, Q1=4.00, Q3=6.00) in the treatment condition (Figure \ref{fig:dg}d). Furthermore, the interaction logs for the treatment condition indicate that participants took advantage of affordances designed for generating and modifying text in alignment with the guidelines for academic blog posts (Appendix \ref{fig:treatActions}). For an average of 6.85 generations per treatment condition, participants often included custom instructions (5.70 times);
though participants may have modified the custom instructions, they were pre-filled by default to better fulfill the guidelines. In addition, participants modified text 6.85 times on average, with the ``condense'' (1.85 times), ``custom'' (1.80 times), and ``simpler terms'' (1.40 times) modifications being the most common.

\subsubsection{Deployment Study - Exploratory}
There was no difference in the median for how well \sysname\ (M=5.00, Q1=4.00, Q3=6.00) versus the baseline (M=5.00, Q1=5.00, Q3=6.00) facilitated production of appropriate text, unlike in the lab study (Figure \ref{fig:dg}e-h).

\subsubsection{Summary}
While lab study participants thought \sysname\ improved their ability to produce appropriate text for the blog post, deployment study participants did not. Perhaps without the pressure of time constraints, deployment study participants found it easier to determine how to prompt the baseline tool or adjust its generated text in order to produce appropriate writing.

\begin{figure*}
    \centering
    \includegraphics[width=\textwidth]{figures/legend.png} 
    
    \begin{subfigure}[b]{0.49\textwidth}
        \centering
        \vspace{-\abovecaptionskip}
        \captionsetup{labelformat=empty}
        \caption{Lab Study}
        \includegraphics[clip,width=\linewidth]{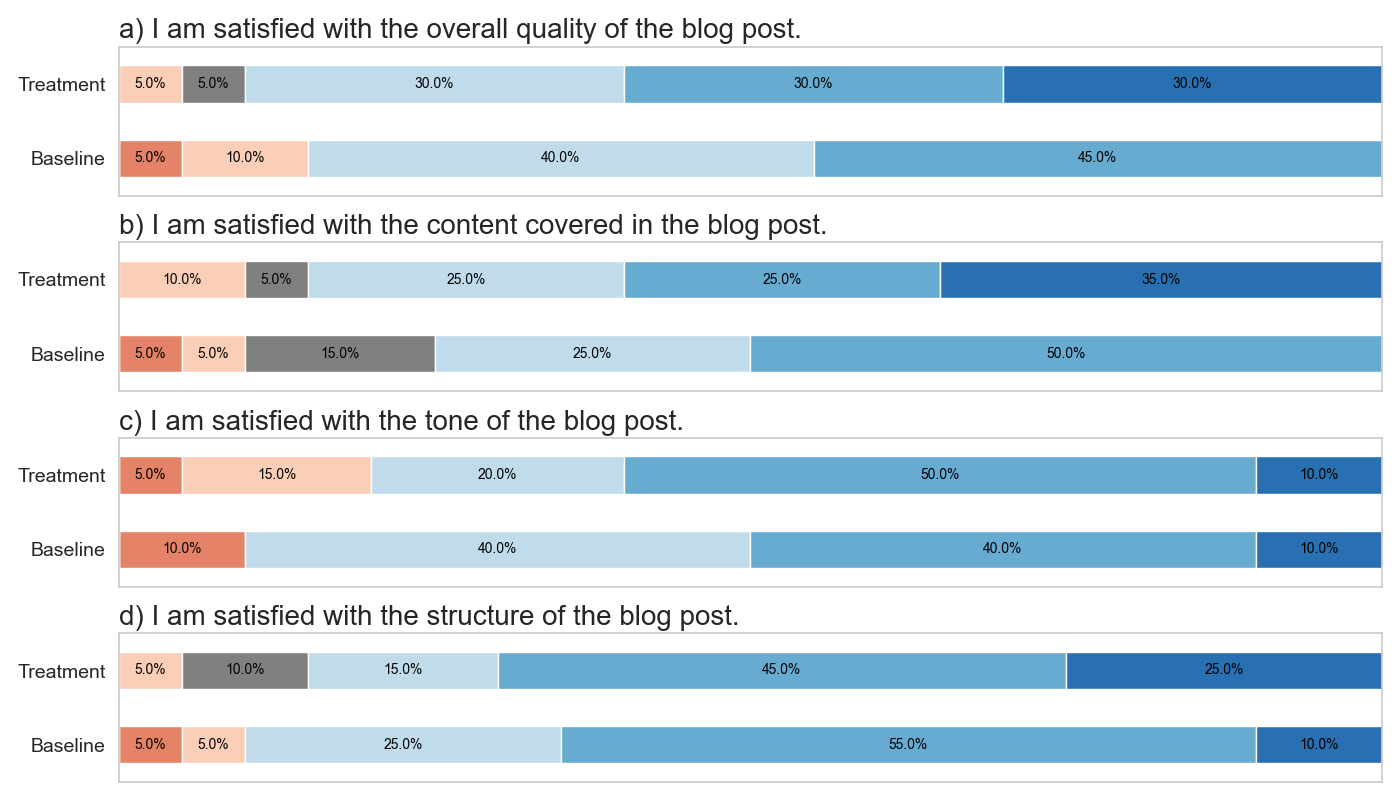}
    \end{subfigure}
    \hfill
    \begin{subfigure}[b]{0.49\textwidth}
        \centering
        \vspace{-\abovecaptionskip}
        \captionsetup{labelformat=empty}
        \caption{Deployment Study}
        \includegraphics[clip,width=\linewidth]{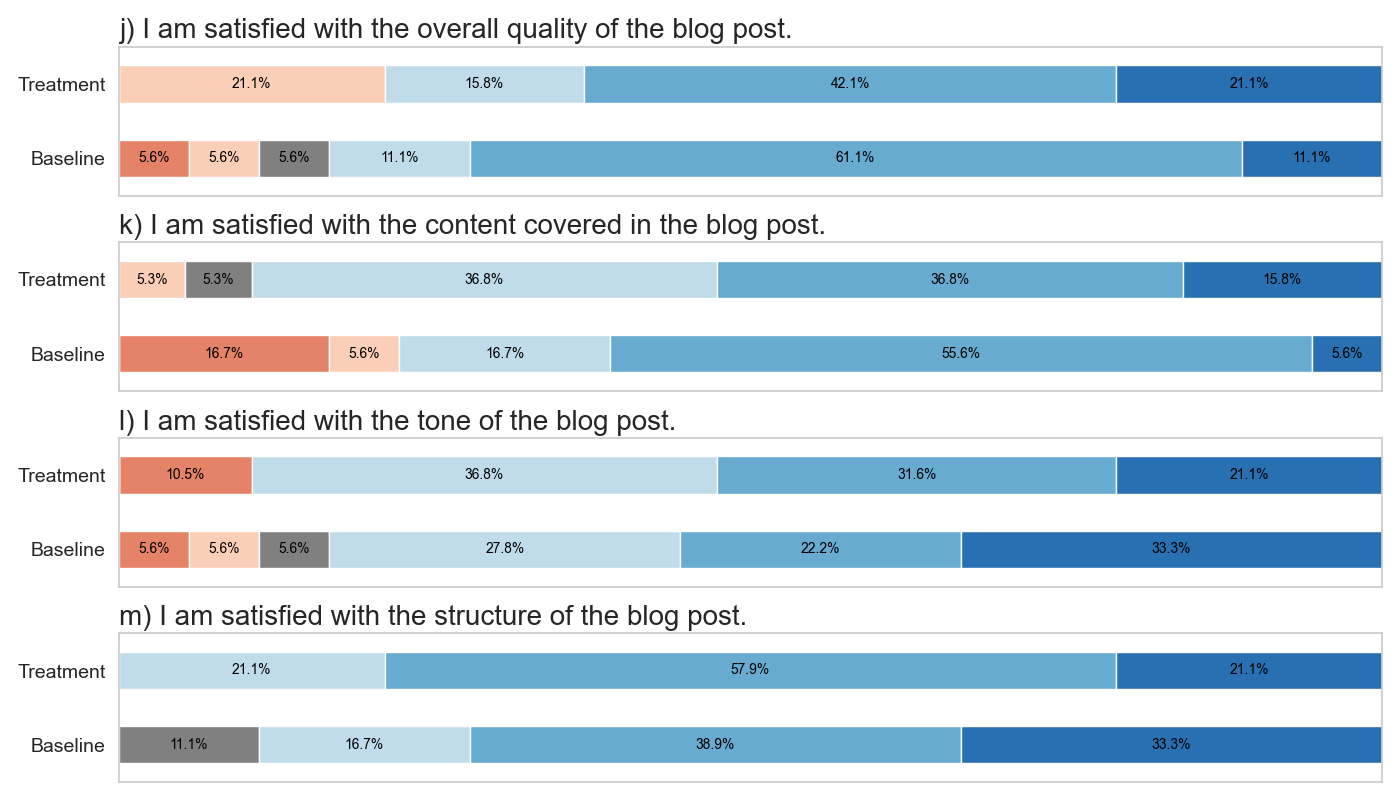}
    \end{subfigure}
    
    \begin{subfigure}[b]{0.49\textwidth}
        \centering
        \includegraphics[clip,width=\linewidth]{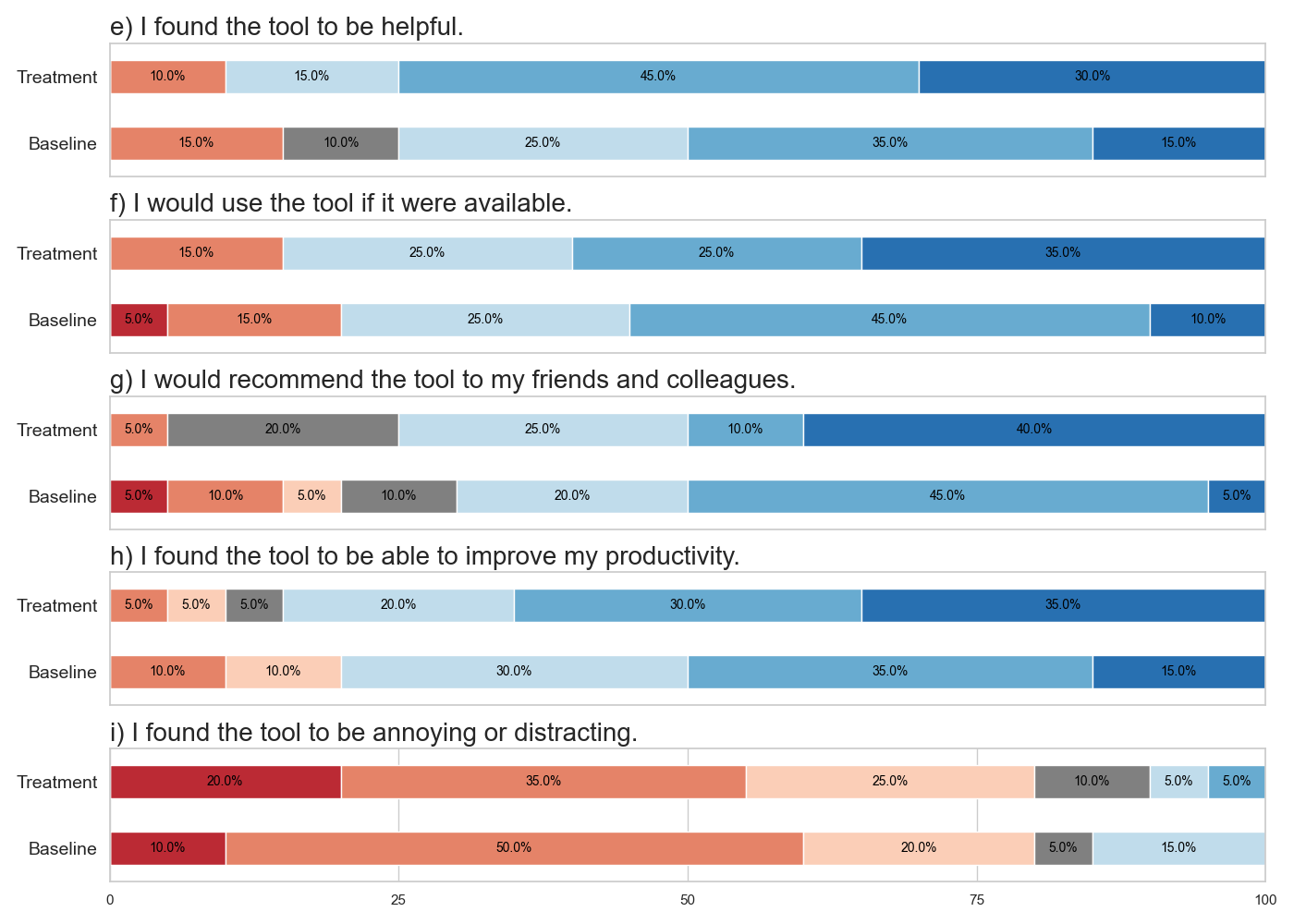}
    \end{subfigure}
    \hfill
    \begin{subfigure}[b]{0.49\textwidth}
        \centering
        \includegraphics[clip,width=\linewidth]{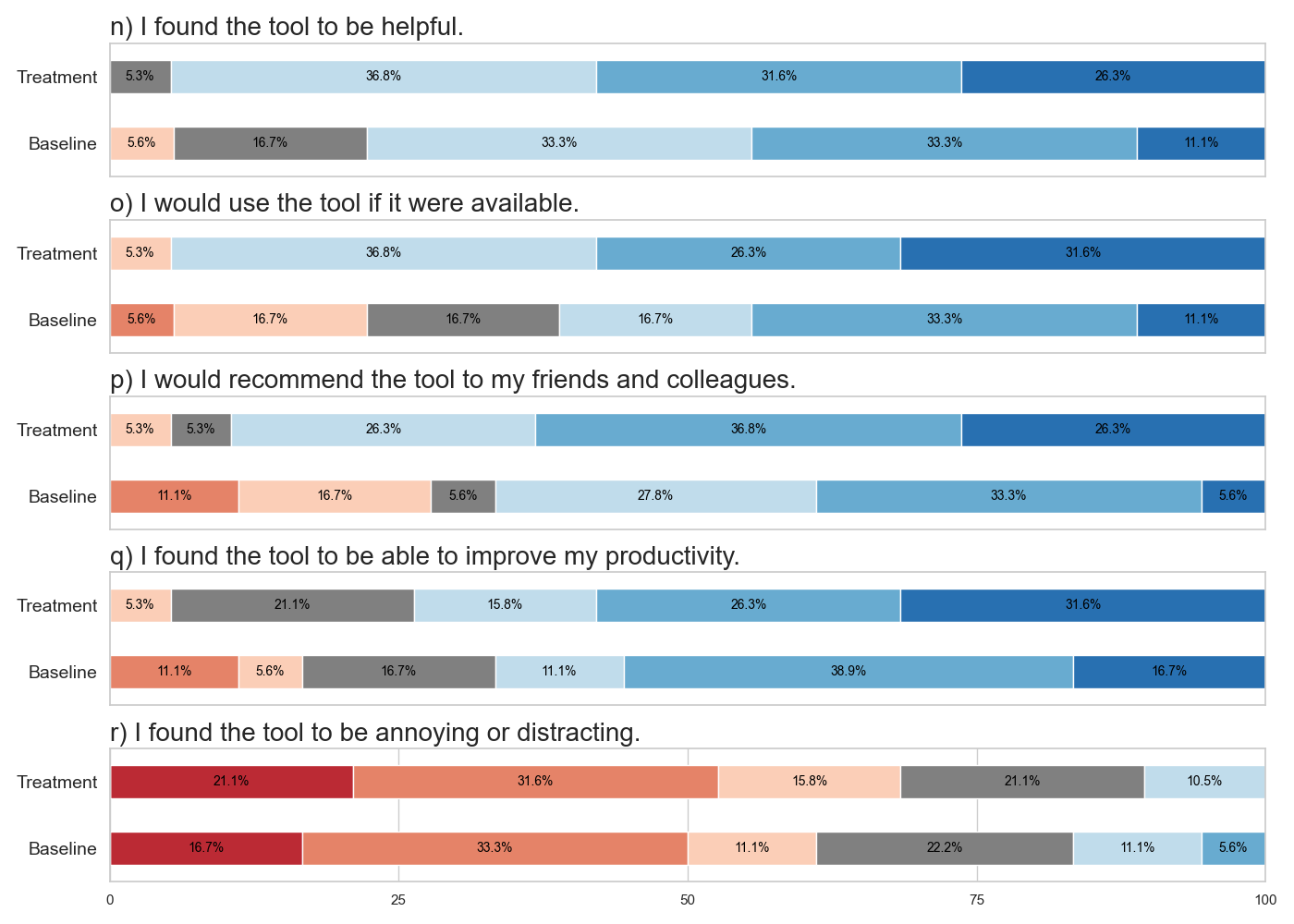}
    \end{subfigure}
    
    \caption{Survey responses to 7-point Likert-type questions regarding satisfaction with the a-d) output blog post in the lab study, e-i) tool in the lab study, j-m) output blog post in the deployment study, and n-r) tool in the deployment study. Responses are shown for both the treatment and baseline conditions.}
    \label{fig:blogtoolsat}
    \Description{The top of the figure shows a legend with the following mapping: dark red = strongly disagree, medium red = disagree, light red = somewhat disagree, gray = neither agree nor disagree, light blue = somewhat agree, medium blue = agree, and dark blue = strongly agree. Below, there are two areas. The left area is labeled "Lab Study," and the right area is labeled "Deployment Study." In each area, there are nine questions labeled "a" through "i" for the lab study and "j" through "r" for the deployment study. The questions are: 1) "I am satisfied with the overall quality of the blog post."; 2) "I am satisfied with the content covered in the blog post."; 3) "I am satisfied with the tone of the blog post."; 4) "I am satisfied with the structure of the blog post."; 5) "I found the tool to be helpful."; 6) "I would use the tool if it were available."; 7) "I would recommend the tool to my friends and colleagues."; 8) "I found the tool to be able to improve my productivity."; 9) "I found the tool to be annoying or distracting." Below each question, there are two horizontal stacked bars, one labeled "treatment" and the other "baseline." In the lab study area, treatment is shown to perform somewhat better than baseline for each question except the last, which is ambiguous. For each question, the medians for treatment followed by baseline are: 1) agree vs somewhat agree, 2) agree vs between somewhat agree and agree, 3) agree vs between somewhat agree and agree, 4) agree vs agree, 5) agree vs between somewhat agree and agree, 6) agree vs agree, 7) between somewhat agree and agree vs between somewhat agree and agree, 8) agree vs between somewhat agree and agree, and 9) disagree vs disagree. In the deployment study area, treatment is shown to perform somewhat better than baseline for the second question and fifth through ninth questions. For the first, third, and fourth question, it is ambiguous. For each question, the medians for treatment followed by baseline are: 1) agree vs agree, 2) agree vs agree, 3) agree vs agree, 4) agree vs agree, 5) agree vs somewhat agree, 6) agree vs somewhat agree, 7) agree vs somewhat agree, 8) agree vs agree, and 9) disagree vs between disagree and somewhat disagree.}
\end{figure*}

\begin{figure*}
    \centering
    \includegraphics[width=.6\textwidth]{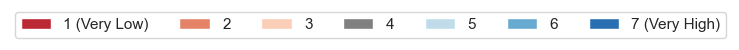} 
    
    \begin{subfigure}[b]{0.49\textwidth}
        \centering
        \vspace{-\abovecaptionskip}
        \captionsetup{labelformat=empty}
        \caption{Lab Study}
        \includegraphics[clip,width=\linewidth]{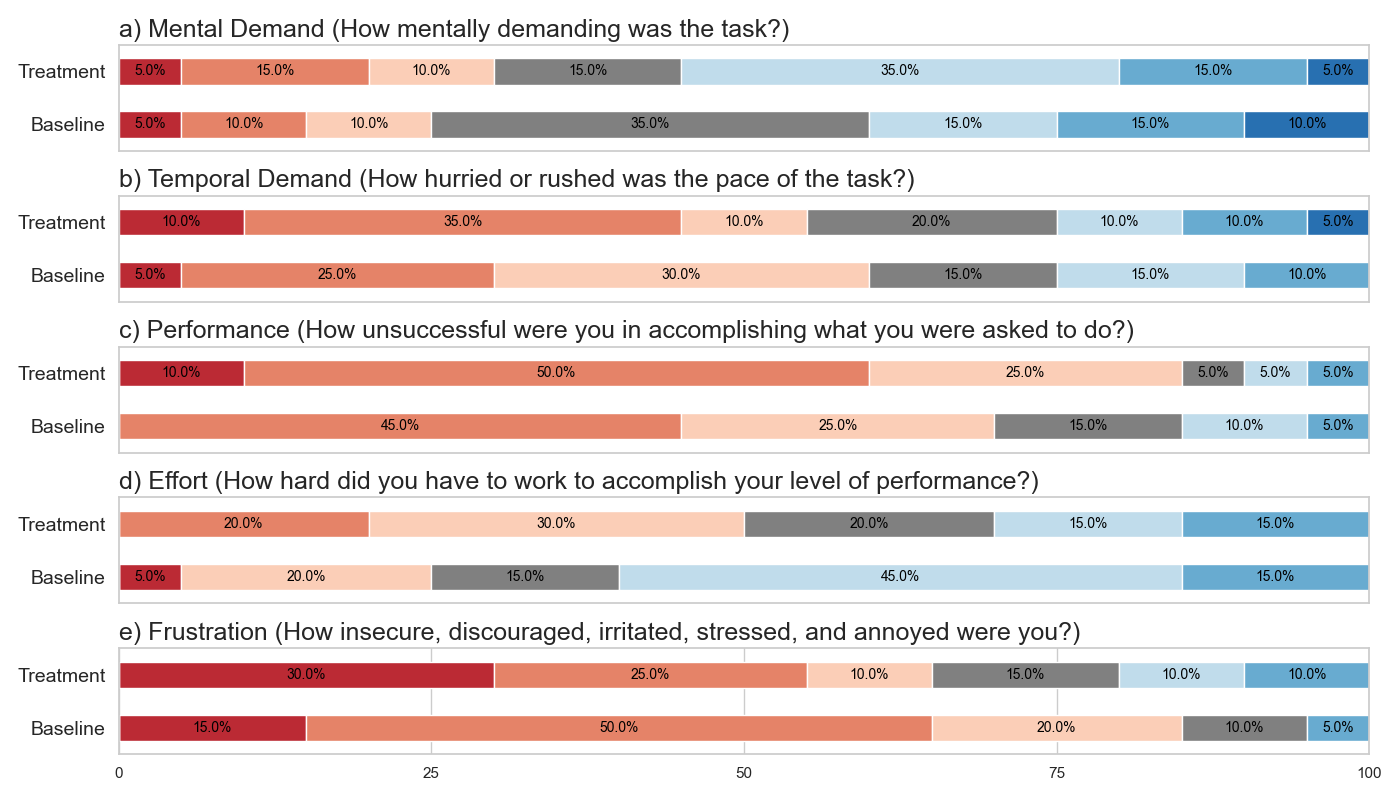}
    \end{subfigure}
    \hfill
    \begin{subfigure}[b]{0.49\textwidth}
        \centering
        \vspace{-\abovecaptionskip}
        \captionsetup{labelformat=empty}
        \caption{Deployment Study}
        \includegraphics[clip,width=\linewidth]{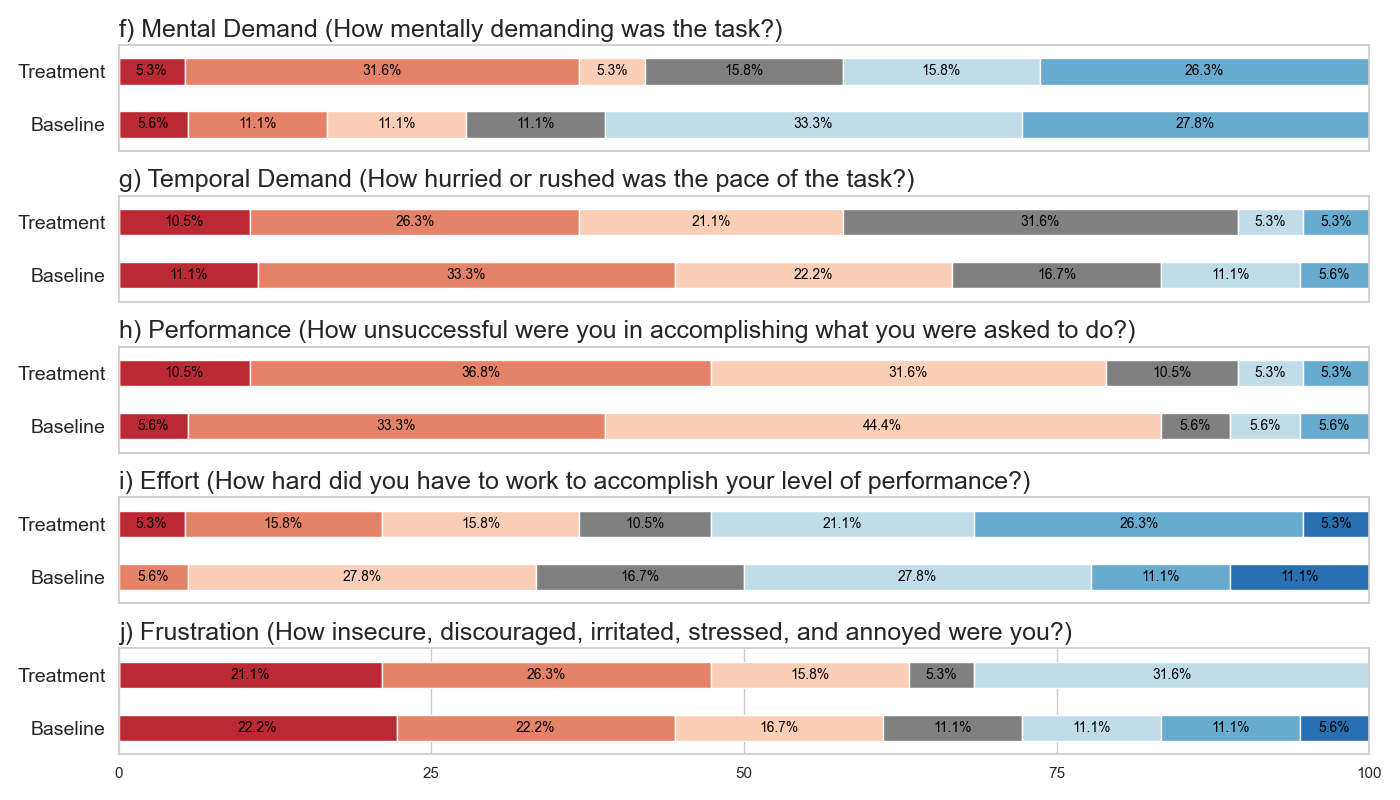}
    \end{subfigure}
    
    \caption{Survey responses to 7-point Likert-type questions regarding cognitive load in the a-e) lab study and f-j) deployment study. Responses are shown for both the treatment and baseline conditions. Please note that for the performance question, the labels for the left and right end of the scale were actually respectively ``Perfect Performance'' and ``Failure Performance.''}
    \Description{The top of the figure shows a legend with the following mapping: dark red = 1 (very low), medium red = 2, light red = 3, gray = 4, light blue = 5, medium blue = 7, and dark blue = 7 (very high). Below, there are two areas. The left area is labeled "Lab Study," and the right area is labeled "Deployment Study." In each area, there are five questions labeled "a" through "e" for the lab study and "f" through "j" for the deployment study. The questions are: 1) "Mental Demand (How mentally demanding was the task?)"; 2) "Temporal Demand (How hurried or rushed was the pace of the task?)"; 3) "Performance (How unsuccessful were you in accomplishing what you were asked to do?)"; 4) "Effort (How hard did you have to work to accomplish your level of performance?)"; and 5) "Frustration (How insecure, discouraged, irritated, stressed, and annoyed were you?)." Below each question, there are two horizontal stacked bars, one labeled "treatment" and the other "baseline." In the lab study area, treatment is shown to perform better than baseline for effort and somewhat better for performance, while baseline is shown to perform somewhat better than baseline for temporal demand and frustration. For mental demand, it is ambiguous which performed better. For each question, the medians for treatment followed by baseline are: 1) 5 vs 4, 2) 3 vs 3, 3) 2 vs 3, 4) between 3 and 4 vs 5, and 5) 2 vs 2. In the deployment study area, treatment is shown to perform somewhat better than baseline for mental demand. For the rest which did better is ambiguous. For each question, the medians for treatment followed by baseline are: 1) 4 vs 5, 2) 3 vs 3, 3) 3 vs 3, 4) 5 vs between 4 and 5, and 5) 3 vs 3.}
    \label{fig:cl}
\end{figure*}

\subsection{Blog Post Satisfaction} 
\label{sec:bpSat}
To measure participants' satisfaction with their final blog posts, we utilized their Likert-type response to the statement ``I am satisfied with the overall quality of the blog post'' (Figure \ref{fig:blogtoolsat}a,j). 

\subsubsection{Lab Study - H1}
\label{sec:bpSatLab}
To compare responses under the two conditions, we employed a non-parametric test, given the data's Likert nature. Furthermore, we used a sign test rather than a Wilcoxon signed rank test, which is normally used to analyze non-parametric within-subjects data for two conditions of one factor, due to the violation of that test's assumption of a symmetric distribution of the paired differences about the median. Participants were significantly more satisfied with the final blog post generated with \sysname\ (M=6.00, Q1=5.00, Q3=7.00) as compared to the baseline tool (M=5.00, Q1=5.00, Q3=6.00) (Dependent-Samples Sign Test, S=11, p<.05). The majority of participants (11) were more satisfied in the treatment condition (equally: 7, less: 2).

Inspecting participants' Likert responses regarding aspects of their blog post satisfaction, we see that participants particularly demonstrated more satisfaction with the content covered in the \sysname\ blog post (M=6.00, Q1=5.00, Q3=7.00) compared to the baseline blog post (M=5.50, Q1=4.75, Q3=6.00) (Figure \ref{fig:blogtoolsat}b). The majority of participants (14) were more satisfied with the content under the treatment condition (equally: 2, less: 4). The median participant was only slightly more satisfied with the blog post's tone under the treatment (M=6.00, Q1=5.00, Q3=6.00) compared to the baseline (M=5.50, Q1=5.00, Q3=6.00) (Figure \ref{fig:blogtoolsat}c).
With respect to the blog post's structure, the median participant was equally satisfied using \sysname\ (M=6.00, Q1=5.00, Q3=6.25) compared to the baseline (M=6.00, Q1=5.00, Q3=6.00) (Figure \ref{fig:blogtoolsat}d). 

In addition, looking at participants' perceived failure to complete their task (Figure \ref{fig:cl}c), we observe that the median participant had lower perceived failure with \sysname\ (M=2.00, Q1=2.00, Q3=3.00) than with the baseline (M=3.00, Q1=2.00, Q3=4.00). The plurality of participants (9) thought that they were more successful in writing a research-paper blog post with \sysname\ compared to the baseline (equally: 7, less: 4). This further corroborates participants' blog post satisfaction under the treatment condition.

\subsubsection{Deployment Study - RQ2a}
In the deployment study, we observed no difference in the median blog post satisfaction under \sysname\ (M=6.00, Q1=5.00, Q3=6.00) versus the baseline (M=6.00, Q1=5.25, Q3=6.00) (Figure \ref{fig:blogtoolsat}j). Again with respect to median satisfaction, participants were equally satisfied with the blog post content (treatment: M=6.00, Q1=5.00, Q3=6.50; baseline: M=6.00, Q1=3.50, Q3=6.00; Figure \ref{fig:blogtoolsat}k), tone (treatment: M=6.00, Q1=5.00, Q3=6.00; baseline: M=6.00, Q1=5.00, Q3=7.00; Figure \ref{fig:blogtoolsat}l), and structure (treatment: M=6.00, Q1=6.00, Q3=6.00; baseline: M=6.00, Q1=5.25, Q3=7.00; Figure \ref{fig:blogtoolsat}m) when using \sysname\ versus the baseline. In terms of perceived failure to complete their task (Figure \ref{fig:cl}h), participants had equal median perceived failure with \sysname\ (M=3.00, Q1=2.00, Q3=3.00) and the baseline (M=3.00, Q1=2.00, Q3=3.00).

\subsubsection{Summary}
We observed a significant increase in blog post satisfaction under the treatment condition during the lab study and found that participants were particularly more satisfied with the blog post content, suggesting that \sysname\ best addressed its design goals to support understanding (DG1) and control (DG2) of the content included in the blog post. We did not observe the same increases in blog post satisfaction in the deployment study. This discrepancy may be related to the fact that participants were not writing under the pressure of a time constraint. With sufficient time to write the blog post to their liking (only three blog posts were not shared with others), participants may not have noticed a pronounced difference in blog post quality between conditions. 

\subsection{Tool Satisfaction} 
For measuring tool satisfaction, we used the five-question Likert scale on tool acceptance from Kocielnik et al. \cite{kocielnik2019will} (Figure \ref{fig:blogtoolsat}e-i). The five answers were averaged to obtain one score. 

\subsubsection{Lab Study - H2} 
We used the sign test to evaluate the lab study data, as it was non-parametric data and violated the Wilcoxon signed rank test's assumption of a symmetric distribution of the paired differences about the median. While the median participant had a higher satisfaction with \sysname\ (M=5.90, Q1=5.20, Q3=6.50) than the baseline tool (M=5.60, Q1=4.80, Q3=6.00), this difference was not significant (Dependent-Samples Sign Test, S=13, p=n.s.).

Based on median results, participants under the treatment condition more often found the tool helpful (treatment: M=6.00, Q1=5.75, Q3=7.00; baseline: M=5.50, Q1=4.75, Q3=6.00) and useful for productivity (treatment: M=6.00, Q1=5.00, Q3=7.00; baseline: M=5.50, Q1=5.00, Q3=6.00). Meanwhile, participants were equally likely under both conditions to use the tool if it were available (treatment: M=6.00, Q1=5.00, Q3=7.00; baseline: M=6.00, Q1=5.00, Q3=6.00), recommend the tool (treatment: M=5.50, Q1=4.75, Q3=7.00; baseline: M=5.50, Q1=4.00, Q3=6.00), and find the tool annoying or distracting (treatment: M=2.00, Q1=2.00, Q3=3.00; baseline: M=2.00, Q1=2.00, Q3=3.00). For each of the first four questions, at least four more participants increased rather than decreased their Likert rating from the baseline to the treatment condition. However, only two more participants indicated decreased rather than increased annoyance with \sysname\ (less: 8, more: 6, equal: 6).

\subsubsection{Deployment Study - RQ2b} 
Similarly to the lab study, we observed little increase in the median tool satisfaction under the treatment condition (M=5.60, Q1=5.00, Q3=6.40) compared to the baseline condition (M=5.40, Q1=4.05, Q3=5.80) (Figure \ref{fig:blogtoolsat}n-r). Based on median results, participants under the treatment condition were more likely to find the tool helpful (treatment: M=6.00, Q1=5.00, Q3=6.50; baseline: M=5.00, Q1=5.00, Q3=6.00), use it if it were available (treatment: M=6.00, Q1=5.00, Q3=7.00; baseline: M=5.00, Q1=4.00, Q3=6.00), and recommend the tool (treatment: M=6.00, Q1=5.00, Q3=6.50; baseline: M=5.00, Q1=3.25, Q3=6.00). They were also slightly less likely to find the tool annoying (treatment: M=2.00, Q1=2.00, Q3=4.00; baseline: M=2.50, Q1=2.00, Q3=4.00). On the other hand, they were equally likely under both conditions to find the tool useful for productivity (treatment: M=6.00, Q1=4.50, Q3=7.00; baseline: M=6.00, Q1=4.00, Q3=6.00). 

\subsubsection{Summary} 
In both studies, we observed little difference in satisfaction with \sysname\ versus the baseline tool. However, participants appeared to find \sysname\ more helpful than the baseline across the two studies.

\subsection{Cognitive Load}
To measure cognitive load, we used the NASA TLX Index questionnaire \cite{hart1988development} and summed the responses for the mental demand, temporal demand, effort, and frustration questions (Figure \ref{fig:cl}a-b,d-e)=.\footnote{See \ref{sec:bpSat} for discussion of the responses to the performance question}

\subsubsection{Lab Study - H3} 
We utilized the sign test to evaluate the lab study data, as it was non-parametric data and violated the Wilcoxon signed rank test's assumption of a symmetric distribution of the paired differences about the median. We saw no significant difference between the baseline and treatment conditions (Dependent-Samples Sign Test, S=8, p=n.s.). The median participant reported slightly lower median cognitive load with \sysname\ (M=13.00, Q1=11.50, Q3=16.30) than the baseline tool (M=14.50, Q1=12.00, Q3=18.30).

We saw the largest difference in median results with respect to perceived effort (treatment: M=3.50, Q1=3.00, Q3=5.00; baseline: M=5.00, Q1=3.75, Q3=5.00). The majority of participants (11) reported using less effort with \sysname\ than with the baseline tool (more: 4, equal: 5). Meanwhile, \sysname\ led to little change in perceived mental demand (treatment: M=5.00, Q1=3.00, Q3=5.00; baseline: M=4.00, Q1=3.75, Q3=5.25). The small plurality of participants (9) perceived more mental demand with \sysname\ (less: 8, equal: 3). We saw no change in the median result for perceived temporal demand (treatment: M=3.00, Q1=2.00, Q3=4.25; baseline: M=3.00, Q1=2.00, Q3=4.25) or frustration (treatment: M=2.00, Q1=1.00, Q3=4.00; baseline: M=2.00, Q1=2.00, Q3=3.00). 

\subsubsection{Deployment Study - RQ2c} 
Once again, we found very slightly lower median cognitive load under the treatment (M=15.00, Q1=9.50, Q3=17.50) in comparison to the baseline (M=15.50, Q1=10.75, Q3=18.75) (Figure \ref{fig:cl}f-g,i-j). Considering the median participant, the treatment led to little change in perceived mental demand (M=4.00, Q1=2.00, Q3=5.50) than the baseline (M=5.00, Q1=3.25, Q3=5.75) but higher perceived effort (treatment: M=5.00, Q1=3.00, Q3=6.00; baseline: M=4.50, Q1=3.00, Q3=5.00). We found no change in the median result for perceived temporal demand (treatment: M=3.00, Q1=2.00, Q3=4.00; baseline: M=3.00, Q1=2.00, Q3=4.00) or frustration (treatment: M=3.00, Q1=2.00, Q3=5.00; baseline: M=3.00, Q1=2.00, Q3=4.75).

\subsubsection{Summary} 
In both studies, we observed little difference in cognitive load from the baseline to the treatment condition. Participants consistently perceived little difference in the tools with respect to mental demand, temporal demand, and frustration. However, their perceptions varied across the studies in terms of effort.

\subsection{Task Completion Time} 
\label{sec:timeSpent}
\subsubsection{Lab Study - H4} 
We recorded two metrics for task completion time. One was the amount of time participants needed \emph{during} the session before they would be comfortable publicly publishing the blog post. The other was the perceived amount of time participants would need \emph{after} the session before they would be comfortable publicly publishing the blog post. As participants reached that point of comfort \emph{during} most sessions, we focused on the former metric. 
The mean participant worked slightly faster with the baseline tool (M=44.32 minutes, SD=12.51) in comparison to \sysname\ (M=45.99 minutes, SD=13.88) (Appendix \ref{fig:labtimespent}), but this difference was not significant (Paired-Samples \emph{t}-Test, t(19)=0.55, p=n.s.). 

\subsection{Qualitative Analysis - DG1 and DG2}
Across the two studies, participants’ survey and interview comments reflected the difference between the two tools in terms of supporting understanding (DG1) and control (DG2) of what content is versus is not included in the blog post. 

Overall, participants using the baseline tool most commonly cited the initial draft (lab: 6/20, deployment: 8/18) as a helpful aspect of the tool. Thus, the most valued baseline feature was not useful for understanding or controlling what content was omitted from the detailed summary. With respect to the initial draft, multiple participants noted editing or adding to the initial draft rather than replacing it. For example, P20-dep explained, ``\textit{It's easier to critique than it is to generate,}'' and P26-dep shared, ``\textit{It was very useful to have a straw-person of what the blog post could look like (generated by the LLM). I mostly used the draft as a place to edit and steer in the direction I wanted the post to take.}'' Without an affordance to help them consider all the possibilities of content to include in the blog post, the participants seem to have focused more on augmenting the content already selected by the tool.

With respect to \sysname, participants often noted the selectable paper bullet points and paragraphs as a helpful feature (lab: 10/20, deployment: 14/19).
Several described the outline as helpful for selecting relevant content to generate text. For instance, P15-lab noted, \emph{``I could rewrite entire sections or direct it with just a few bullets.''} Seven participants 
mentioned that the outline was useful even just for reviewing the paper's content or structure. P4-lab commented, \emph{``The outline was really helpful in providing a quick overview of what content I had written [in the paper].''} For two lab participants, the bullet points provided transparency around the origins of the initial LLM draft and generations.
P12-lab explained, \emph{``I liked the bullet point features to see what was being fed in vs not.''} Thus, participants' interviews reflected how \sysname' outline affordance facilitated review and selection of content to include in the blog post.

That said, participants also specified difficulties related to the outline mechanism. For instance, two lab participants were overwhelmed with the number of bullet points. P9-lab described the outline as \emph{``really bloated and hard to navigate.''} Meanwhile, three lab participants were confused or encumbered by aspects of utilizing the bullet points. For example, P3-lab \emph{``felt stressed by choosing between the entire paragraph at hand and the (presumably) AI-generated summary bullets.''} Future work may investigate how to make the outline mechanism more usable.

\subsection{Qualitative Analysis - DG3}
Across the two studies, participants’ survey and interview comments reflected the difference between the two tools in the tradeoff between supporting flexibility and providing scaffolding for alignment with academic blog post guidelines (DG3).

A number of participants noted the baseline tool's flexibility as a benefit (lab: 6/20, deployment: 4/18). For example, P18-lab shared, ``\textit{I was able to provide instructions that fit my mental model of constructing a blog post and it being there as a `copilot' is just what I need.}'' Flexibility in drafting text may help writers obtain the edits they want. However, the writer holds more responsibility in steering the LLM appropriately; P7-lab reflected, ``\textit{As a CS student, I like this way more [than the treatment's way of supporting writing], but I think it needs more thinking to use it creatively.}"

Regarding \sysname, participants frequently mentioned the utility of its text modification abilities, which were designed with the guidelines in mind (lab: 12/20, deployment: 9/19). For example, P18-dep reflected, ``\textit{The `MODIFY WRITING' section is helpful. I like that the interface contains this particular section to support iterating on existing text.}'' All of the preset modification buttons (``simpler terms'', ``condense'', ``expand'', ``more dramatic'', ``less dramatic'') were specifically mentioned as helpful affordances at least once in the lab study alone. The most commonly noted one was the ``condense'' button. For example, P12-lab commented, \emph{``I sometimes put the modify text through multiple rounds of expand and contract, focusing on different parts. I found that helpful.''} Three participants commented that they desired more preset modification buttons. For instance, P16-lab wanted a ``more technical'' button, while P14-dep wanted \textit{``more or less casual''} options.
P7-lab then added, \emph{``Or even let user to add their favorite rewrite button.''} Interestingly, P2-lab found the preset modification buttons useful for understanding the tool's capabilities: \emph{``It was nice to know what options the tool already knew of, and would perhaps have success with. For example, I would probably not have gone to a blank-box tool and said, 'please make the following more dramatic'. I would have stuck to more basic things like `please summarize the following'.''} Several participants (lab: 5/20, deployment: 10/19) also noted the usefulness of at least one aspect of the customizable instructions for either generating or modifying text using \sysname. P9-shared, \emph{``...being able to include explicit instructions for the model to generate text from was helpful in being able to control the information in the text that it generated.''} Thus, participants' comments indicate that \sysname's preset yet flexible LLM instructions for generating academic blog post text provided utility. 

Participants also discussed difficulties that they encountered in devising instructions to \sysname. Four lab participants expressed wanting to take specific actions that were not explicitly supported by \sysname, contrasting with the baseline's flexibility that participants appreciated. Two wanted assistance with checking for redundancies in the blog post, and the other two wanted feedback on the blog post writing. Regarding the affordances for modifying text, 8/20 lab and 5/19 deployment participants noted a difficulty. Four lab participants raised confusion about how or when to use certain affordances. As an example,
P6-lab was unsure of \emph{``when to use which one of the six [modification] options.''} Another recurring issue noted by five lab participants was a modification affordance not working as desired. For P20-lab, setting the desired length of the modified writing to one paragraph still resulted in three, as the back-end prompts are not guaranteed to work as planned. For P12-lab, modifications were sometimes too exaggerated, while for P16-lab, a custom modification to make writing more technical was too subdued. A couple deployment participants also noted that modifying text sometimes introduced errors. Future work may address these points of restriction and confusion noted by the participants using \sysname.

\section{Discussion}
Based on our results, we reflect on design implications for other mixed-initiative tools for detailed long-document summarization.

\subsection{Post-Editing is Not Enough}
Both \sysname\ and the baseline tool included initial drafts generated with powerful LLMs. 
Nevertheless, while participants had the option to simply post-edit the initial draft and refrain from interacting further with the LLM, only two baseline and two treatment participants across the lab and deployment studies did so. This suggests that \textbf{writers may benefit from additional support other than an initial draft when writing detailed summaries of long-documents.} 

\subsection{Increasing Understanding and Control of Content Selection for LLM-Generated Detailed Summaries}
We presented a novel mechanism to support understanding and controlling the content of detailed long-document summaries-- an interactive LLM-generated outline of the long document. This mechanism was implemented in \sysname. Compared to a strong baseline, \sysname\ made incorporating research-paper content in a blog post summary easier for researchers. Furthermore, under time constraints, researchers were more satisfied with their blog posts written using \sysname, particularly with respect to their content coverage. Researchers also found it easier to iterate on their blog posts with \sysname. Although the scaffolded and preset LLM instructions for adhering to academic blog post guidelines may have also facilitated iteration, researchers consistently noted the interactive paper outline as one of the most useful features of the tool. In alignment with increased ease of iteration, researchers experienced increased editing power (i.e., change in the blog post per minute or writing action) with \sysname. We therefore conclude that \textbf{LLM-generated interactive outlines of long documents show promise for making detailed summaries easier to write}. 

On the other hand, researchers often described the baseline as more flexible and allowing them more freedom. Thus, utilizing the two tools in combination may prove useful. A tool like \sysname\ may be used when the writer is unsure of what content to include in their detailed long-document summary, and a tool like the baseline tool could be used when the writer already has a specific set of points in mind or wants to give themselves creative space to use the LLM in a variety of manners.

\section{Limitations and Future Work}
Most participants were PhD students in computer science, who may benefit from raising awareness of their early-career work through blog posts; future work may investigate how more senior researchers utilize a tool like \sysname. Future studies can also examine how researchers in other fields, who are generally less familiar with LLMs and may have different traditional paper formats, may benefit from a tool like \sysname. 
Future work may additionally investigate how different aspects of the input paper such as length, recency, and author position affect the user experience and outcome.

\sysname\ itself has several limitations. For one, multiple deployment study participants commented on how the automatic paper parsing makes mistakes and misses parts of the paper text. The tool also does not support incorporating figures and tables in the blog post. 
Moreover, how detailed the paper outline should be remains uncertain. While several participants appreciated the bullet points, a couple mentioned that the outline was overwhelming. In addition, \sysname' outline only had one level of bullet points, but future work could explore the effects of a hierarchical outline. Lastly, future work may look into how writers' expertise may be utilized in supporting their blog post writing; a recent tool for Q-and-A research-paper summarization, for example, utilizes the researcher's background to personalize its output \cite{lim2024co}.

\section{Conclusion}
We introduce \textit{interactive reverse source outlines}, a novel mechanism for summarizing long-documents in detailed articles. The mechanism consists of an LLM-generated reverse outline of the source document with pre-selected bullet points for an initial draft of the detailed summary, which the user manipulates to control the content covered in the summary. We implement this mechanism in \sysname, an LLM-powered tool for writing research-paper blog posts. We validated that \sysname\ helps users to understand and control their detailed summaries' content through two user studies: a within-subjects lab study (N=20 participants) and a between-subjects deployment study (N=37 blog posts, 26 participants). The studies compared how researchers write about their own papers with \sysname\ versus a strong baseline (LLM blog post draft and free-form LLM prompting). In the time-constrained within-subjects study, \sysname\ significantly increased participants' satisfaction with their final blog posts, especially in terms of the content covered. Furthermore, across both studies, participants found incorporating document content in their blog posts easier and demonstrated increased editing power. We hope that this work informs future research and tools related to human-LLM long-document summarization.

\begin{acks}
This research was supported by the Allen Institute for Artificial Intelligence (AI2). The authors thank the Semantic Scholar team for helpful feedback on this work and the participants who made this work possible.
\end{acks}

\bibliographystyle{ACM-Reference-Format}
\bibliography{refs}

\appendix

\begin{figure*}[tb]
    \centering
    \includegraphics[width=\textwidth]{figures/legend.png} 
    
    \begin{subfigure}[b]{0.49\textwidth}
        \centering
        \vspace{-\abovecaptionskip}
        \captionsetup{labelformat=empty}
        \caption{Lab Study}
        \includegraphics[clip,width=\linewidth]{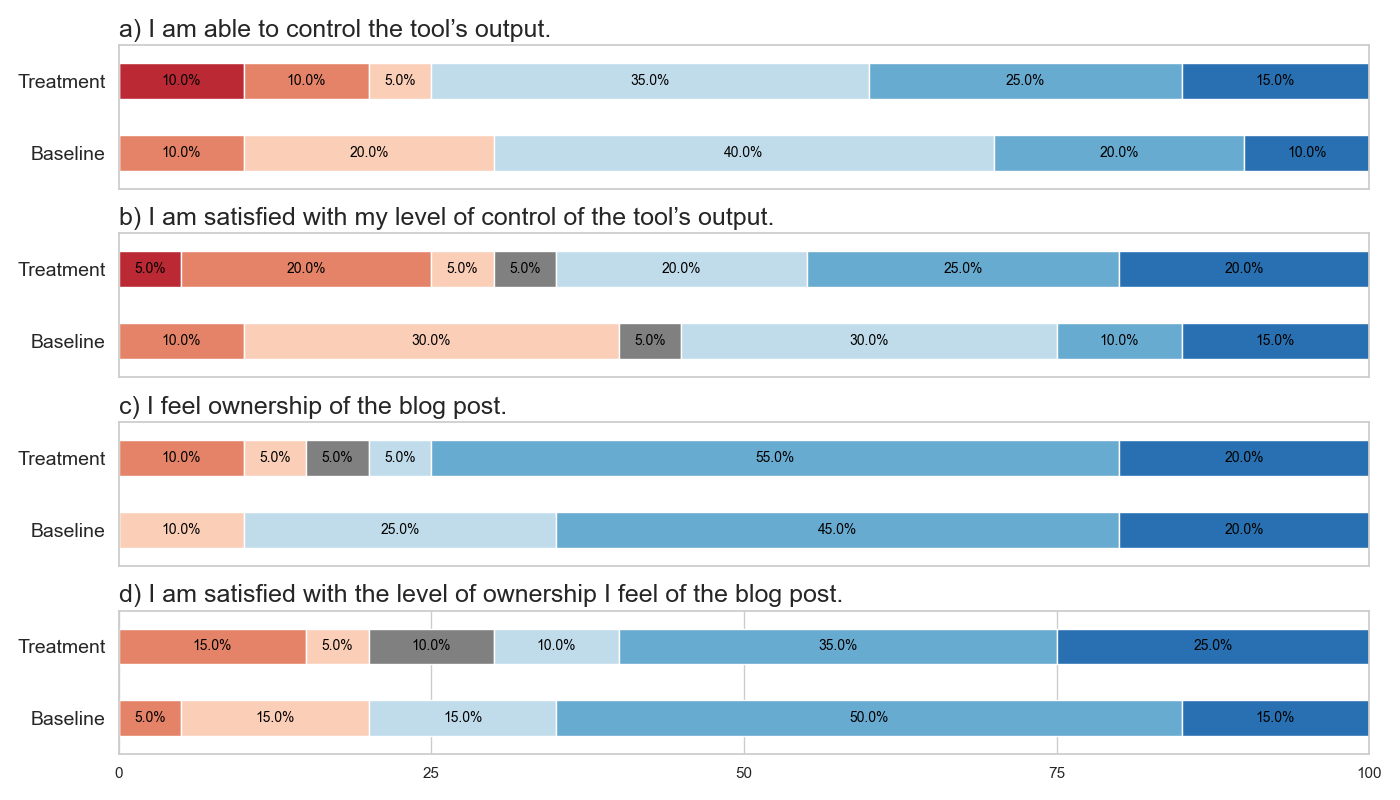}
    \end{subfigure}
    \hfill
    \begin{subfigure}[b]{0.49\textwidth}
        \centering
        \vspace{-\abovecaptionskip}
        \captionsetup{labelformat=empty}
        \caption{Deployment Study}
        \includegraphics[clip,width=\linewidth]{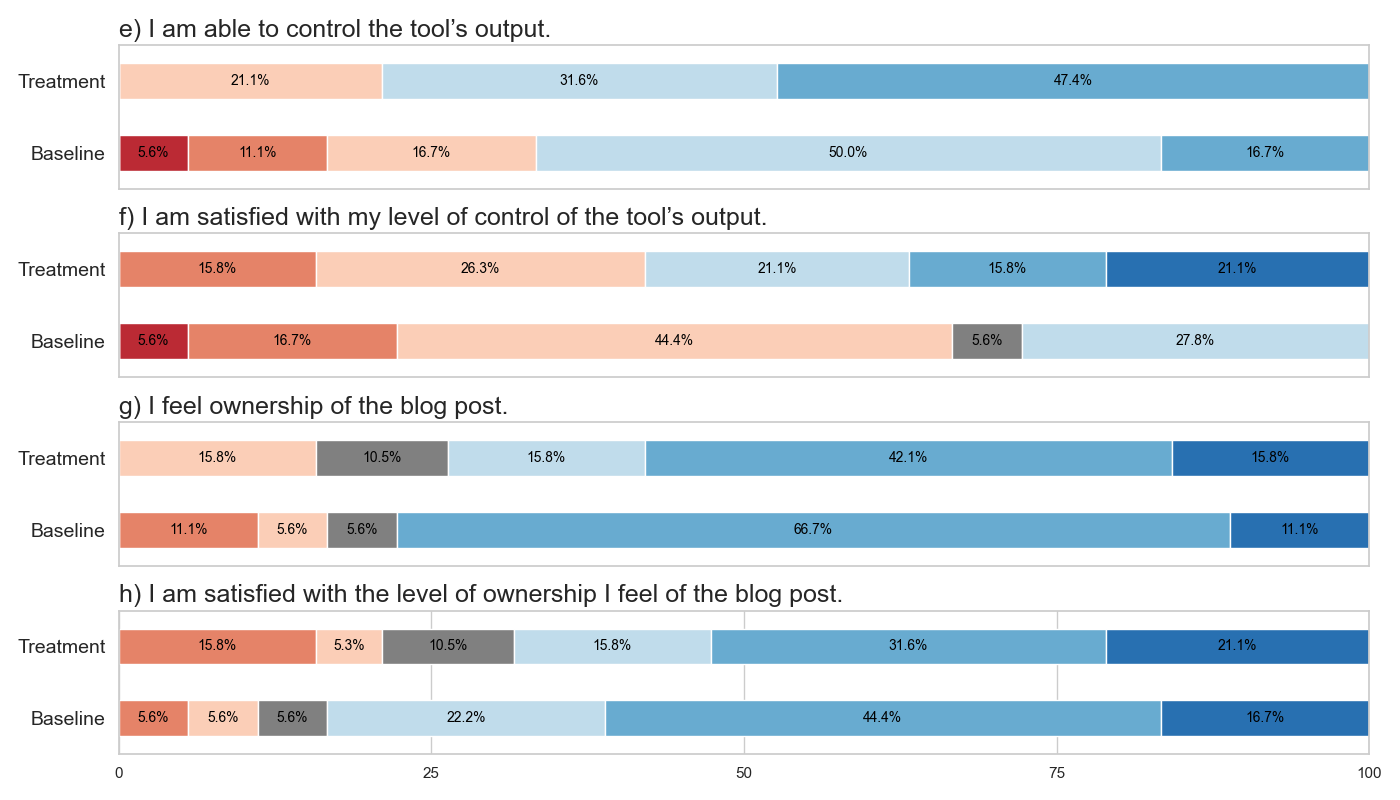}
    \end{subfigure}
    
    \caption{Survey responses to 7-point Likert-type questions regarding perceived control and ownership in the a-d) lab study and e-h) deployment study. Responses are shown for both the treatment and baseline conditions.}
    \Description{}
    \label{fig:control}
\end{figure*}

\section{Exploratory Analysis of Perceived Control and Ownership}
\label{sec:control}
We analyzed participants' Likert-type responses to questions related to control and ownership (Figure \ref{fig:control}).

\subsection{Lab Study} 
The median participant did not indicate higher perceived control (treatment: M=5.00, Q1=4.50, Q3=6.00; baseline: M=5.00, Q1=3.00, Q3=6.00) or satisfaction with their control (treatment: M=5.00, Q1=2.75, Q3=6.00; baseline: M=5.00, Q1=3.00, Q3=5.25) under the treatment condition (Figure \ref{fig:control}a-b). However, if we look within-subjects, 11 participants had higher perceived control with \sysname\ than the baseline tool (equal: 2, lower: 7), and 10 participants had higher satisfaction with their control using \sysname\ (equal: 4, lower: 6). On the other hand, participants did not demonstrate increased perceived ownership (treatment: M=6.00, Q1=5.75, Q3=6.00; baseline: M=6.00, Q1=5.00, Q3=6.00) or satisfaction with ownership (treatment: M=6.00, Q1=4.00, Q3=6.25; baseline: M=6.00, Q1=5.00, Q3=6.00) under the treatment condition (Figure \ref{fig:control}c-d). Although they felt that they had more control over \sysname, participants may not necessarily have felt that the tool contributed more or less to the blog post's creation.

\subsection{Deployment Study} 
As shown in Figure \ref{fig:control}e-h, participants indicated higher median satisfaction with control in the treatment condition (treatment: M=5.00, Q1=3.00, Q3=6.00; baseline: M=3.00, Q1=3.00, Q3=4.75), but they showed no difference in median perceived control (treatment: M=5.00, Q1=5.00, Q3=6.00; baseline: M=5.00, Q1=3.00, Q3=5.00), perceived ownership (treatment: M=6.00, Q1=4.50, Q3=6.00; baseline: M=6.00, Q1=6.00, Q3=6.00) or satisfaction with ownership (treatment: M=6.00, Q1=4.00, Q3=6.00; baseline: M=6.00, Q1=5.00, Q3=6.00) in the treatment condition. 

\subsection{Summary}
While lab study participants had higher median perceived control, deployment study participants did not. However, we saw some evidence in both studies of higher perceived satisfaction with control under the treatment condition. Participants may have been more satisfied with their control without perceiving a higher control in the sense that they were more comfortable with letting the tool have a certain amount of control in exchange for more editing power. Across the two studies, we did not see a change in median perceived ownership or satisfaction with ownership.

\section{Prompts}
\label{sec:prompts}

\subsection{Prompt for Initial LLM Draft for Baseline Tool}
\label{sec:baselinePrompts}

\begin{figure}[H]
    \centering
    \includegraphics[width=\linewidth]{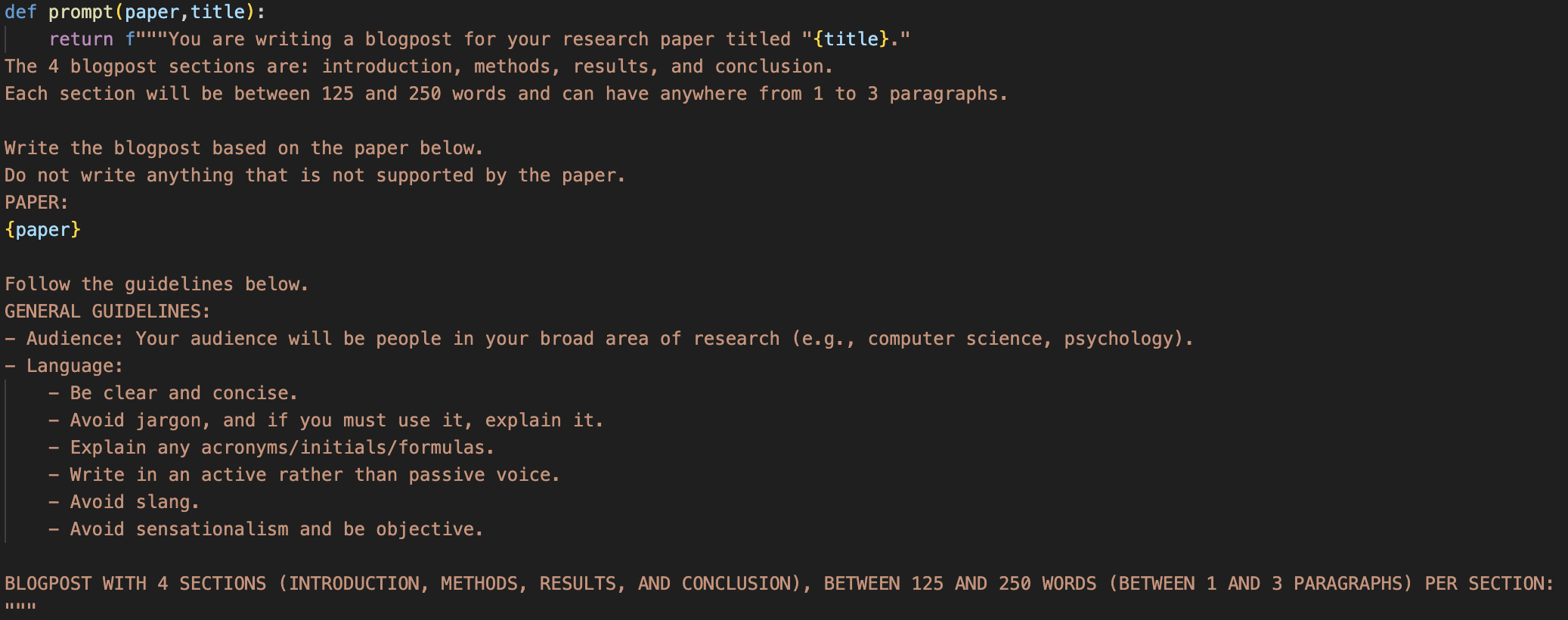}
    \caption{Prompt for initial LLM draft for baseline tool.}
    \Description{}
    \label{fig:baselinePrompt}
\end{figure}

\subsection{Prompts for the Warm Start Step of \sysname}
\label{sec:startupPrompts}
\begin{figure}[H]
    \centering
    \includegraphics[width=\linewidth]{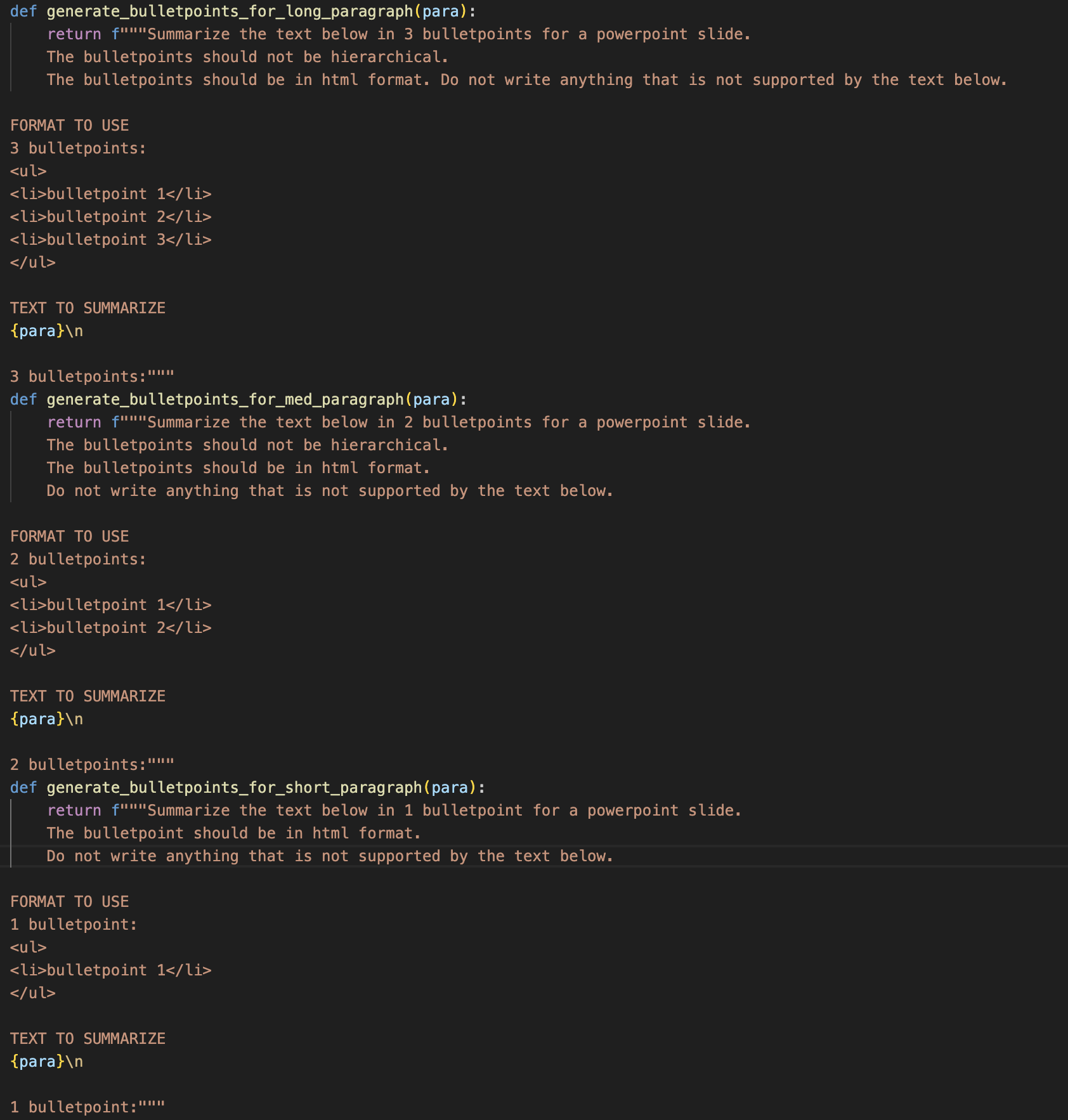}
    \caption{Prompts for generating bullet points for each paragraph in full paper being summarized. Each prompt corresponds to different length paragraphs. The top prompt generates three bullet points for paragraphs of more than 100 words, the second generates two bullet points for paragraphs of between 51 and 100 words, and the third generates one bullet point for paragraphs less than 51 words long.}
    \Description{}
    \label{fig:draftPrompt1}
\end{figure}

\begin{figure}[H]
    \centering
    \includegraphics[width=\linewidth]{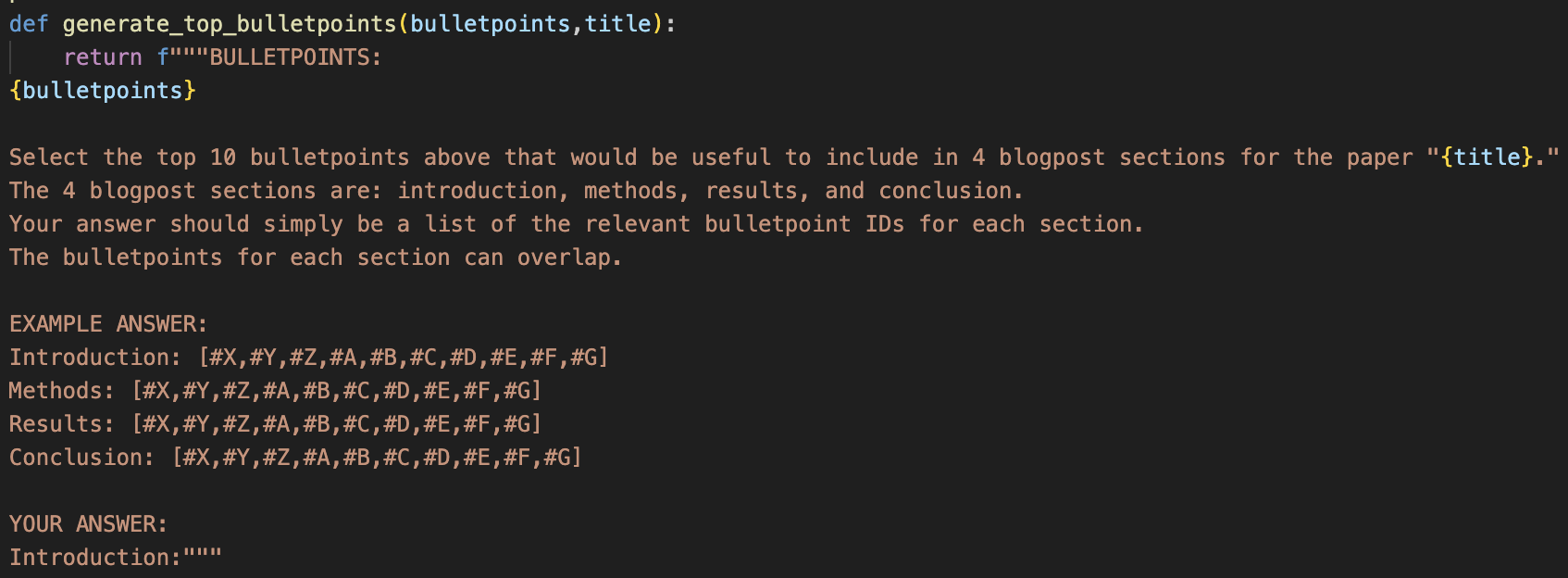}
    \caption{Prompt to select top 10 bullet points relevant for each of the initial draft's blog post sections (introduction, methods, results, conclusion).}
    \Description{}
    \label{fig:draftPrompt2}
\end{figure}

\begin{figure}[H]
    \centering
    \includegraphics[width=\linewidth]{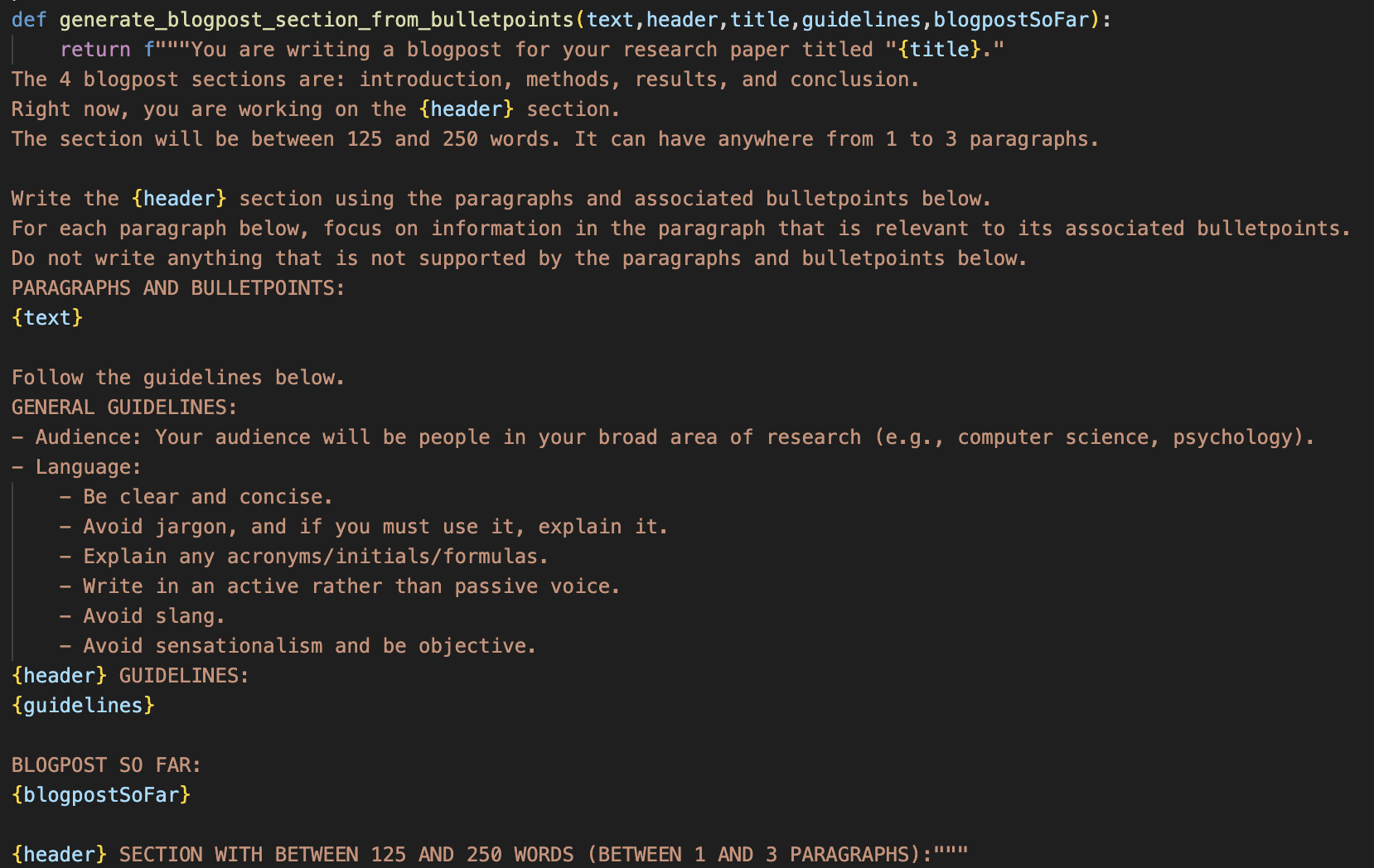}
    \caption{Prompt to generate each of the initial draft's blog post sections (introduction, methods, results, conclusion) using the 10 relevant selected bullet points. The section-specific guidelines are as follows. Introduction: ``-Present hook (e.g., interesting fact, quote, promise of change in knowledge, illustrating example of the topic). If context allows, visual or sensory elements are helpful anchors.-Provide high-level description of problem being solved.-Explain why work is interesting and a solution to the problem matters.-Do not repeat information from prior blogpost sections.'' Methods: ``-Focus on methods and do NOT discuss results.-Do not repeat information from prior blogpost sections.'' Results: ``-State key takeaway.-Discuss up to 3 most interesting aspects of work.-Do not repeat information from prior blogpost sections.'' Conclusion: ``-Restate key takeaway in new way.-Present future work ideas. [optional]-Loop back to hook. [optional]``}
    \Description{}
    \label{fig:draftPrompt3}
\end{figure}

Figures~\ref{fig:draftPrompt1}--\ref{fig:draftPrompt3} show the prompts for the initial LLM draft for \sysname.

\subsection{Prompts for Drafting Step of \sysname}
\label{sec:draftPrompts}
\begin{figure}[H]
    \centering
    \includegraphics[width=\linewidth]{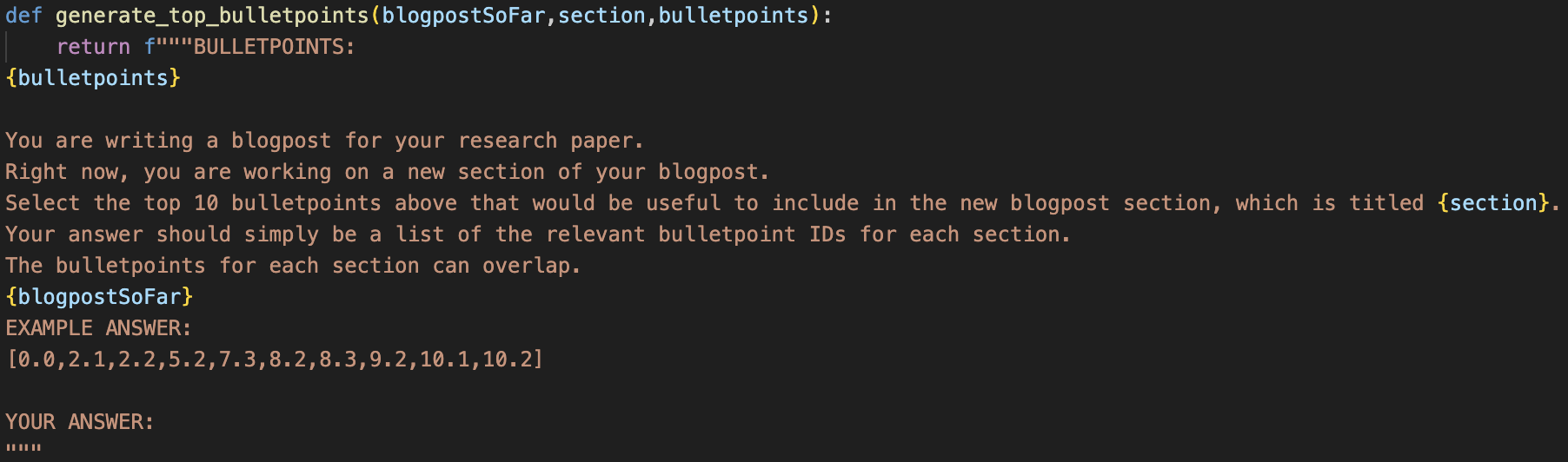}
    \caption{Prompt for selecting bullet points for generating a new section based on the section header provided by the user.}
    \Description{}
    \label{fig:genBullets}
\end{figure}
\begin{figure}[H]
    \centering
    \includegraphics[width=\linewidth]{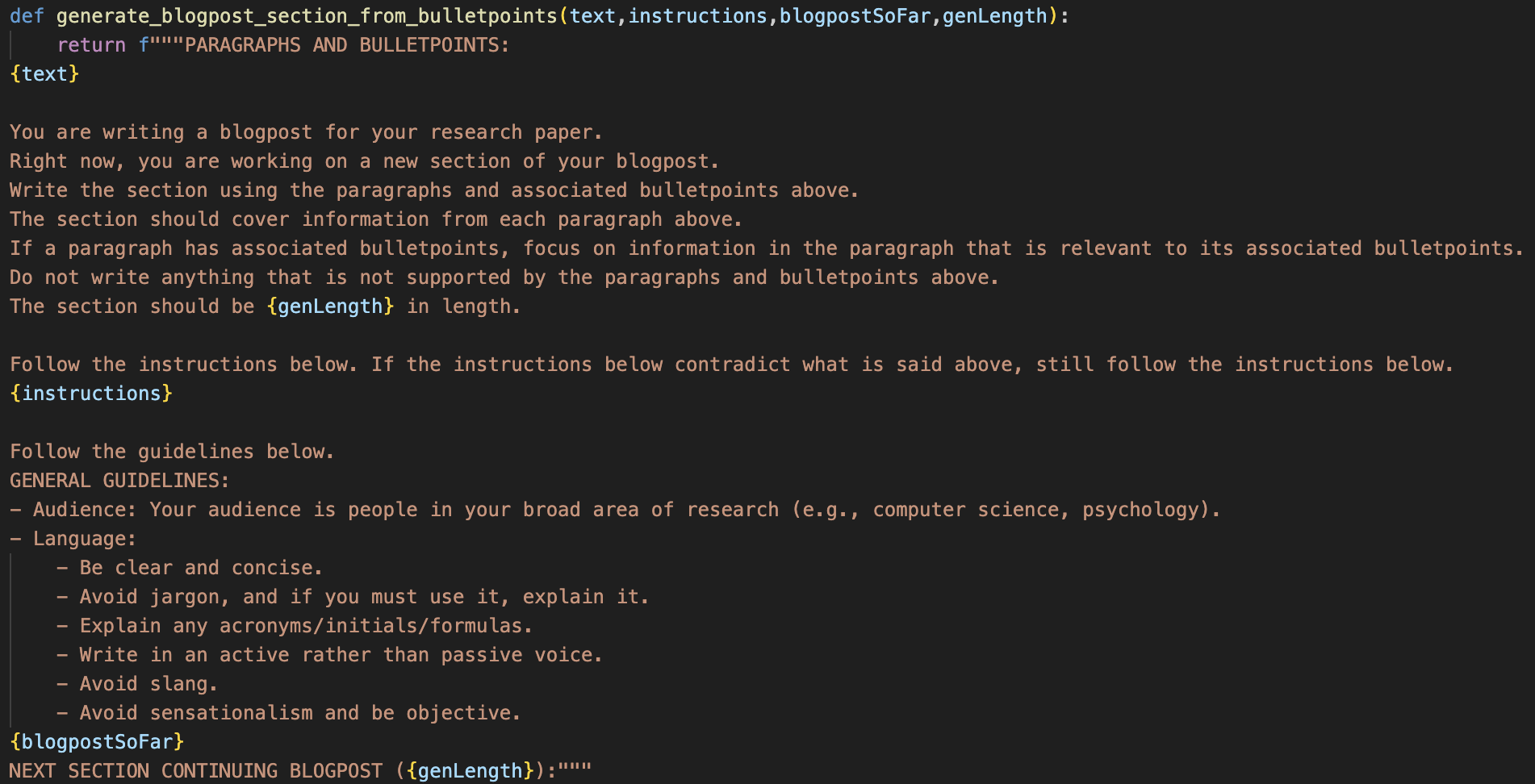}
    \caption{Prompt for generating text for a blog post section when there are selected paragraphs or bullet points as well as custom bullet points, custom instructions, or starting text.}
    \Description{}
    \label{fig:genSec}
\end{figure}
\begin{figure}[H]
    \centering
    \includegraphics[width=\linewidth]{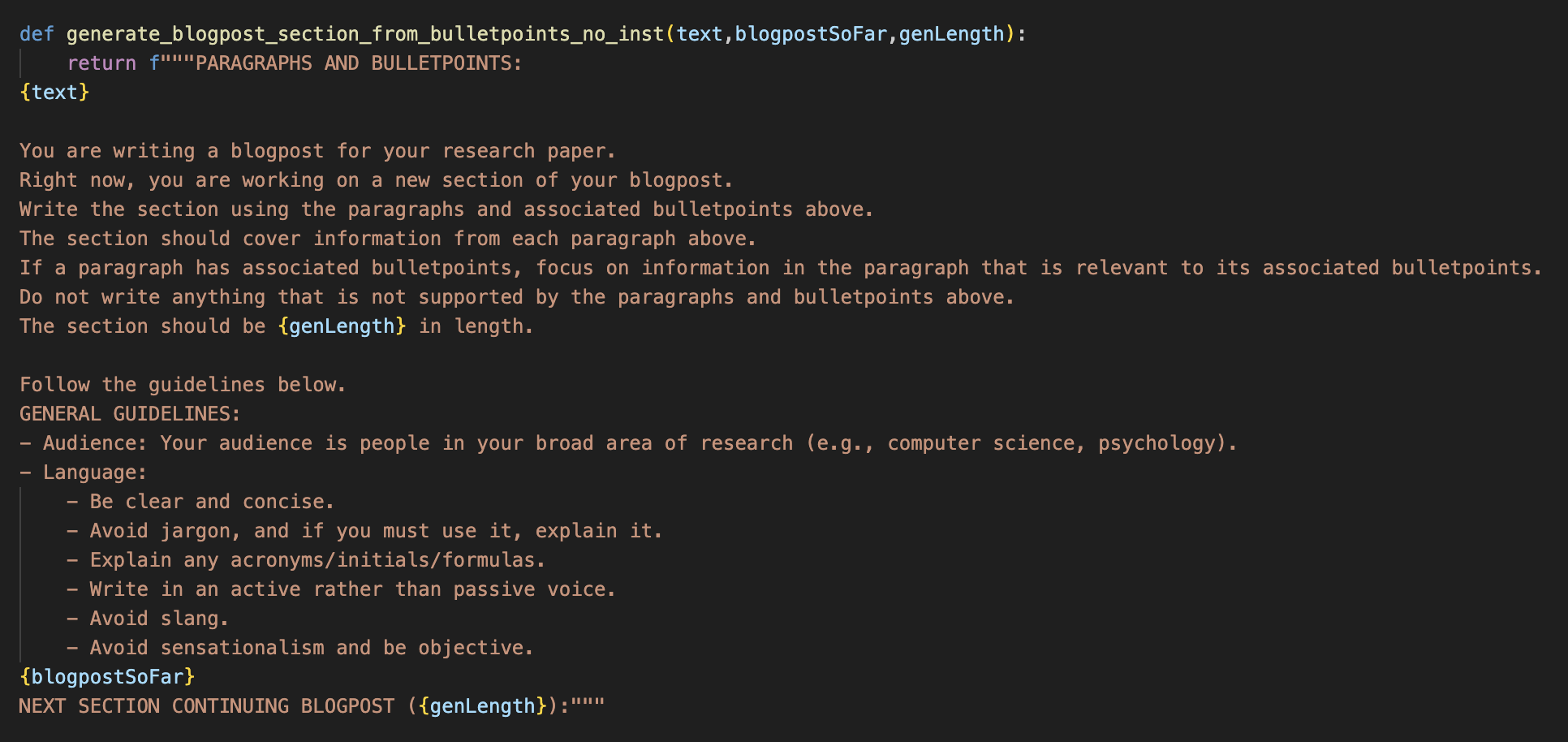}
    \caption{Prompt for generating text for a blog post section when there are selected paragraphs or bullet points but no custom bullet points, custom instructions, or starting text.}
    \Description{}
    \label{fig:genSecNoInst}
\end{figure}
\begin{figure}[H]
    \centering
    \includegraphics[width=\linewidth]{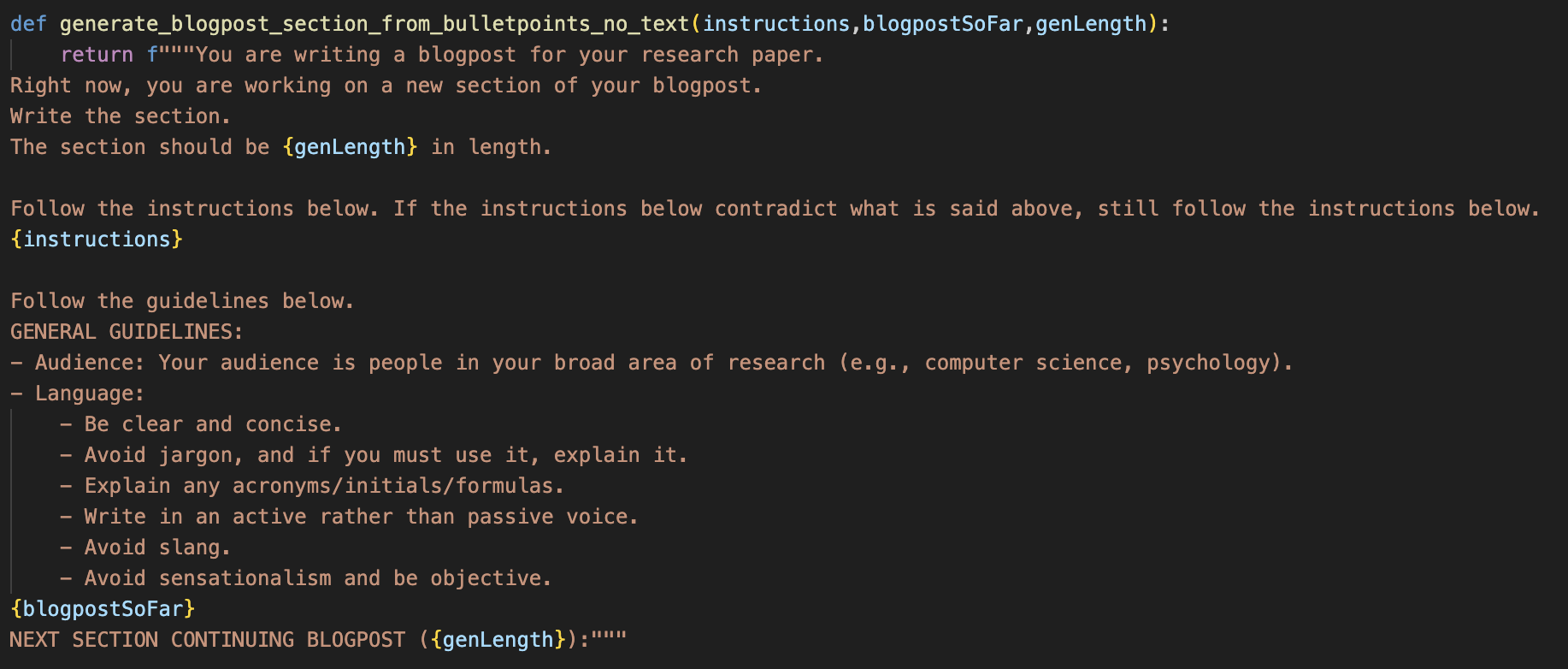}
    \caption{Prompt for generating text for a blog post section when there are no selected paragraphs or bullet points, but there are custom bullet points, custom instructions, or starting text.}
    \Description{}
    \label{fig:genSecNoSelection}
\end{figure}

\subsection{Prompts for Revising Step of \sysname}
\label{sec:revisePrompts}
\begin{figure}[H]
    \centering
    \includegraphics[width=\linewidth]{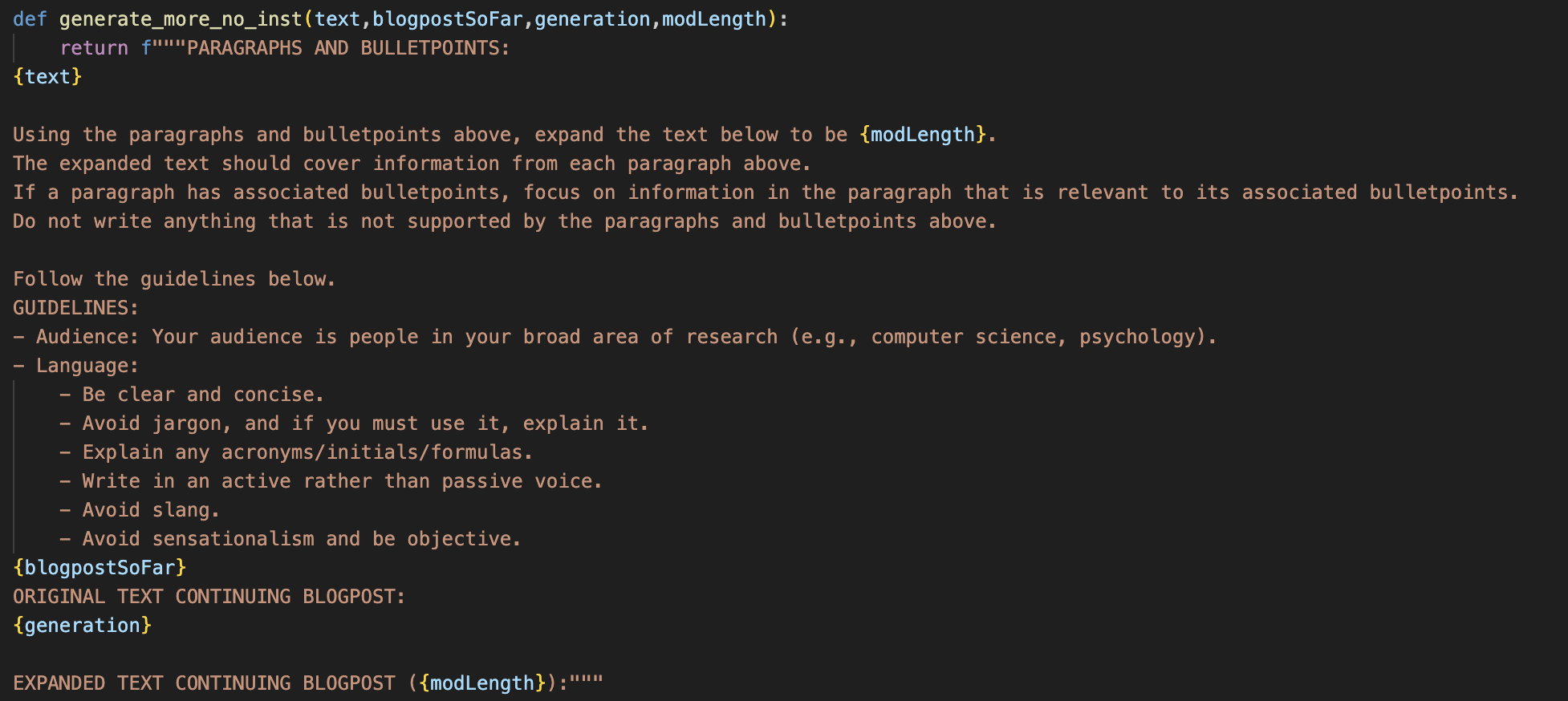}
    \caption{Prompt for the ``expand'' modification when there are no custom modification instructions. The automatic modLength variable for this modification is ``twice the length that it currently is.''}
    \Description{}
    \label{fig:expandNoInst}
\end{figure}
\begin{figure}[H]
    \centering
    \includegraphics[width=\linewidth]{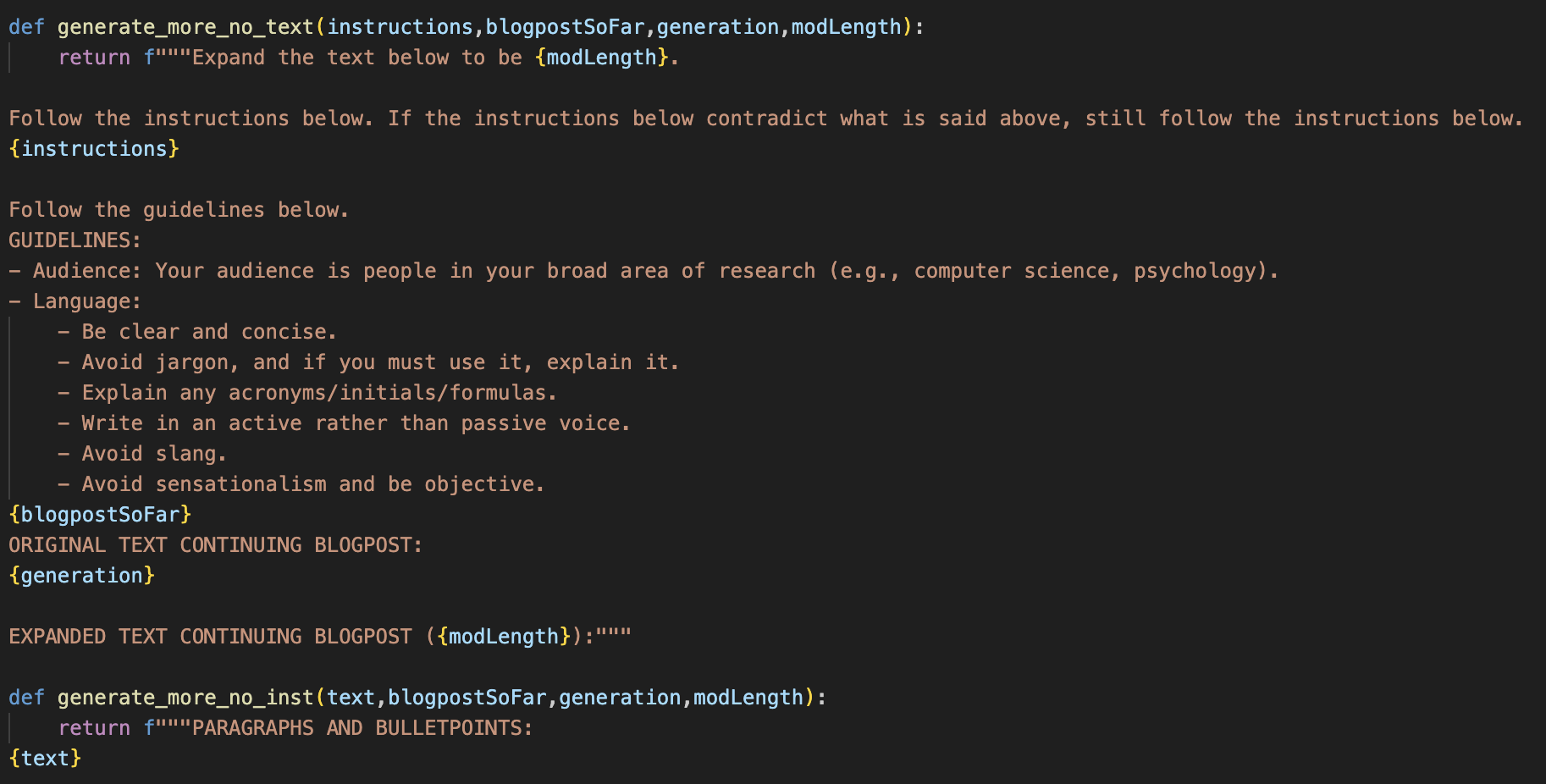}
    \caption{Prompt for the ``expand'' modification when there are custom modification instructions but no selected paragraphs or bullet points of which to be aware. The automatic modLength variable for this modification is ``twice the length that it currently is.''}
    \Description{}
    \label{fig:expandNoSelection}
\end{figure}
\begin{figure}[H]
    \centering
    \includegraphics[width=\linewidth]{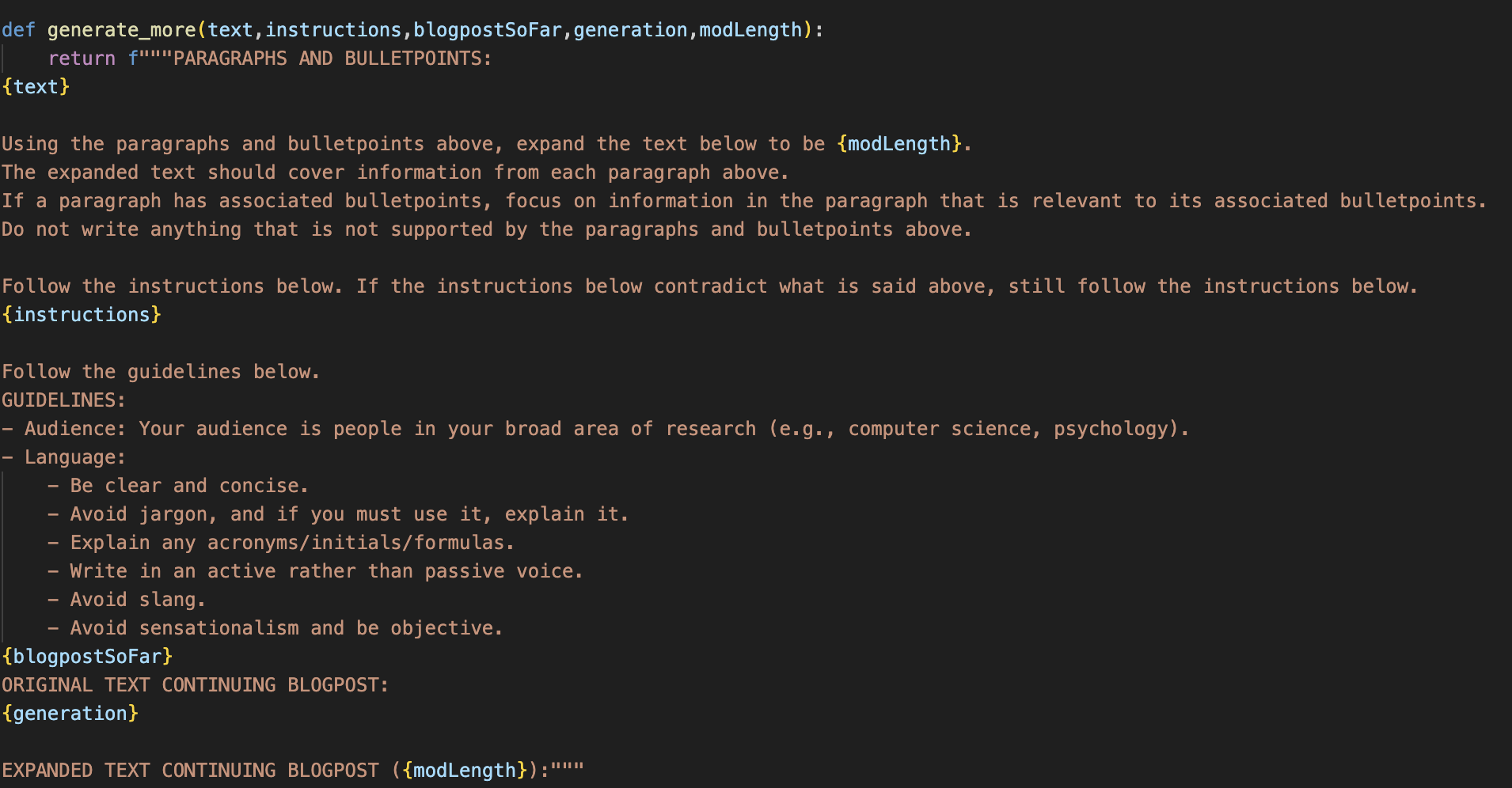}
    \caption{Prompt for the ``expand'' modification when there are selected paragraphs or bullet points of which to be aware and custom modification instructions. The automatic modLength variable for this modification is ``twice the length that it currently is.''}
    \Description{}
    \label{fig:expand}
\end{figure}
\begin{figure}[H]
    \centering
    \includegraphics[width=\linewidth]{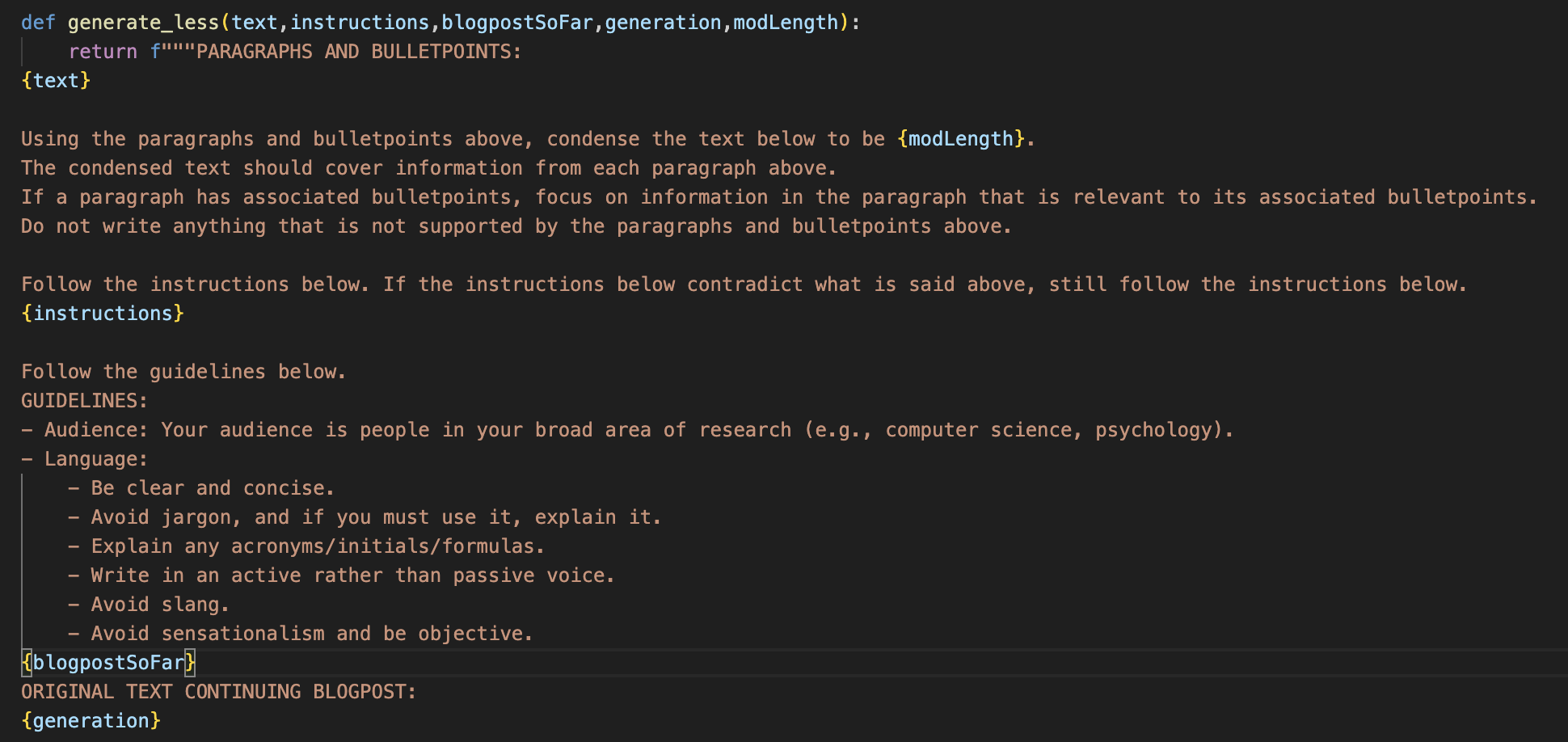}
    \caption{Prompt for the ``condense'' modification when there are selected paragraphs or bullet points of which to be aware and custom modification instructions. The automatic modLength variable for this modification is ``half the length that it currently is.'' The other prompts for this modification mirror those for ``expand.''}
    \Description{}
    \label{fig:condense}
\end{figure}
\begin{figure}[H]
    \centering
    \includegraphics[width=\linewidth]{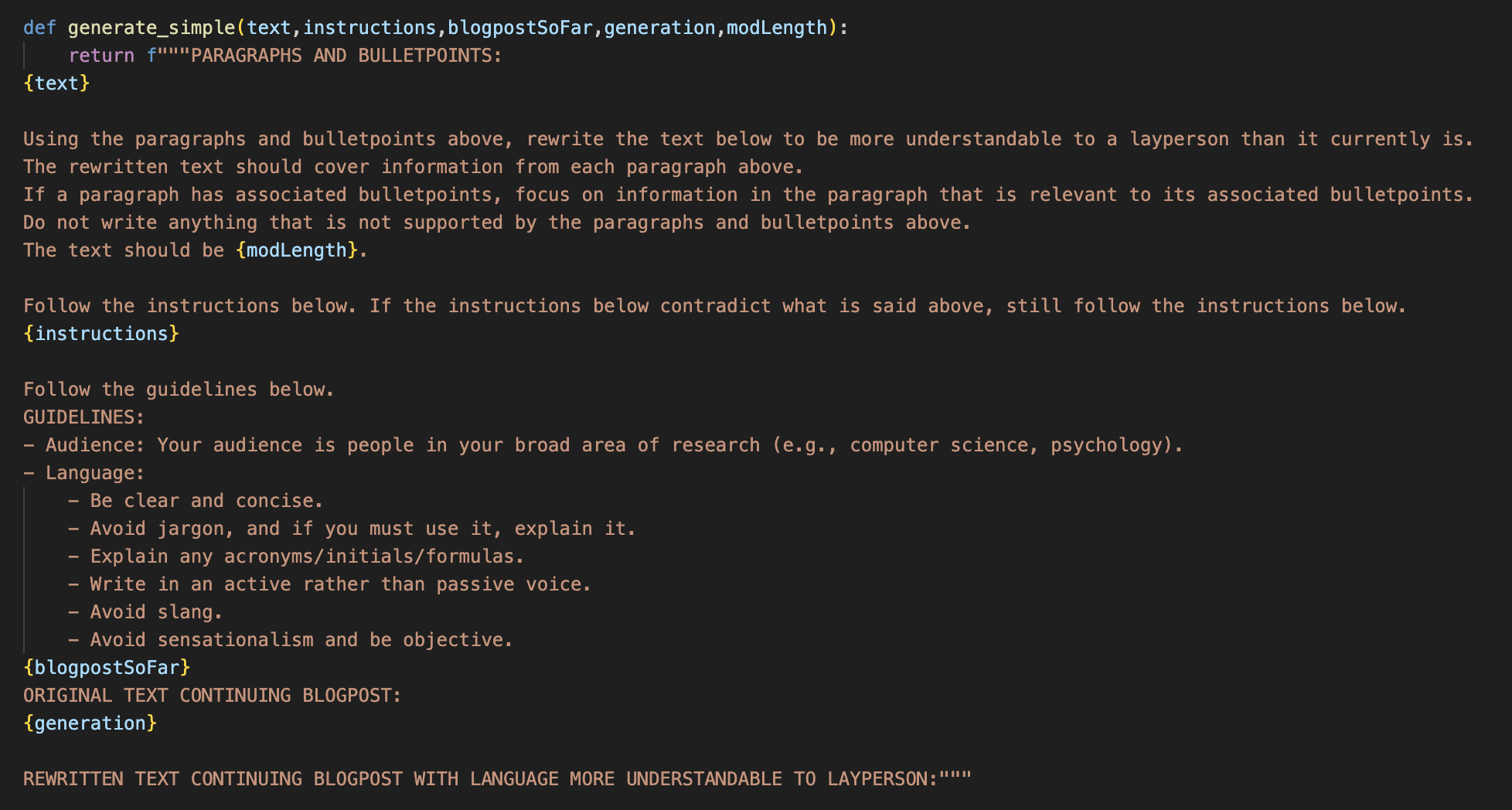}
    \caption{Prompt for the ``simpler terms'' modification when there are selected paragraphs or bullet points of which to be aware and custom modification instructions. The automatic modLength variable for this modification is ``about the same length that it currently is (no more than 25 words longer or shorter).'' The other prompts for the this modification mirror those for ``expand.''}
    \Description{}
    \label{fig:simple}
\end{figure}
\begin{figure}[H]
    \centering
    \includegraphics[width=\linewidth]{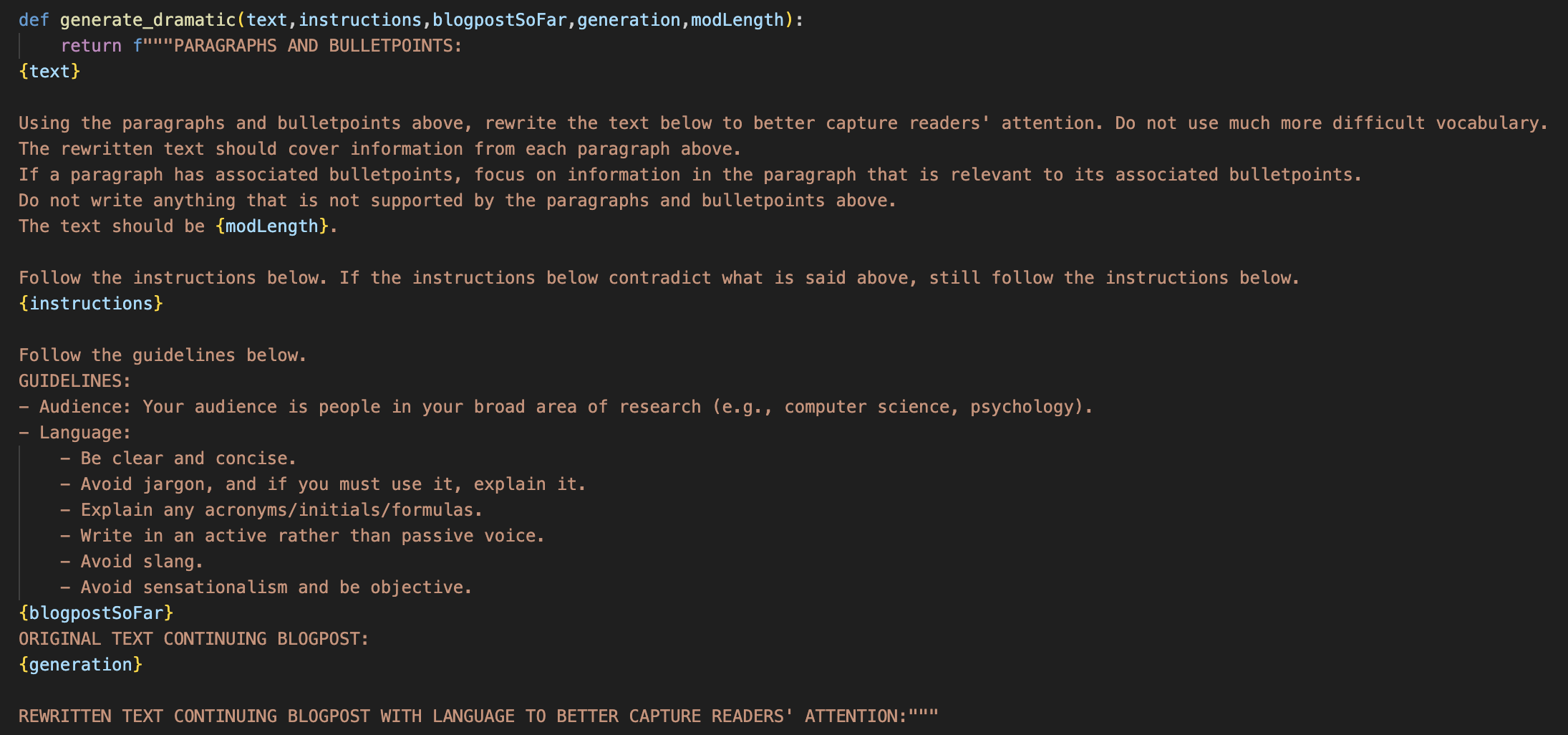}
    \caption{Prompt for the ``more dramatic'' modification when there are selected paragraphs or bullet points of which to be aware and custom modification instructions. The automatic modLength variable for this modification is ``about the same length that it currently is (no more than 25 words longer or shorter).'' The other prompts for the this modification mirror those for ``expand.''}
    \Description{}
    \label{fig:dramatic}
\end{figure}
\begin{figure}[H]
    \centering
    \includegraphics[width=\linewidth]{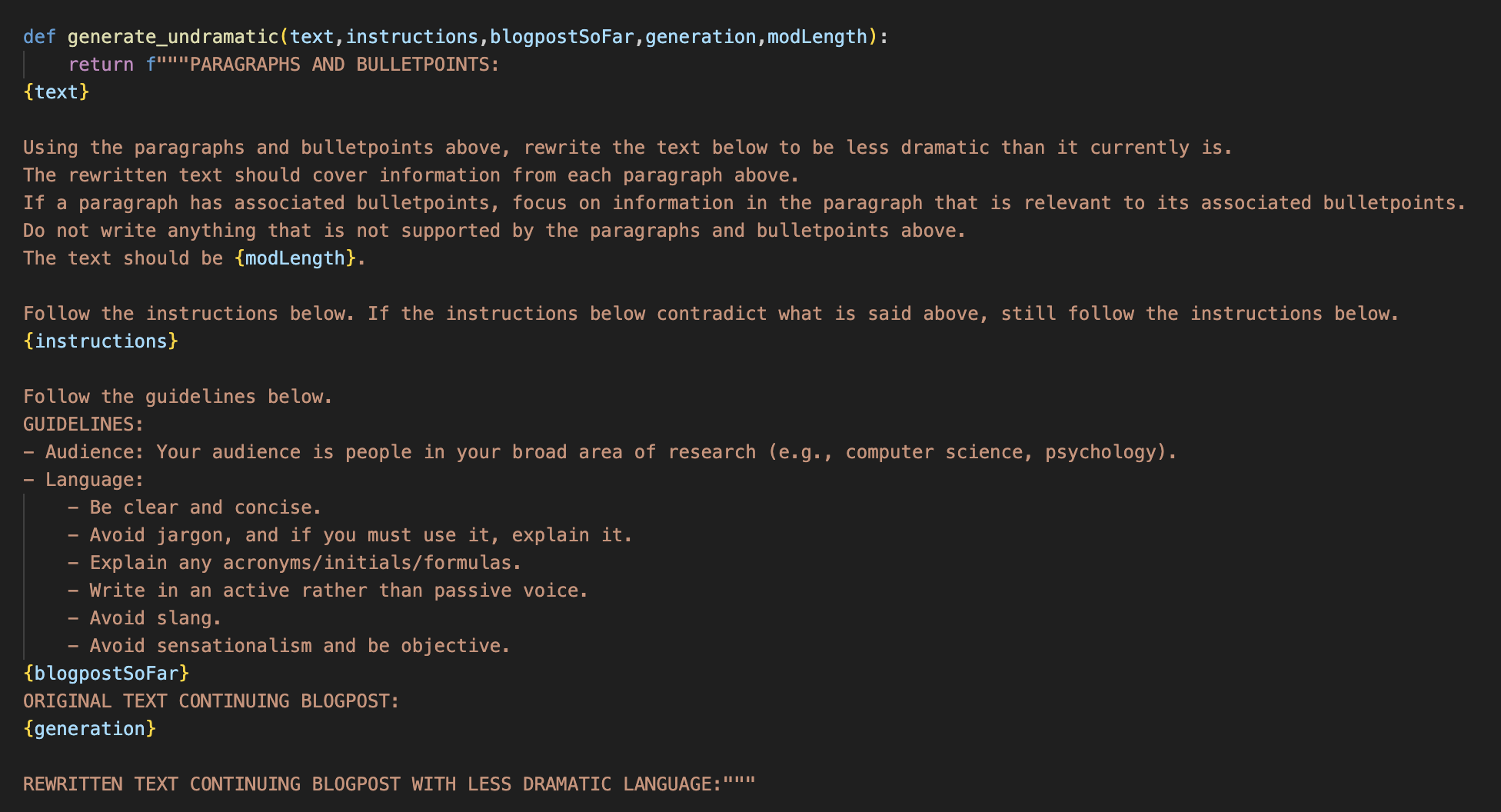}
    \caption{Prompt for the ``less dramatic'' modification when there are selected paragraphs or bullet points of which to be aware and custom modification instructions. The automatic modLength variable for this modification is ``about the same length that it currently is (no more than 25 words longer or shorter).'' The other prompts for the this modification mirror those for ``expand.''}
    \Description{}
    \label{fig:undramatic}
\end{figure}

\section{Additional Log Analysis Plots}
\begin{figure}[H]
    \centering
    \includegraphics[width=\linewidth]{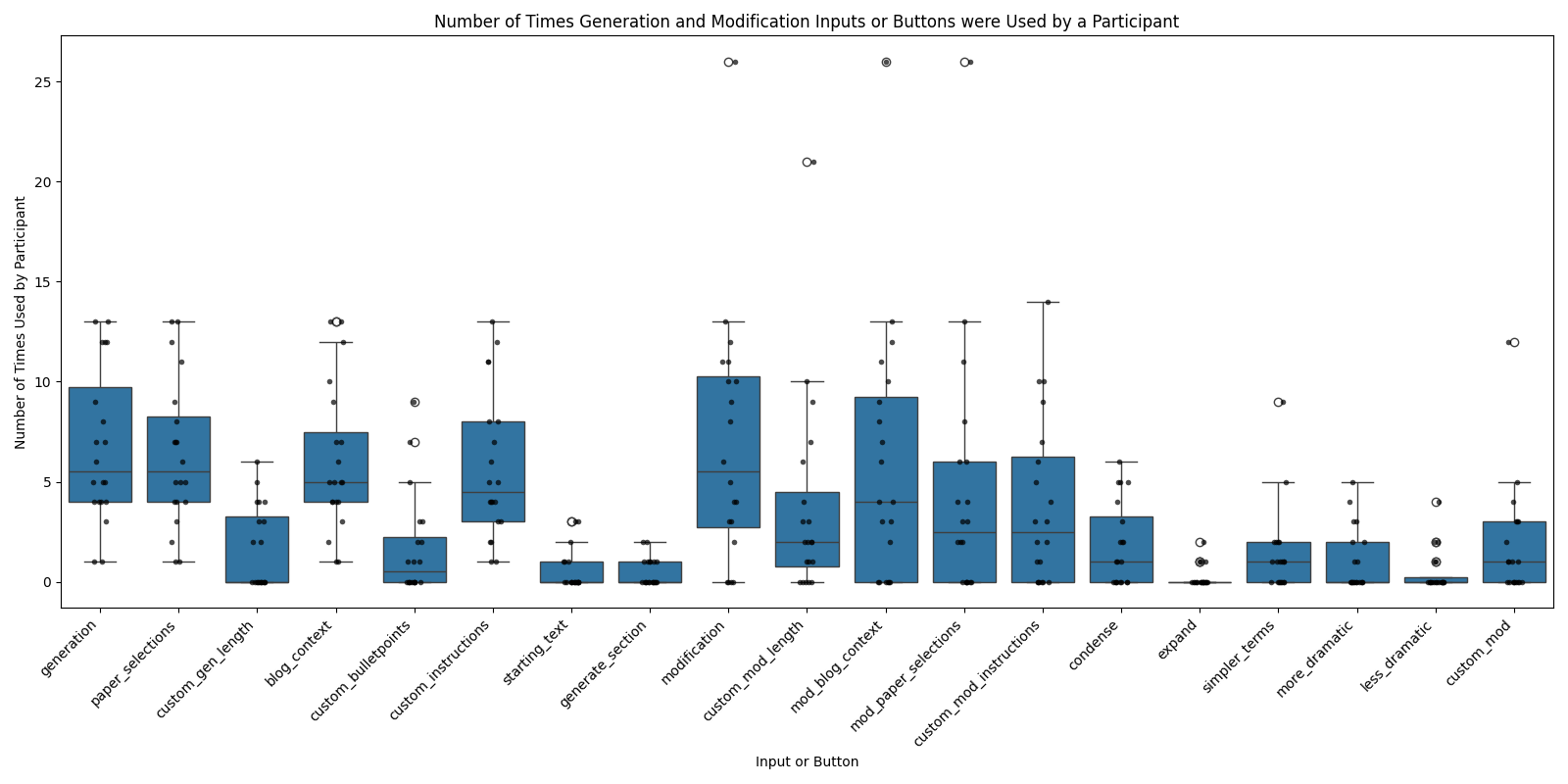}
  \caption{Number of times each input or button related to generating and modifying text was used by each participant in lab study. From left to right: the act of generating text, inputs/buttons related to generating text, the act of modifying text, inputs/buttons related to modifying text.}
  \Description{}
  \label{fig:treatActions}
\end{figure}

\begin{figure}[H]
    \centering
    \begin{subfigure}[b]{0.33\textwidth}
        \centering
        \includegraphics[width=\textwidth]{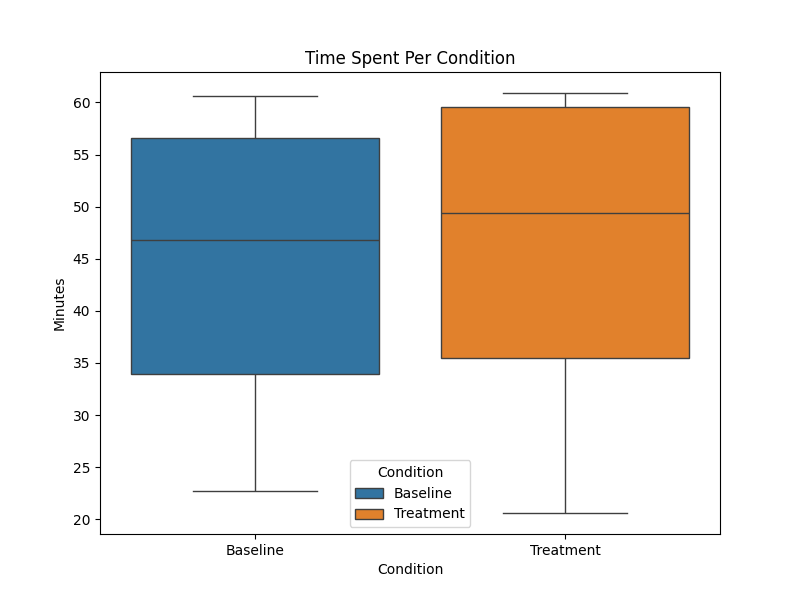}
        \caption{}
        \label{fig:labtimespent}
    \end{subfigure}
    \begin{subfigure}[b]{0.33\textwidth}
        \centering
        \includegraphics[width=\textwidth]{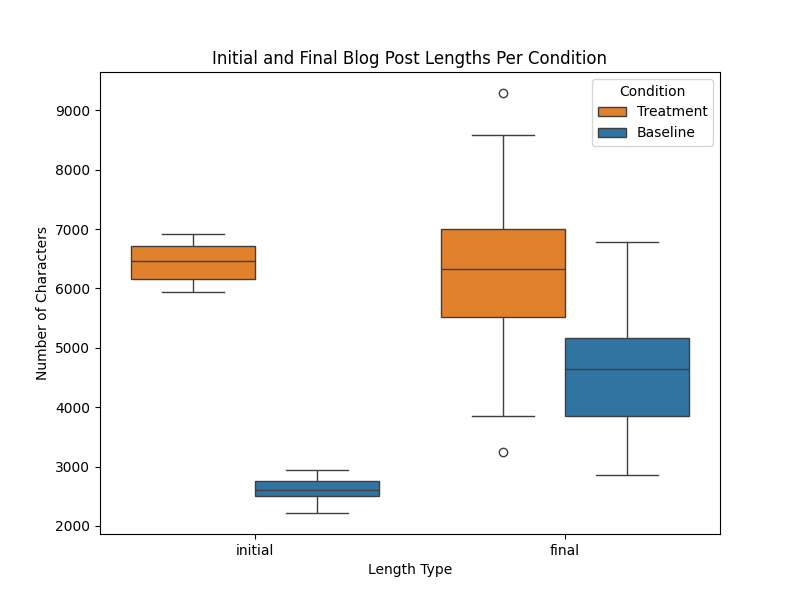}
        \caption{}
        \label{fig:lablengths}
    \end{subfigure}
    \begin{subfigure}[b]{0.33\textwidth}
        \centering
        \includegraphics[width=\textwidth]{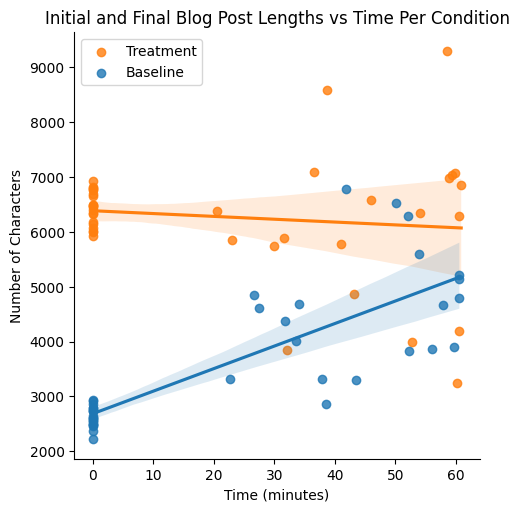}
        \caption{}
        \label{fig:lablengths-bytime}
    \end{subfigure}
    \caption{Additional plots for lab study.}
    \Description{}
    \label{fig:labLevExtra}
\end{figure}

\begin{figure}[H]
    \centering
    \begin{subfigure}[b]{0.33\textwidth}
        \centering
        \includegraphics[width=\textwidth]{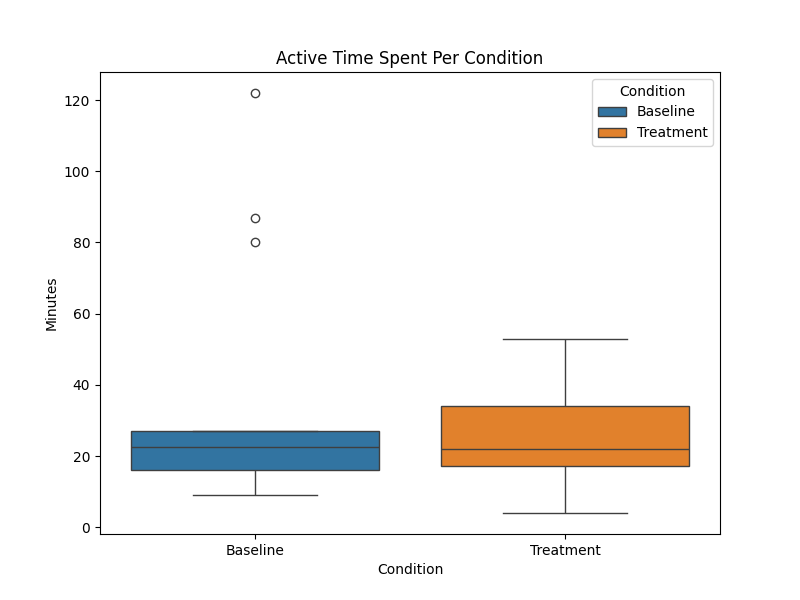}
        \caption{}
        \label{fig:deploytimespent}
    \end{subfigure}
    \begin{subfigure}[b]{0.33\textwidth}
        \centering
        \includegraphics[width=\textwidth]{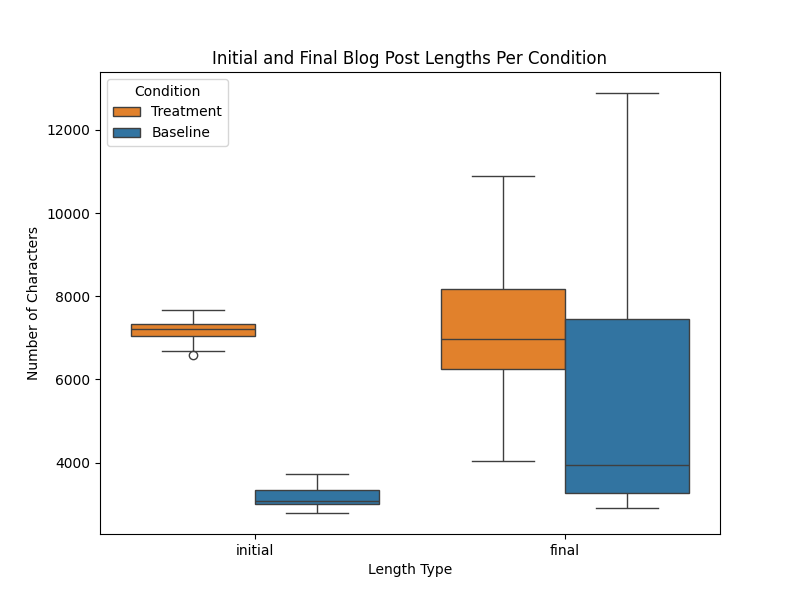}
        \caption{}
        \label{fig:deploylengths}
    \end{subfigure}
        \begin{subfigure}[b]{0.33\textwidth}
        \centering
        \includegraphics[width=\textwidth]{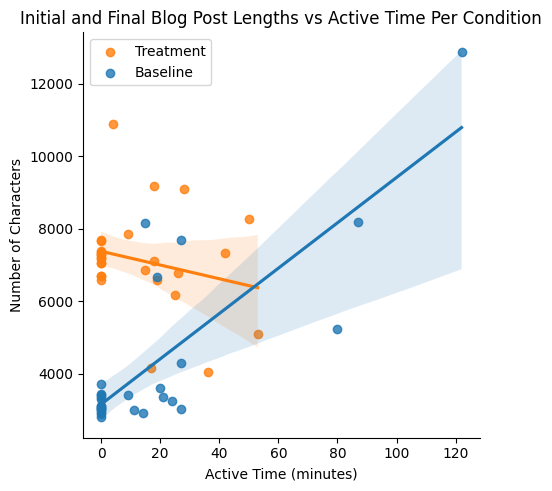}
        \caption{}
        \label{fig:deploylengths-bytime}
    \end{subfigure}
    \caption{Additional plots for deployment study.}
    \Description{}
    \label{fig:deployLevExtra}
\end{figure}

Figures~ ~\ref{fig:treatActions} and \ref{fig:labLevExtra} show additional log analysis plots for the lab study, while Figure \ref{fig:deployLevExtra} shows additional log analysis plots for the deployment study.

\clearpage

\section{Lab Study Version of Tool}
\begin{figure*}[htbp]
    \centering
    \begin{subfigure}[b]{0.7\textwidth}
        \centering
        \includegraphics[width=\textwidth]{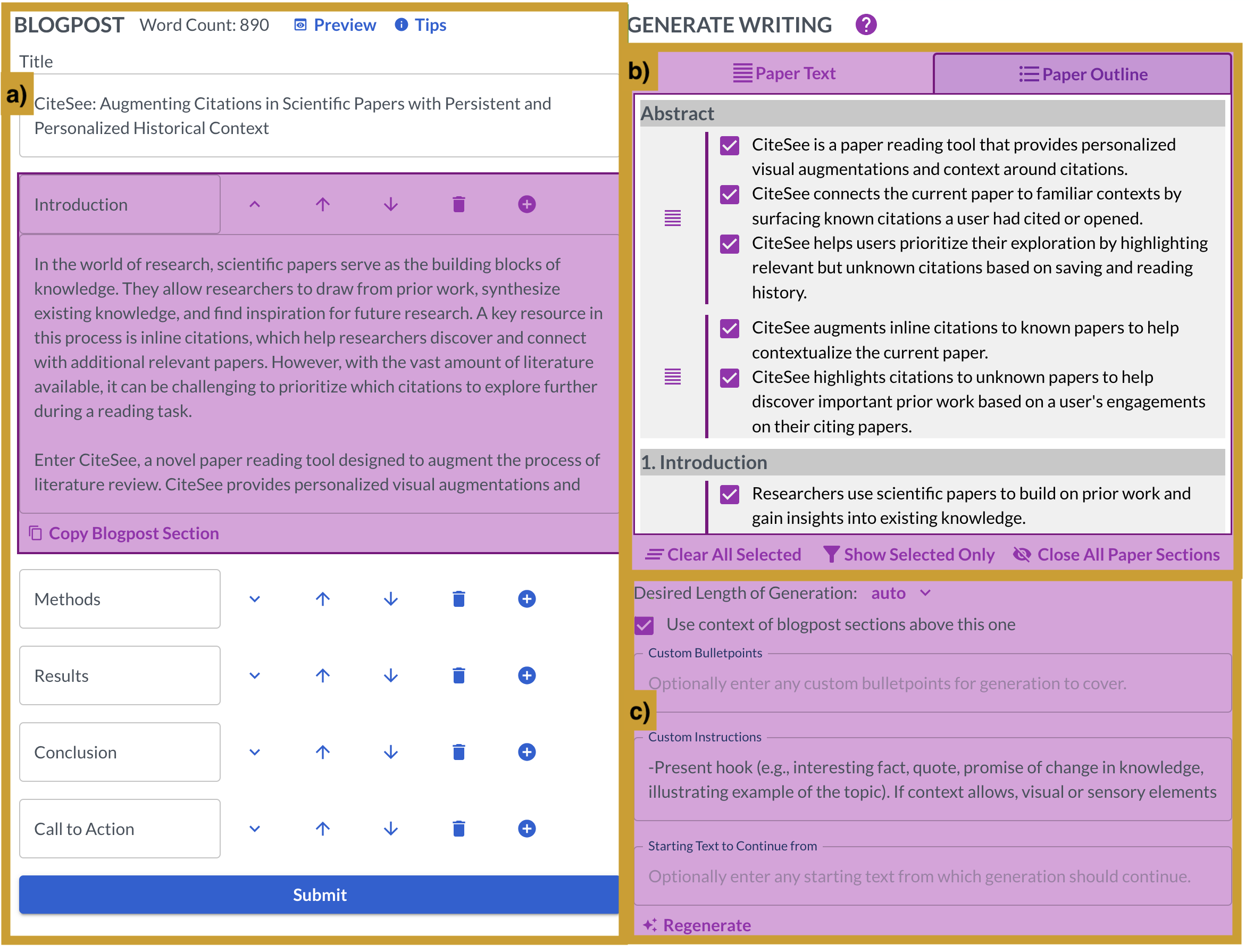}
        \caption{}
        \label{fig:firstTool1}
    \end{subfigure}
    \begin{subfigure}[b]{0.28\textwidth}
        \centering
        \includegraphics[width=\textwidth]{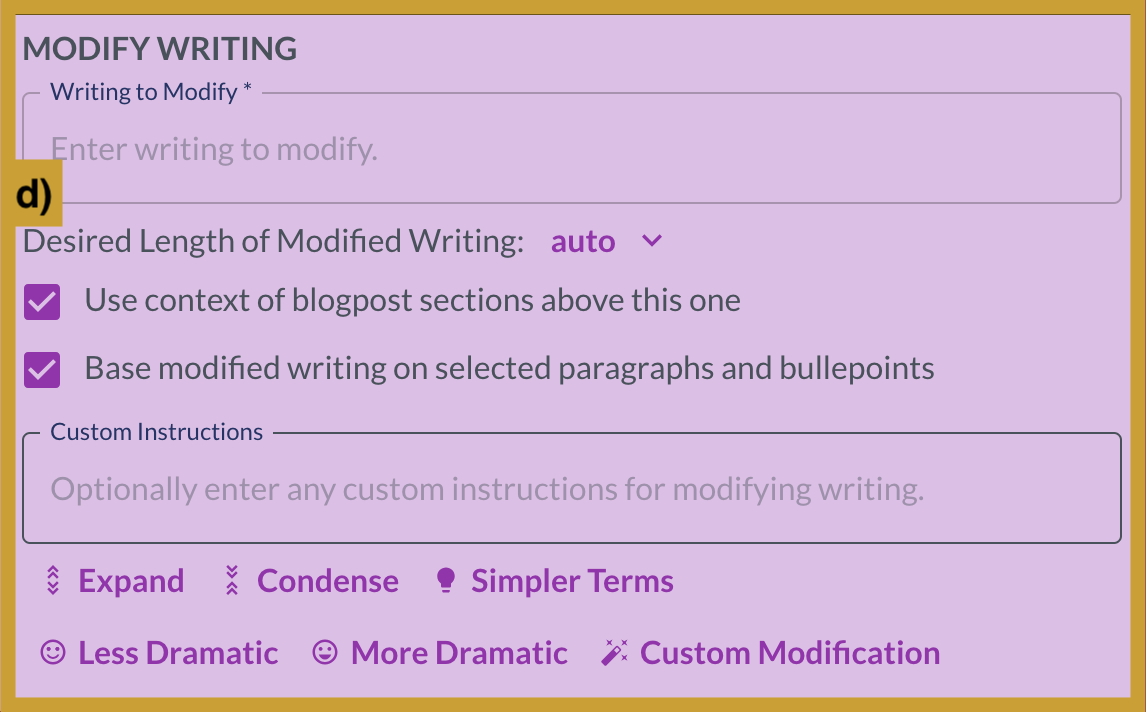}
        \caption{}
        \label{fig:firstTool2}
    \end{subfigure}
    \caption{\sysname\ in the lab study. a) Blog-Post Area, b) Planning-Support Area, c) Drafting-Support Area, and d) Revising-Support Area.}
    \Description{}
    \label{fig:firstTool}
\end{figure*}

Figure~\ref{fig:firstTool} shows how the tool looked for the lab study. 

\section{Example Baseline Initial Draft}
\label{sec:baselineDraftExample}
Below is example output from the baseline tool for the paper ``ConceptEVA: Concept-Based Interactive Exploration and Customization of Document Summaries'' \cite{zhang2023concepteva}. Note that this output is for the deployment study version of \sysname, which used automatic paper parsing without manual fixes. This example was generated in October 2024.

\subsection{Introduction}
\textit{In the realm of academic research, distilling the essence of long, multi-topic documents into concise summaries is a formidable challenge. This task becomes even more daunting when considering the diverse backgrounds of readers who may seek to understand these documents. Traditional summarization techniques, both extractive and abstractive, often fall short in addressing the nuances of such complex texts. Recognizing this gap, our research introduces ConceptEVA, a novel system designed to revolutionize the way we approach the summarization of academic papers. ConceptEVA leverages the latest advancements in natural language processing (NLP) and artificial intelligence (AI) to offer a mixed-initiative approach, enabling users to generate, evaluate, and customize summaries tailored to their specific interests. This system not only promises to enhance the accessibility of academic literature but also to enrich the reader's engagement with the material.}

\subsection{Methods}
\textit{ConceptEVA is built upon a multi-task Longformer Encoder Decoder (LED), fine-tuned to adeptly handle the summarization of lengthy, multi-domain documents. This innovative system employs attention mechanisms to efficiently process extensive texts while maintaining high information fidelity. A key feature of ConceptEVA is its interactive visualization of document concepts, presented as a network that illustrates semantic relatedness and co-occurrence. Users can interact with this visualization to select concepts of interest, which ConceptEVA then uses to automatically update the summary, emphasizing the chosen concepts. Our evaluation of ConceptEVA involved two iterations: an initial expert review and a subsequent within-subjects study, comparing ConceptEVA's performance against manually-generated summaries.}

\subsection{Results}
\textit{The findings from our study indicate a clear preference among participants for ConceptEVA's customized summaries over their manually-generated counterparts. Participants appreciated the ability to tailor summaries to their interests, resulting in higher satisfaction levels with the summaries produced by ConceptEVA. However, incorporating critique into summaries remained a challenge, highlighting an area for future improvement. Despite this, the overall response to ConceptEVA was positive, with users finding the system helpful for exploring and verifying concepts within documents, especially those outside their domain of expertise.}

\subsection{Conclusion}
\textit{ConceptEVA represents a significant step forward in the field of document summarization, particularly for academic literature that spans multiple topics and domains. By integrating interactive visual analysis with advanced NLP techniques, ConceptEVA offers a user-friendly platform for generating, evaluating, and customizing summaries. This mixed-initiative approach not only improves the summarization process but also enhances the user's engagement with the document. While challenges remain, particularly in integrating critique into summaries, the positive feedback from our study participants underscores ConceptEVA's potential as a valuable tool for researchers, educators, and anyone seeking to navigate the complexities of academic literature. Future developments will focus on refining ConceptEVA's capabilities, further bridging the gap between comprehensive document analysis and user-centric summarization.}

\section{Example \sysname\ Output}
\label{sec:treatmentOutputExample}
Below is example output from \sysname\ for the paper ``ConceptEVA: Concept-Based Interactive Exploration and Customization of Document Summaries'' \cite{zhang2023concepteva}. Note that this output is for the deployment study version of \sysname, which used automatic paper parsing without manual fixes. These examples were generated in October 2024.

\subsection{Initial Draft}
\label{sec:treatmentDraftExample}
\subsubsection{Introduction}
\textit{In the rapidly evolving field of natural language processing and artificial intelligence, the challenge of effectively summarizing long, complex documents that span multiple topics remains a significant hurdle. This is particularly true for academic papers, which often contain dense, multi-disciplinary content that can be difficult to distill into a concise summary. Traditional methods of extractive summarization, which rely on pulling key sentences directly from the text, fall short in capturing the nuanced essence of such documents. The advent of abstractive summarization, which paraphrases the original text to produce a more human-like summary, offers a promising solution but is still plagued by challenges, especially when dealing with documents that cover a broad range of knowledge domains.}

\textit{Enter ConceptEVA, a novel mixed-initiative approach designed to tackle the complexities of summarizing long, multi-topic academic documents. By integrating advanced machine learning techniques with interactive visual analysis, ConceptEVA allows users to generate, evaluate, and customize summaries in a way that is both efficient and tailored to their specific interests. This system represents a significant leap forward in automated text summarization, offering a more dynamic and user-centric solution to the problem of digesting extensive academic literature.}

\textit{The significance of ConceptEVA lies not only in its ability to produce concise summaries of complex documents but also in its interactive nature, which empowers users to steer the summarization process. This human-in-the-loop approach ensures that the generated summaries are relevant and valuable to the reader, addressing the long-standing challenge of creating universally applicable summaries for documents that traverse multiple domains. As we continue to push the boundaries of what's possible in automated text summarization, ConceptEVA stands out as a beacon of innovation, promising to transform how we engage with and understand academic literature.}

\subsubsection{Methods}
\textit{In developing ConceptEVA, we employed a multi-faceted approach to address the challenges of summarizing long, multi-topic documents. At the core of ConceptEVA is a custom multi-task Longformer encoder-decoder, specifically designed for the summarization of extensive documents. This model leverages the power of attention mechanisms from transformers to efficiently manage memory usage while ensuring high fidelity in information preservation across longer texts. The attention mechanisms operate at both local and global levels, facilitating a nuanced understanding and processing of the document's content.}

\textit{To enhance user interaction and customization, ConceptEVA integrates interactive visualizations of document concepts as a network. This visualization reflects semantic relatedness and co-occurrence, allowing users to focus on and select concepts of interest. These selected concepts can then automatically update the summary to emphasize them, supported by a "focus-on" function that surfaces and highlights related concepts. This interactive component is underpinned by a multi-domain ontology for concept identification and a force-directed layout for visual representation, ensuring a user-friendly interface for customizing summaries.}

\textit{The iterative development of ConceptEVA was informed by an initial survey of eight research practitioners and underwent two stages of refinement based on expert reviews and a within-subjects study. This process included fine-tuning a pre-trained Longformer encoder-decoder for scientific document summarization, incorporating paraphrasing and semantic sentence embedding techniques, and developing an interactive summary editor to facilitate better human-AI collaboration. Additionally, we utilized Faiss, a similarity search library for dense vectors, to enhance the summarizer's context understanding, thereby improving the quality and relevance of the generated summaries.}

\subsubsection{Results}
\textit{Our study revealed that ConceptEVA significantly enhances user satisfaction with the summarization process, particularly when compared to manually-generated summaries. Specifically, 8 out of 12 participants preferred the summaries generated using ConceptEVA's human-in-the-loop approach over the baseline, indicating a clear preference for the customized summaries provided by our system. This preference was especially pronounced among participants reviewing documents outside their domain of expertise, underscoring ConceptEVA's utility in helping users explore unfamiliar topics through efficient and focused summarization.}

\textit{Moreover, the integration of ConceptEVA's "focus-on" function, which allows users to select and emphasize specific concepts of interest, was instrumental in achieving higher satisfaction levels. This feature not only facilitated a more interactive and personalized summarization experience but also highlighted the system's capability to adapt summaries based on user input. However, it was noted that participants with less experience in the subject matter found it challenging to incorporate critique, such as limitations and implications, into their summaries, suggesting an area for further improvement in guiding users through more nuanced aspects of summarization.}

\textit{A significant association was found between the summarization approach employed by ConceptEVA and the summary satisfaction rating, as evidenced by a chi-squared test of independence. This statistical validation, along with user feedback, underscores the effectiveness of ConceptEVA in enhancing the summarization process through mixed-initiative interactions and customized content, paving the way for future developments in automated text summarization technologies.}

\subsubsection{Conclusion}
\textit{In conclusion, ConceptEVA represents a significant advancement in the field of automated text summarization, particularly for academic documents. By leveraging a human-in-the-loop approach, ConceptEVA not only enhances the relevance and quality of summaries but also fosters a more collaborative interaction between users and AI. This system's ability to integrate interactive visual analysis with natural language processing techniques allows for a more nuanced and customizable summarization process, addressing the critical need for summaries that cater to individual user interests and requirements. The positive feedback from our user study, especially regarding the system's content-specific assistance and the preference for ConceptEVA-generated summaries over manual efforts, underscores the potential of this approach to significantly improve how we engage with and digest complex academic literature.}

\textit{Looking forward, we aim to expand ConceptEVA's capabilities by testing it on a broader range of documents and incorporating additional features based on user feedback. This will include enhancing the system's flexibility to adapt summaries more dynamically to user priorities and further refining the AI's ability to maintain logical and narrative coherence throughout the summary. By continuing to focus on the synergy between human expertise and AI efficiency, ConceptEVA is poised to set new standards in the automation of document summarization, making it an invaluable tool for researchers, scholars, and anyone looking to efficiently extract insights from extensive academic writings.}

\subsection{Initial Draft Bullet Point Selection}
\label{sec:treatmentBulletpointsExample}
\subsubsection{Introduction Section's Selected Bullet Points}
\begin{itemize}
  \item \textit{ConceptEVA is a mixed-initiative approach for summarizing long and multi-topic documents.}
  \item \textit{CHI 2023 conference in Hamburg, Germany featured a multi-disciplinary research paper with an auto-generated summary}
  \item \textit{Automated text summarization aims to present key concepts in a concise manner and reduce repetition.}
  \item \textit{Challenges in abstractive summarization include summarizing long, complex documents spanning multiple knowledge domains.}
  \item \textit{A human-in-the-loop approach is needed for interactive generation of relevant summaries.}
  \item \textit{ConceptEVA integrates visual analysis and NLP for evaluating and customizing long document summaries.}
  \item \textit{ConceptEVA introduces a human-in-the loop, mixed-initiative approach to evaluate and customize document summary generation}
  \item \textit{Our work focuses on interactively exploring concept focus in generated summaries and emphasizing concepts of interest in academic publications.}
  \item \textit{ConceptEVA customizes summaries by updating them with user-selected concepts of interest.}
  \item \textit{ConceptEVA is an interactive document summarization system for long, multi-domain documents.}
\end{itemize}

\subsubsection{Methods Section's Selected Bullet Points}
\begin{itemize}
  \item \textit{It incorporates a custom multi-task longformer encoder decoder for summarizing longer documents.}
  \item \textit{Interactive visualizations of document concepts as a network in ConceptEVA help users focus on concepts of interest}
  \item \textit{Attention mechanisms from transformers used at local levels to reduce memory usage}
  \item \textit{ConceptEVA design informed by initial survey of eight research practitioners and refined through two stages of development and evaluation}
  \item \textit{ConceptEVA includes fine-tuning an LED model, identifying concepts using an ontology, and providing interactive visualization for customization.}
  \item \textit{Maintain the user\&\#x27;s mental map of the original document by preserving its layout}
  \item \textit{We use Faiss to implement the approach, a similarity search library for dense vectors in large scale.}
  \item \textit{Extended ConceptEVA with an interactive summary editor for better human-AI collaboration}
  \item \textit{We constructed a set of guidelines for participants to follow when generating a summary manually or using ConceptEVA.}
  \item \textit{The second iteration includes a longformer encoder-decoder pre-trained for scientific documents, fine-tuned for paraphrasing and sentence embedding, and concepts visualized using a force-directed network.}
\end{itemize}

\subsubsection{Results Section's Selected Bullet Points}
\begin{itemize}
  \item \textit{Participants\&\#x27; satisfaction with ConceptEVA customized summaries is higher than their own manually-generated summary}
  \item \textit{ConceptEVA is helpful for examining and verifying ideas}
  \item \textit{ConceptEVA is more useful for participants outside their domain of interest}
  \item \textit{Inexperienced participants found it difficult to incorporate critique into the summary}
  \item \textit{Text embeddings are pre-computed for each sentence in the source document and retrieved when users select concepts.}
  \item \textit{8 out of 12 participants rated the summary generated using ConceptEVA\&\#x27;s human-in-the-loop approach (task T3) higher than the baseline}
  \item \textit{Study used 6 participants to summarize ConceptScope and 6 participants to summarize BodyVis}
  \item \textit{A significant association was found between summarization approach and summary satisfaction rating}
  \item \textit{Higher rating for human-in-the-loop approach led to more efficient location and focus on concepts}
  \item \textit{Participants preferred summaries created through ConceptEVA\&\#x27;s human-in-the-loop approach}
\end{itemize}

\subsubsection{Conclusion Section's Selected Bullet Points}
\begin{itemize}
  \item \textit{ConceptEVA is a mixed-initiative system for academic document readers and writers to generate, evaluate, and customize automated summaries.}
  \item \textit{ConceptEVA integrates visual analysis and NLP for evaluating and customizing long document summaries.}
  \item \textit{ConceptEVA\&\#x27;s approach better exploits human-AI collaboration and iteratively improves the summary}
  \item \textit{Saliency, fluency, novelty, and coherence are key factors in evaluating automated summarization techniques}
  \item \textit{The technique should be flexible enough to change the summaries based on the priority of the user.}
  \item \textit{Research shows that regular exercise can improve mental health by reducing symptoms of depression and anxiety.}
  \item \textit{The AI-generated summarization approach prioritizes content over the flow of the summary, compromising logical and narrative connections between sentences}
  \item \textit{Participants examining their own summaries can reveal insights into the effectiveness of the human-in-the-loop approach.}
  \item \textit{We will eliminate the limitation by testing ConceptEVA on more diverse papers in the future}
  \item \textit{ConceptEVA was found helpful for content-specific aspects of summarization}
\end{itemize}

\subsection{Example Modifications}
\label{sec:modificationExamples}
We present modifications of the first sentence in the initial draft, using the preset modification macros and default settings for the LLM instructions (except we toggled off ``Check if want modified writing to be based on selected paragraphs and bullet points'').

\subsubsection{Input Text to be Modified}
\textit{In the rapidly evolving field of natural language processing and artificial intelligence, the challenge of effectively summarizing long, complex documents that span multiple topics remains a significant hurdle.}

\subsubsection{Expanded Version}
\textit{In the rapidly evolving field of natural language processing (NLP) and artificial intelligence (AI), the challenge of effectively summarizing long, complex documents that span multiple topics remains a significant hurdle. This is particularly true for academic papers, which often contain dense, multi-disciplinary content that can be difficult to distill into a concise summary. Traditional summarization techniques, which may work well for shorter, single-topic documents, often fall short when applied to these more complex texts.}

\subsubsection{Condensed Version}
\textit{In the fast-paced domains of natural language processing and artificial intelligence, summarizing lengthy, multi-topic documents is a major challenge.}

\subsubsection{Simpler-Terms Version}
\textit{In the fast-changing world of computer programs that understand and use human language, finding a way to briefly explain long, complicated texts that cover many subjects is still a big challenge.}

\subsubsection{Less Dramatic Version}
\textit{In the field of natural language processing and artificial intelligence, summarizing lengthy, multi-topic documents is a notable challenge.}

\subsubsection{More Dramatic Version}
\textit{Title: Tackling the Complexity of Summarizing Multi-Topic Documents in AI}

\textit{In the dynamic world of AI and natural language processing, we're constantly facing the tough challenge of summarizing lengthy, intricate documents covering various topics.}

\end{document}